\documentclass{aastex631}
\usepackage{csquotes}
\usepackage{amsmath}
\usepackage{mathtools}
\usepackage[inline]{enumitem} 
\usepackage{float}

\graphicspath{{./}{figures/}}
\usepackage{filecontents}

\usepackage{hyperref}

\newcommand{\dotdeg}{\rlap{.}^{\circ}}
\newcommand{\dotarcsec}{\rlap{.}\arcsec}

\usepackage{float}
\usepackage{graphicx}
\usepackage{booktabs}

\shortauthors{liu et al.}

\newcommand{\emailaddress}{liu.yuhua.519@s.kyushu-u.ac.jp}

\begin{document}


\title{Dust Polarization of Prestellar and Protostellar Sources in OMC-3}

\author[0009-0009-2263-5502]{yuhua liu}
\affiliation{\textnormal{Department of Earth and Planetary Sciences, Faculty of Sciences, Kyushu University, Fukuoka 819-0395, Japan; \href{mailto:\emailaddress}{\emailaddress}}}

\author[0000-0002-7287-4343]{satoko takahashi}
\affiliation{\textnormal{National Astronomical Observatory of Japan, 2-21-1 Osawa, Mitaka, Tokyo 181-8588, Japan}}
\affiliation{\textnormal{Department of Astronomical Science, School of Physical Sciences, The Graduate University for Advanced Studies, SOKENDAI, 2-21-1 Osawa, Mitaka, Tokyo 181-8588, Japan}}

\author[0000-0002-0963-0872]{masahiro machida}
\affiliation{\textnormal{Department of Earth and Planetary Sciences, Faculty of Sciences, Kyushu University, Fukuoka 819-0395, Japan; \href{mailto:\emailaddress}{\emailaddress}}}

\author[0000-0003-2726-0892]{kohji tomisaka}
\affiliation{\textnormal{National Astronomical Observatory of Japan, 2-21-1 Osawa, Mitaka, Tokyo 181-8588, Japan}}
\affiliation{\textnormal{Department of Astronomical Science, School of Physical Sciences, The Graduate University for Advanced Studies, SOKENDAI, 2-21-1 Osawa, Mitaka, Tokyo 181-8588, Japan}}

\author[0000-0002-3829-5591]{Josep Miquel Girart}
\affiliation{\textnormal{Institut de Ciències de l’Espai (ICE, CSIC), Can Magrans s/n, E-08193 Cerdanyola del Vallès, Catalonia, Spain}}
\affiliation{\textnormal{Institut d’Estudis Espacials de Catalunya (IEEC), E-08034 Barcelona, Catalonia, Spain}}

\author[0000-0002-3412-4306]{paul t. p. ho}
\affiliation{\textnormal{Institute of Astronomy and Astrophysics, Academia Sinica, 11F of Astronomy-Mathematics Building, AS/NTU No. 1, Sec. 4, Roosevelt
Rd, Taipei 10617, Taiwan, R.O.C.}}
\affiliation{\textnormal{East Asian Observatory, 660 N. A’ohoku Place, Hilo, HI 96720, USA}}

\author[0000-0002-6939-0372]{kouichiro nakanishi}
\affiliation{\textnormal{National Astronomical Observatory of Japan, 2-21-1 Osawa, Mitaka, Tokyo 181-8588, Japan}}
\affiliation{\textnormal{Department of Astronomical Science, School of Physical Sciences, The Graduate University for Advanced Studies, SOKENDAI, 2-21-1 Osawa, Mitaka, Tokyo 181-8588, Japan}}

\author[0000-0001-5817-6250]{asako sato}
\affiliation{\textnormal{Department of Earth and Planetary Sciences, Faculty of Sciences, Kyushu University, Fukuoka 819-0395, Japan; \href{mailto:\emailaddress}{\emailaddress}}}


\begin{abstract}
We present the Atacama Large Millimeter/submillimeter Array (ALMA) observations of linearly polarized 1.1\,mm continuum emission at ${\sim}0\dotarcsec14$ (55 au) resolution and CO~($J$=2$-$1) emission at ${\sim}1\dotarcsec5$ (590 au) resolution towards one prestellar (MMS\,4), four Class 0 (MMS\,1, MMS\,3, MMS\,5, and MMS\,6), one Class I (MMS\,7), and one flat-spectrum (MMS\,2) sources in the Orion Molecular Cloud\,3 region. The dust disk-like structures and clear CO outflows are detected towards all sources except for MMS\,4. The diameters of these disk-like structures, ranging from 16 au to 97 au, are estimated based on the deconvolved full width half maximum (FWHM) values obtained from the multi-Gaussian fitting. Polarized emissions are detected towards MMS\,2, MMS\,5, MMS\,6, and MMS\,7, while no polarized emission is detected towards MMS\,1, MMS\,3, and MMS\,4. MMS\,2, MMS\,5, and MMS\,7 show organized polarization vectors aligned with the minor axes of the disk-like structures, with mean polarization fractions ranging from 0.6\% to 1.2\%. The strongest millimeter source, MMS\,6, exhibits complex polarization orientations and a remarkably high polarization fraction of $\sim$10$\%$ around the Stokes $I$ peak, and 15$-$20$\%$ on the arm-like structure, as reported by \cite{takahashi2019alma}. The origins of the polarized emission, such as self-scattering and dust alignment due to the magnetic field or radiative torque, are discussed for individual sources. Some disk-like sources exhibit a polarized intensity peak shift towards the nearside of the disk, which supports that the polarized emission originates from self-scattering.
\end{abstract}


\section{Introduction} \label{sec:intro}
Magnetic fields are considered to play an important role in regulating star formation \citep[e.g.,][]{shu1987star,hildebrand1988magnetic,crutcher2012magnetic}. The role of magnetic fields could change over the size scale and the timescale in the star formation process \citep{koch2022multiscale}. However, the specific role of magnetic fields during star formation has not been fully understood. 

Polarized emission from dust grains at (sub)millimeter wavelengths is commonly used as a tracer to probe the configuration of magnetic fields, as the non-spherical dust grains tend to align their long axes perpendicular to the magnetic fields \citep[e.g.,][]{hildebrand1988magnetic,lazarian2007tracing,andersson2015,hull2019interferometric}. The configuration of magnetic fields has been studied through the dust polarization observations on large-scale, spanning from the molecular cloud-scale (a few pc) down to the protostellar envelope-scale (a few 1000 au) by using the Submillimeter Common-User Bolometer Array (SCUBA) camera on the James Clerk Maxwell Telescope \citep[JCMT; e.g.,][]{matthews2000magnetic}, the Berkeley-Illinois-Maryland Association (BIMA) millimeter array \citep[e.g.,][]{rao1998bima,lai2001bima,matthews2005multiscale}, the Submillimeter Array \citep[SMA; e.g.,][]{girart2006sma,rao2009sma,rao2014sma,girart2013sma,tang2013sma,zhang2014sma}, the Combined Array for Research in Millimeter-wave Astronomy \citep[CARMA; e.g.,][]{stephens2013carma,hull2014carma}, and the Atacama Large Millimeter/submillimeter Array \citep[ALMA; e.g.,][]{koch2018alma, koch2022multiscale,liu2020alma}.

Recently, ALMA has achieved higher sensitivity and resolution enabling the dust polarization observations on the disk-scale \citep[$\sim$10$-$100 au; e.g.,][]{kataoka2016submillimeter,kataoka2017evidence,stephen2017,ohashi2018,lee2018alma,cox2018alma,sadavoy2019}. These observations suggest that there are alternative mechanisms producing polarized emission such as: (i) Self-scattering of thermal dust emission \citep[e.g.,][]{kataoka2015millimeter,kataoka2017evidence,yang2017scattering}. When the dust grains have the comparable sizes with the observing wavelengths ($a_{\textnormal{\scriptsize{max}}}\sim\lambda/2\pi$, where $a_{\textnormal{\scriptsize{max}}}$ is the maximum size of the dust grain and $\lambda$ is the observing wavelength) and the radiation field has an anisotropic distribution \citep{kataoka2015millimeter}, the thermal emission can be scattered by the dust grains, which induces polarized emission; (ii) Dust grains alignment due to radiative torque with a different direction from the magnetic field \citep[e.g.,][]{Lazarian&Hoang2007RAT,tazaki2017radiative,kataoka2017evidence}. The dust grains are expected to align their minor axes to the radiative flux and the polarization orientation traces the direction of radiation anisotropy or the radiative flux; in addition to the observational studies, theoretical studies also proposed
(iii) Grain alignment by gas flow due to the mechanical alignment introduced by \cite{gold1952MNRAS.112..215G}, \cite{lazarian2007mechanical}, \cite{hoang2018}, and \cite{kataoka2019gasflow}. With this mechanism, dust grains are aligned with the gas flow. The orientations of the polarization vectors will depend on the Stokes numbers of the dust grains \citep{kataoka2019gasflow}; (iv) Reversal of self-scattering proposed by \cite{yang2020reversal}. \cite{yang2020reversal} found the reversed polarization orientation, where the polarization vectors are parallel to the major axis of the disk. This pattern has not been directly detected by ALMA due to the low polarization fraction of $\sim$0.1\%; (v) Magnetic field alignment in the Mie regime ($2\pi a/\lambda\gtrsim1$, where $a$ is the radius of the particle and $\lambda$ is the wavelength of the light) proposed by \cite{guillet2020mie}. In contrast to the magnetic field alignment mentioned earlier (e.g., Rayleigh regime: $2\pi a/\lambda\ll1$), the polarization vectors of the dust thermal emission in the Mie regime tend to be parallel to the magnetic field, also known as negative polarized emission.

The observations towards HD\,142527, HL\,Tau, and some sources from Ophiuchus (e.g., GSS\,30\,IRS\,1, Oph-emb-6, and IRS\,63) suggest that polarization on disk-scale is more consistent with that expected from self-scattering and radiative torque than other mechanisms \citep[e.g.,][]{kataoka2016submillimeter,kataoka2017evidence,stephen2017,cox2018alma,sadavoy2019}. Moreover, polarization due to self-scattering could be detected at the inner region of the disk \citep{kataoka2016submillimeter}. Therefore, dust polarization due to self-scattering is also an effective tool to constrain the dust grain size in the optically thick region \citep{kataoka2016submillimeter}. Grain size constraints are important because they are related to the dust growth, which is the first step towards understanding the composition and properties of protoplanetary disks and the formation of planets within them. 

Our targets are the seven prestellar and protostellar sources, MMS\,1$-$7, in the Orion Molecular Cloud\,3 (OMC-3) region, which are located in the northern part of the nearest giant molecular cloud, the Orion-A Giant Molecular Cloud, at a distance of 393 pc \citep{tobin2020vla}. Previous studies have identified several millimeter sources associated with prestellar and early stage of Class 0 sources embedded within the OMC-3 region \citep{chini1997dust}. The Spatial structure of the dust continuum sources was revealed in the size scale down to $\sim$1800 au by the SMA observations \citep{takahashi2013hierarchical}. The evolutionary stages of the sources are primarily identified based on the spectral energy distribution (SED) study by \cite{furlan2016herschel}. 

The dust polarization of the OMC-3 region on large-scale was firstly observed by \cite{matthews2000magnetic} and \cite{matthews2001magnetic} in 850\,$\mu$m continuum with the SCUBA on the JCMT, and then by \cite{houde2004tracing} in 350\,$\mu$m continuum with the Hertz polarimeter and Submillimeter High Angular Resolution Camera II (SHARC II), showing that the polarization orientations are highly ordered and aligned along the filament. Since the dust grains are considered to align with the magnetic field on this scale, this indicates that the interstellar magnetic field is perpendicular to the long axis of OMC-3.

Although polarization studies on the cloud- and envelope-scale have been done extensively in the OMC-3 region, the polarization on disk-scale towards the individual sources in this region has not been widely studied except for MMS\,6 \citep{takahashi2019alma,Gouellec2020,liubaobab2021}. In this paper, we present polarization observations on disk-scale along with the molecular line observations towards seven millimeter sources, MMS\,1$-$7, located within the OMC-3 region. The source names including those from other catalogues, coordinates, Herschel Orion Protostar Survey (HOPS) IDs, bolometric luminosity ($L_{\textnormal{bol}}$), outflow associations, and evolutionary stages (classifications) for all observed sources are summarized in Table~\ref{table:targets-summary}. Our aim is to examine the origins of the polarization for these young embedded sources and also to study how the polarization characteristics change at different evolutionary stages.

We describe the ALMA observations, data reduction, and imaging in Section~\ref{sec:obser}. The results are summarized in Section~\ref{results}. We discuss the polarization origins and examine how the source size, the outflow size, and the polarization fraction change at different evolutionary stages in Section~\ref{discussion}. Finally, we present the conclusions in Section~\ref{conslusion}.

\begin{table*}[ht!]
{\scriptsize
\begin{center}
\caption{\small Source Summary}
\label{table:targets-summary}
\begin{tabular}{ccccccccc}
\hline\hline \noalign {\smallskip}
Source Name & R.A.$^a$ & Decl.$^a$ & Other Source Name&HOPS ID$^d$&$L_{\textnormal{bol}}$$^d$&Outflow & Classification$^e$ & Comments\\
 & (J2000) & (J2000) &(350\,$\mu$m$^b$, 850\,$\mu$m$^c$) & &($L_{\odot}$)&   [Yes/No] &  & \\
\hline
MMS\,1 & 5 35 18.03 & $-$5 00 17.77 &CSO\,5, SMM\,2  &     &$<55^f$&Yes & Class 0$^g$ &\\
MMS\,2 & 5 35 18.30 & $-$5 00 33.01 &CSO\,6, SMM\,3  &HOPS-92&17.6 &Yes & Flat-spectrum & Triple source  \citep{tobin2020vla} \\
MMS\,3 & 5 35 18.98 & $-$5 00 51.63 &CSO\,7, SMM\,4  &HOPS-91&3.6  &Yes & Class 0 & \\   
MMS\,4 & 5 35 20.88 & $-$5 00 56.25 &CSO\,8, SMM\,5  &     &$<56^f$&No & Prestellar &\\
MMS\,5 & 5 35 22.47 & $-$5 01 14.43 &CSO\,9, SMM\,6  &HOPS-88&13.8 &Yes & Class 0 &\\
MMS\,6 & 5 35 23.42 & $-$5 01 30.35 &CSO\,10, SMM\,8 &HOPS-87&31.6 &Yes & Class 0 &\\
MMS\,7 & 5 35 26.56 & $-$5 03 55.04 &CSO\,12, SMM\,11&HOPS-84&42.9 &Yes & Class I & Binary source  \citep{tobin2020vla}\\
\hline \noalign {\smallskip}
\end{tabular}
\end{center}}
\footnotesize $^a${The location of the source denotes the phase center of the ALMA observations.}\\
$^b${CSO sources reported by \cite{lis1998350}.}\\
$^c${SMM sources reported by \cite{takahashi2013hierarchical}.}\\
$^d${HOPS ID and $L_{\textnormal{bol}}$ are obtained from \cite{furlan2016herschel}, which can also be obtained from HOPS catalog (\url{https://planetstarformation.iaa.es/OMC-3}).}\\
$^e${The classification of the protostellar sources are defined based on the SED study presented by \cite{furlan2016herschel}.}\\
$^f${No counterpart source is identified in the HOPS catalog, hence $L_{\textnormal{bol}}$ is referenced from \cite{chini1997dust}.}\\
$^g${We classify MMS\,1 as a Class 0 source since the detection of a very compact and collimated jet by Takahashi et al.\,(2023a, submitted to ApJ).}\\
\end{table*}

\section{Observations and Data Reduction} \label{sec:obser}

\subsection{1.1 mm continuum}
The ALMA 1.1\,mm continuum observations at Band 6 were conducted during Cycle 4 (2015.1.00341.S; P.I. S. Takahashi) on 2016 September 18, October 9, and October 11. The ALMA observations were obtained in full polarization, targeting MMS\,1$-$7 in the OMC-3 region. The observational parameters are listed between the second and fourth columns of Table~\ref{table:parameters}. The total on-source time of all sources was between 33 and 35 minutes. Between 38 and 44 of the 12-m antennas were operated during the observations. The full width at half maximum (FWHM) of the ALMA primary beam is $\sim$22$\arcsec$. The projected baselines range from 15 to 3144 m, and the minimum baseline yields in a maximum recoverable size of $\sim$9$\dotarcsec1$. Four spectral windows (spws) in Time Division Mode (TDM), each with a total bandwidth of 2 GHz, were allocated for the continuum observations. The number of the effective channels varies for each spw because we flagged out some of the channels where molecular line emissions were detected. After subtraction of those, the obtained total effective bandwidth for the continuum observation is $\sim$6.4$-$6.9 GHz. All the raw data were reduced with the Common Astronomy Software Applications \citep[CASA;][]{bean2022casa} package version 4.7.0. 

In order to enhance the image quality and suppress the noise, we performed self-calibration on phase by using CASA tasks ``\texttt{gaincal}'' and ``\texttt{applycal}'' for MMS\,2, MMS\,5, MMS\,6, and MMS\,7, respectively. The solution interval for self-calibration was optimized for each source based on its structure and signal-to-noise ratio (S/N). We did not apply self-calibration for MMS\,1, MMS\,3, and MMS\,4 because there were no significant improvements in both image quality and S/N after applying self-calibration. The CLEANed images of Stokes \textit{I}, \textit{Q}, and \textit{U} for all sources were made using a CASA task ``\texttt{tclean}'' with Briggs weighting (robust = 0.5). The image of Stokes $I$ shows the total intensity. The linearly polarized intensity (PI) was derived from Stokes $Q$ and $U$ as $\textnormal{PI}=\sqrt{Q^2+U^2-\sigma_{\textnormal{PI}}}$, where $\sigma_{\textnormal{PI}}=\sqrt{\sigma_{\textnormal{Q}}^2+\sigma_{\textnormal{U}}^2}\sim26\mu$Jy\,beam$^{-1}$. Each of the PI images obtained using the pixels with an emission detection is $\ge3\sigma_{\textnormal{PI}}$. The polarization fraction \({P_{\textnormal{frac}}}=\textnormal{PI}/I\) was obtained using the debiased PI and Stokes $I$. The position angle of polarization vector (also polarization angle) \(\chi=\frac{1}{2}\textnormal{arctan}(U/Q)\) was derived where the detection is $\ge3\sigma_{\textnormal{PI}}$. The PI, ${P_{\textnormal{frac}}}$, and $\chi$ were computed using a CASA task ``\texttt{immath}''. The 1$\sigma$ error of the polarization fraction is calculated as
\begin{equation}
\sigma_{P_{\textnormal{frac}}}=\sqrt{\frac{Q^2\Delta Q^2}{(Q^2+U^2)}+\frac{U^2\Delta U^2}{(Q^2+U^2)}+\frac{(Q^2+U^2)\Delta I^4}{I^4}}.
\end{equation}
The average value of $\sigma_{P_{\textnormal{frac}}}$ among our sources is estimated to be 0.01$^{+0.01}_{-0.01}$\%, where $I$, $Q$, and $U$ are the flux values, and $\Delta I$, $\Delta Q$, and $\Delta U$ are their respective errors. The 1$\sigma$ error of the polarization angle is calculated as 
\begin{equation}
{\sigma_{\chi}}=\frac{180}{\pi}\times\sqrt{\left(\frac{\Delta U^2}{4Q^2} + \frac{U^2\Delta Q^2}{4Q^4}\right)\times \left(1+\left(\frac{U}{Q}\right)^2\right)^{-2}}.
\end{equation}
The average value of $\sigma_{\chi}$ among our sources is estimated to be 3$\dotdeg$3$^{+3.6}_{-2.1}$. According to ALMA Cycle 4 Technical Handbook\footnote{\url{https://almascience.org/documents-and-tools/cycle4/alma-technical-handbook}}, the 1$\sigma$ systematic error of the linear polarization fraction is 0.03\% for the compact sources and 0.1\% for the extended sources within one-third of the primary beam (which corresponds to a minimum detectable polarization fraction of 0.1\% for the compact sources and 0.3\% for the extended sources). Table~\ref{tab:continuum statistics} summarizes the rms noise level measured in each Stokes component, synthesized beam, position angle (P.A.), and solution interval for self-calibration. The achieved rms noise level in Stokes $Q$ and $U$ is close to the theoretical noise level of $\sim$21 $\mu$Jy\,beam$^{-1}$, and Stokes $I$ images are dynamic range limited.

 
\subsection{Molecular line}
The ALMA molecular line observations towards MMS\,1$-$7 at Band 6 were performed on 2016 January 29 during Cycle 4 (2015.1.00341.S; P.I. S. Takahashi). Four spectral lines include CO~($J$ = 2$-$1; 230.538 GHz), N$_{2}$D$^{+}$~($J$ = 3$-$2; 231.322 GHz), SiO~($J$ = 5$-$4; 217.105 GHz), and C$^{18}$O~($J$ = 2$-$1; 219.560 GHz), respectively. In this paper, we present CO~($J$ = 2$-$1) data to show the molecular outflow and the extremely high-velocity (EHV) jet associated with each observed source\footnote{Some of the molecular line data were already presented in our previous studies \citep{takahashi2019alma,matsushita2019very,morii2021revealing}.}. Here, ``outflows'' refer to the CO emissions detected in the low-velocity range of $v_{\textnormal{LRS}}-v_{\textnormal{sys}}\lesssim\pm$50 km\,s$^{-1}$, while ``EHV jets'' refer to the CO emissions detected in the high-velocity range of $v_{\textnormal{LRS}}-v_{\textnormal{sys}}\gtrsim\pm$50 km\,s$^{-1}$. The observational parameters are listed in the fifth column of Table~\ref{table:parameters}. The raw data were reduced with the CASA \citep{bean2022casa} package version 4.6.0. 

The CLEANed images of CO~($J$ = 2$-$1) emissions were made using a CASA task ``\texttt{tclean}'' with Briggs weighting (robust = 0.5). The velocity interval of 2 km\,s$^{-1}$ was used for imaging the outflows associated with MMS\,2 and MMS\,7. The velocity interval of 5 km\,s$^{-1}$ was used for imaging both the outflows and the EHV jets associated with MMS\,5 and MMS\,6. The velocity interval of 1 km\,s$^{-1}$ was used for imaging the CO emission associated with MMS\,4. The CLEANed images of CO emissions for MMS\,3 and MMS\,1 were obtained from \cite{morii2021revealing} and Takahashi et al. (2023a, submitted to ApJ), respectively. The integrated intensity maps were made using a CASA task ``\texttt{immoment}'' with the 3$\sigma$ threshold. Table~\ref{tab:line statistics} summarizes the rms noise level, synthesized beam, position angle (P.A.), and velocity interval for the CO images.

\begin{table*}[ht!]
{\scriptsize
\begin{center}
\caption{\small ALMA Observations}
\label{table:parameters}
\begin{tabular}{ccccc}
\hline\hline \noalign {\smallskip}
Parameters &  \multicolumn{3}{c}{Full Polarization Continuum}  & CO ($J$ = 2$-$1) \\
\hline
Observed sources & \multicolumn{2}{c}{MMS\,1, MMS\,2, MMS\,5, MMS\,6, and MMS\,7}   & MMS\,3 and MMS\,4  & MMS\,1$-$7\\
Observing date (YYYY-MM-DD)	& 2016-10-09	& 2016-10-11	& 2016-09-18  & 2016-01-29\\
Number of antennas			&44  			& 42			& 38	& 40	\\
FWHM of Primary beam (arcsec)  & 22			& 22	    & 22	& 26	\\
PWV (mm)  & 0.44$-$0.90 	  &0.55$-$0.57     &0.84$-$0.91    & 2.46$-$2.60   	\\
Phase stability rms (degree)$^a$ &16$-$56 		&10$-$15 		& 20   & 9$-$11  	\\
Polarization calibrator		& J0522-3627	& J0522-3627    &J0522-3627   &	- 	\\
Bandpass calibrators  	    & J0510+1800    & J0510+1800    & J0510+1800  &J0522-3627 \\
Flux calibrator  		    & J0423-0120    & J0423-0120    & J0423-0120  &J0522-3627 \\
Phase calibrators (separation from the target)$^b$  &J0532-0307(2$\dotdeg$1)  &J0541-0541(1$\dotdeg$7)   &J0541-0211(3$\dotdeg$2) 	&J0541-0541(1$\dotdeg$7) \\
{Central frequency USB/LSB (GHz)} & {257/273} & {257/273} & {257/273} & {220/231}\\
Total continuum bandwidth; USB+LSB (GHz)	&6.875   &6.875	&6.438 	& - \\
Projected baseline ranges (m)   &    19$-$3144	& 19$-$3144 &  15$-$2483 & 15$-$331\\
Maximum recoverable size (arcsec)$^c$	& 7.3	& 7.3 	& 9.1	  & 9.1      \\
On-source time (minutes)      	& 20    & 13			& 35	  & 4		\\

\hline \noalign {\smallskip}
\end{tabular}

\end{center}
}
\footnotesize $^a${Antenna-based phase differences measured on the bandpass calibrator.}\\
\footnotesize $^b${The phase calibrator was observed every 8 minutes for the observations.}\\
\footnotesize $^c${Our observations were insensitive to emission more extended than this size scale structure at the 10\% of the total flux density (ALMA Cycle 4 Technical Handbook).}\\ 
\end{table*}

\begin{table*}[ht!]
{\small
\begin{center}
\caption{\small Summary of the 1.1\,mm Continuum Image }
\label{tab:continuum statistics}
\begin{tabular}{ccccccc}
\hline\hline \noalign {\smallskip}
Source Name & rms Noise Level  & Synthesized Beam$^b$ & P.A. & Is Self-calibration$^c$  &  Solution Interval for\\
& Stokes \textit{I $^a$, Q, U} & & & Applied? & Self-calibration  \\
&($\mu$Jy\,beam$^{-1}$)& (arcsec $\times$ arcsec) &(degree)&(Yes/No) & (min)\\
\hline
MMS\,1 & 35, 19, 18 & 0.14$\times$0.13& $-$53 &No &-\\
MMS\,2 & 38, 18, 18	& 0.14$\times$0.13& $-$54 &Yes &15 \\
MMS\,3 & 25, 19, 19	& 0.17$\times$0.14&  15 &No &-\\
MMS\,4 & 24, 19, 19	& 0.17$\times$0.14&  15 &No &-\\
MMS\,5 & 60, 19, 19 & 0.14$\times$0.13& $-$52 &Yes &60, 30$^d$	\\
MMS\,6 & 76, 19, 19 & 0.14$\times$0.13& $-$55 &Yes &16 \\
MMS\,7 & 49, 18, 19 & 0.14$\times$0.13& $-$56 &Yes &17 \\
\hline \noalign {\smallskip}
\end{tabular}
\end{center}
}
\footnotesize $^a${Stokes $I$ images are dynamic range limited.}\\
\footnotesize $^b${Briggs weighting (robust = 0.5) was used for the imaging.}\\
\footnotesize $^c${Self-calibration was applied for the phase.}\\
 $^d${We applied two rounds self-calibration on MMS\,5 with long solution interval of 60 minutes and 30 minutes, respectively, because the short solution interval gives large phase fluctuation.}
\end{table*}

\begin{table*}[ht!]
{\scriptsize 
\begin{center}
\caption{Summary of the CO~($J$ = 2$-$1) Image}
\label{tab:line statistics}
\begin{tabular}{lccccc}
\hline\hline \noalign {\smallskip}
Image Name& rms Noise Level & Synthesized Beam$^a$& P.A. &Velocity Interval &Figure\\
&(Jy\,beam$^{-1}$)&(arcsec $\times$ arcsec)&(degree)& (km\,s$^{-1}$)\\
\hline \noalign {\smallskip}
MMS\,2 Integrated intensity&0.32&1.5$\times$0.9&$-$78&-&\ref{fig-co-1}\,(a)\\
MMS\,5 Integrated intensity&0.39, 0.087$^b$&1.5$\times$0.9&$-$77&-&\ref{fig-co-1}\,(b)\\
MMS\,6 Integrated intensity&0.0925&1.5$\times$0.9&$-$78&-&\ref{fig-co-1}\,(c)\\
MMS\,7 Integrated intensity&0.11&1.5$\times$0.9&$-$77&-&\ref{fig-co-1}\,(d)\\
\hline \noalign {\smallskip}
MMS\,2 Channel map&0.0058&1.5$\times$0.9& $-$78&2&\ref{channel-mms2}\\
MMS\,5 Channel map&0.0039&1.5$\times$0.9& $-$77&5&\ref{channel-mms5} \citep{matsushita2019very}\\
MMS\,6 Channel map&0.0039&1.5$\times$0.9& $-$78&5&\ref{channel-mms6}\\
MMS\,7 Channel map&0.0058&1.5$\times$0.9& $-$77&2&\ref{channel-mms7}\\
MMS\,1 Integrated intensity&0.12&1.5$\times$0.9&$-$78&-&\ref{mom0-mms1} (Takahashi et al. 2023a, submitted to ApJ)\\
MMS\,3 Integrated intensity&0.145&1.7$\times$1.0&$-$74&-&\ref{mom0-mms3} \citep{morii2021revealing}\\
MMS\,4 Integrated intensity&0.12&1.6$\times$0.9&$-$78&-&\ref{mom0-mms4}\\
\hline \noalign {\smallskip}
\end{tabular}
\end{center}
\footnotesize $^a${Briggs weighting (robust = 0.5) was used for the imaging.}\\
\footnotesize $^b${Different rms noise levels correspond to the CO integrated intensities obtained by integrating over different velocity ranges in the CO channel maps, which are associated with the outflow and the EHV jet, respectively. }\\
}
\end{table*}

\section{Results} \label{results}
This section is divided into three subsections. In Section~\ref{continuum and co}, we summarize the observations of the 1.1\,mm continuum (Stokes $I$, $Q$, and $U$) and CO~($J$ = 2$-$1) emissions for individual sources. In Section~\ref{relation}, we briefly describe the relation between the cloud- and disk-scale polarization observations. In Section~\ref{mass}, we present the estimated physical properties of each millimeter source, such as the dust mass, number density, and column density.

\subsection{The 1.1 mm continuum emission and CO~(J = 2$-$1) emission} \label{continuum and co}

Figure~\ref{fig-stokesIQU-3} shows the Stokes $I$, $Q$, and $U$ for each source. We have detected all the seven sources in Stokes $I$. Out of the seven sources, MMS\,2 (Flat-spectrum), MMS\,5 (Class 0), MMS\,6 (Class 0), and MMS\,7 (Class I) are detected in Stokes $Q$ and $U$, while MMS\,1 (Class 0), MMS\,3 (Class 0), and MMS\,4 (Prestellar) are not detected in Stokes $Q$ and $U$. These three sources that are not detected in Stokes $Q$ and $U$ have the weakest Stokes $I$ fluxes among the observed seven sources. In addition, we estimate the 3$\sigma$ upper limits of the polarization fractions for these three sources by assuming that PI is detected at the Stokes $I$ peak and is comparable to the 3$\sigma$ level of the Stokes $Q$ and $U$. The estimated 3$\sigma$ upper limits of the polarization fractions for MMS\,1, MMS\,3, and MMS\,4 are derived to be $\sim$0.5$\%$, $\sim$2.1$\%$, and $\sim$3.7$\%$, respectively. These 3$\sigma$ upper limits are higher than the minimum detectable linear polarization fraction of 0.1\% for the compact sources and 0.3\% for the extended sources (ALMA Cycle 4 Technical Handbook). This suggests that our observations are likely limited in the sensitivity.

Table~\ref{tab:Summary of the fitting results} describes the fitting results of the 1.1\,mm continuum sources. In order to characterize the structures of the sources detected in Stokes $I$, we performed the two-dimensional (2D) Gaussian fitting. The total flux density, peak flux, and deconvolved size of each source were obtained by using a CASA task ``\texttt{imfit}''. The residual after subtracting the 2D Gaussian fitted model is $\lesssim2\%$ with respect to the Stokes $I$ peak flux for each source. Our 2D Gaussian fitting analysis indicates that some compact sources have deconvolved sizes smaller than the synthesized beam with considerably small errors, we consider these sources are spatially resolved. Fitting reliability of the compact sources are discussed in Appendix~\ref{fitting}. The geometric mean diameters of the fitted compact components range from $\sim$0.04$\arcsec$ to $\sim$0.93$\arcsec$, which correspond to the linear sizes of $\sim$16 to $\sim$364 au, respectively. The geometric mean diameters of the fitted extended components range from $\sim$0.45$\arcsec$ to $\sim$4.02$\arcsec$, which correspond to the linear sizes of $\sim$177 to $\sim$1581 au. For simplicity, we will mainly refer to the linear size of each fitted component in the subsequent sections. We estimate the inclination angle using the axis ratio of each fitted component, assuming that they are geometrically thin disk-like structures. The inclination angles ($i$) are listed in the last column of Table~\ref{tab:Summary of the fitting results}. Here, the face-on and edge-on configurations are defined with an inclination angle of $i=0^{\circ}$ and $i=90^{\circ}$, respectively.

The left panels of Figure~\ref{fig-co-1} show the CO~($J$ = 2$-$1) emission overlaid with PI. In these panels, we present the images of the integrated CO emissions along with the PI for MMS\,2, MMS\,5, MMS\,6, and MMS\,7. The outflows associated with MMS\,2, MMS\,5, MMS\,6, and MMS\,7, which are traced by the CO emissions detected in the low-velocity range of $v_{\textnormal {LSR}}-v_{\textnormal {sys}}$ $\lesssim \pm 50$ km\,s$^{-1}$, show a cavity-like structure with a wide opening angle. Furthermore, we have detected CO emissions in the EHV jets associated with MMS\,5 and MMS\,6 in the high-velocity range of $v_{\textnormal {LSR}}-v_{\textnormal {sys}}$ $\gtrsim \pm$50 km\,s$^{-1}$ \citep{matsushita2019very,gomez2019warm,takahashi2019alma}. The channel maps of the CO emissions for MMS\,2, MMS\,5, MMS\,6, and MMS\,7, as well as the integrated intensity images of the CO emissions for the non-polarized detected sources MMS\,1, MMS\,3, and MMS\,4 are presented in Appendix~\ref{co-image}.

The right panels of Figure~\ref{fig-co-1} show the PI overlaid with the Stokes $I$ and the polarization vectors (hereafter $E$-vectors). The outflow axes are also shown in the right panels of Figure~\ref{fig-co-1}, which are determined by eye from the integrated intensity CO images in the left panels. PI of each source shows a compact structure concentrated around the Stokes $I$ peak. In addition to the compact PI structures, we have also detected an extended arm-like structure in MMS\,6 that extends up to $\sim$1400 au \citep{takahashi2019alma}. PI peak positional shifts towards the red-shifted CO emissions are observed in MMS\,2, MMS\,5, MMS\,6, and MMS\,7. The positional accuracy is given by $\sim$$(1/2)\times(\theta/S/N)$, where $\theta$ is the synthesized beam size, and S/N is the signal-to-noise ratio of the PI image. Note that this positional accuracy is mainly based on the S/N. We are aware that phase noise effects may also induce the systematic positional uncertainties, possibly at different levels in the X and Y linearly orthogonal polarization components of the received signal. These X and Y orthogonal polarization components are correlated to produce the cross-correlation visibilities (ALMA Cycle 4 Technical Handbook). Since the Stokes parameters are derived from different combinations of these cross-correlation visibilities, we expect that the systematic positional uncertainties induced by the phase noise effects may vary slightly for the different Stokes parameters. Therefore, the PI peak positional shift might not only be influenced by the positional accuracy based on S/N but also by the systematic positional uncertainties induced by the phase noise effects. For simplicity, we only discuss the PI peak positional shift with the positional accuracy in this paper. The PI peak shifts with respect to the Stokes $I$ peaks among MMS\,2, MMS\,5, MMS\,6, and MMS\,7 are larger than their respective positional accuracies by a factor of $\sim$4$-$110. $E$-vector orientations differ among the individual sources, such as the direction parallel to the minor axis of the disk and the azimuthal direction. The position angle distributions of the $E$-vectors are presented in Figure~\ref{hist-pa}. To plot these histograms, hyper-Nyquist sampling \citep[e.g.,][]{hull2020} was performed to regrid the images of Stokes $I$, $Q$, and $U$ to 4 pixels per synthesized beam. The regridded images of Stokes $Q$ and $U$ were used to derive the $E$-vector position angles for each source. All histograms include the pixels whose $\sqrt{Q^2+U^2}$ is $\ge3\sigma$. The polarization characteristics will be explained separately by each source in the following subsections.

Figure~\ref{fig-fraction} shows polarization fraction overlaid with Stokes $I$ and the inferred magnetic field vectors (hereafter $B$-vectors). Except for MMS\,6, low polarization fractions of $\sim$0.4$-$2.4\% are detected around the Stokes $I$ peaks of MMS\,2, MMS\,5, and MMS\,7. High polarization fractions of $\sim$0.4$-$10.4\% and $\sim$2.7$-$47.4\%, are detected around the Stokes $I$ peak and on the arm-like structure of MMS\,6, respectively \citep{takahashi2019alma}. Figure~\ref{hist-pfrac} presents the histograms plotted from the regridded images of the polarization fractions. We performed hyper-Nyquist sampling using a similar approach as explained for Figure~\ref{hist-pa} to create the regridded images of the polarization fractions. We also derived the mean polarization fractions from these histograms. We calculated the weighted mean value of the polarization fractions from each histogram. We then derived the standard deviation of the mean value, and calculated the statistical errors, in which the statistical errors were obtained by using the standard deviation and dividing it by the square root of the total count in each histogram. The final errors shown in the figure include both statistical and systematic errors, where the systematic error is 0.03\% for the compact sources and 0.1\% for the extended sources (ALMA Cycle 4 Technical Handbook). The mean polarization fractions for all sources range between $\sim$0.6\% and $\sim$15.7\%.


In this paper, we do not present the line tracers that kinematically define the disk and envelope. Hence the definition of the disk and envelope is based on the morphological structure measured from the continuum emission. We define the disk as a structure with the linear size of $\sim$10$-$200 au, which corresponds to the most compact component fitted by 2D Gaussian, and we define the envelope as a structure with the linear size of $\sim$300$-$3000 au, which corresponds to the fitted extended component. In this paper, we consider a disk to be nearly face-on when the inclination angle is $i\le30^{\circ}$, a disk to be nearly edge-on when the inclination angle is $i\ge60^{\circ}$, and a disk to be inclined when the inclination angle is $31^{\circ}\le i\le59^{\circ}$.

\renewcommand{\thefigure}{\arabic{figure} (Continued)}
\begin{figure*}[]
\gridline{\hspace{-1.5\baselineskip}
          \fig{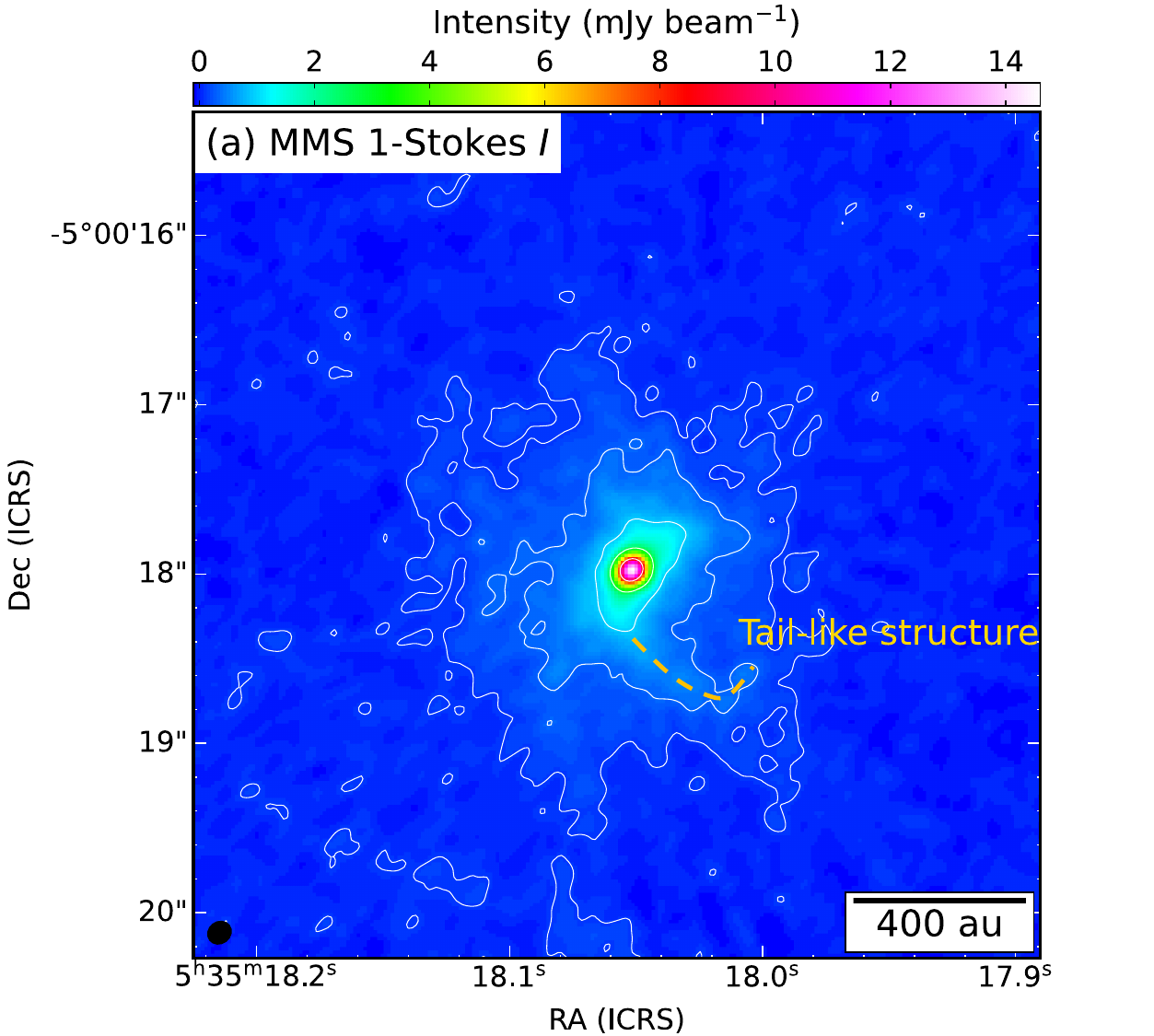}{0.43\textwidth}{}
          {}\hspace{-2.1\baselineskip}
          \fig{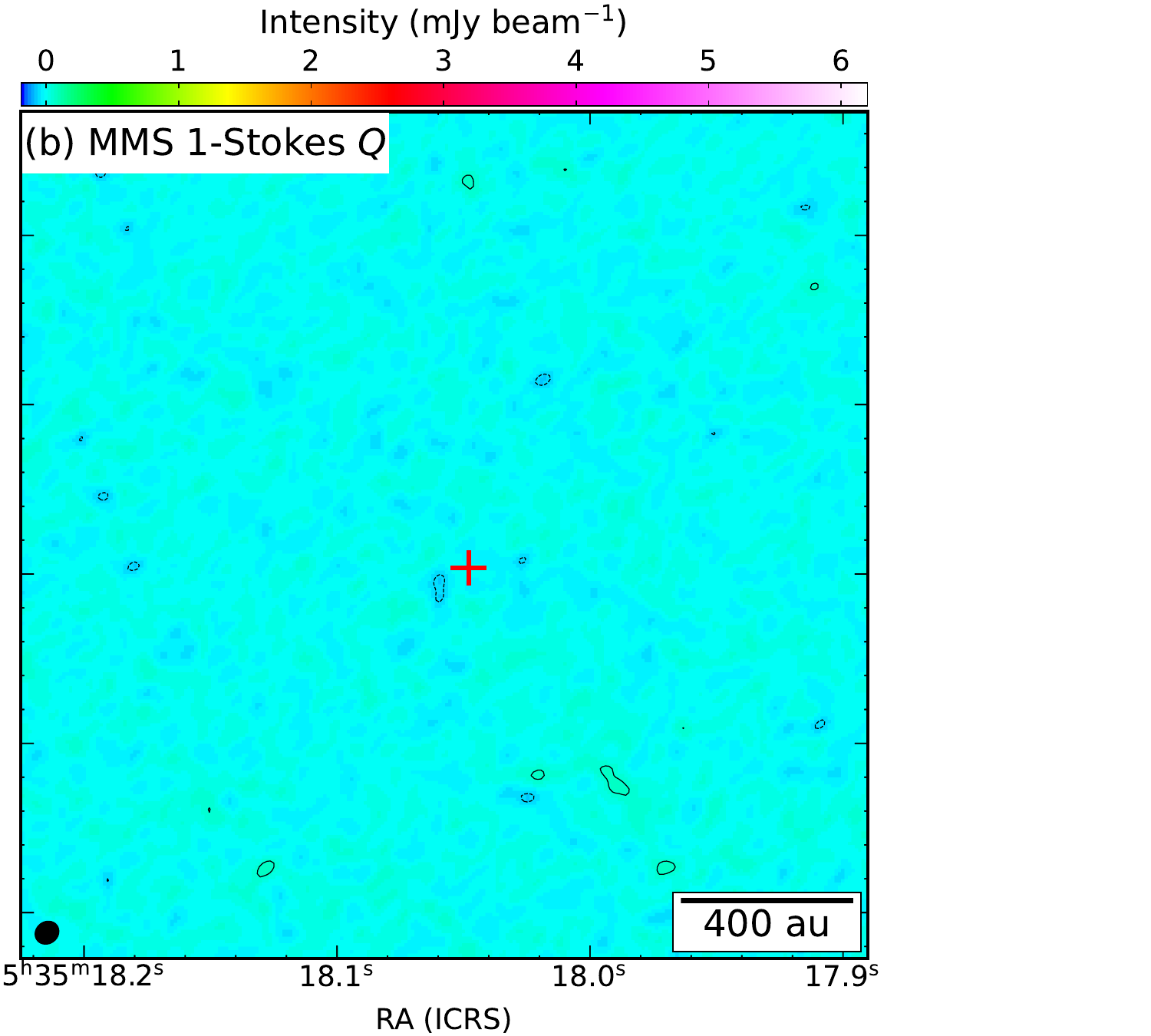}{0.43\textwidth}{}
          {}\hspace{-5\baselineskip}
          \fig{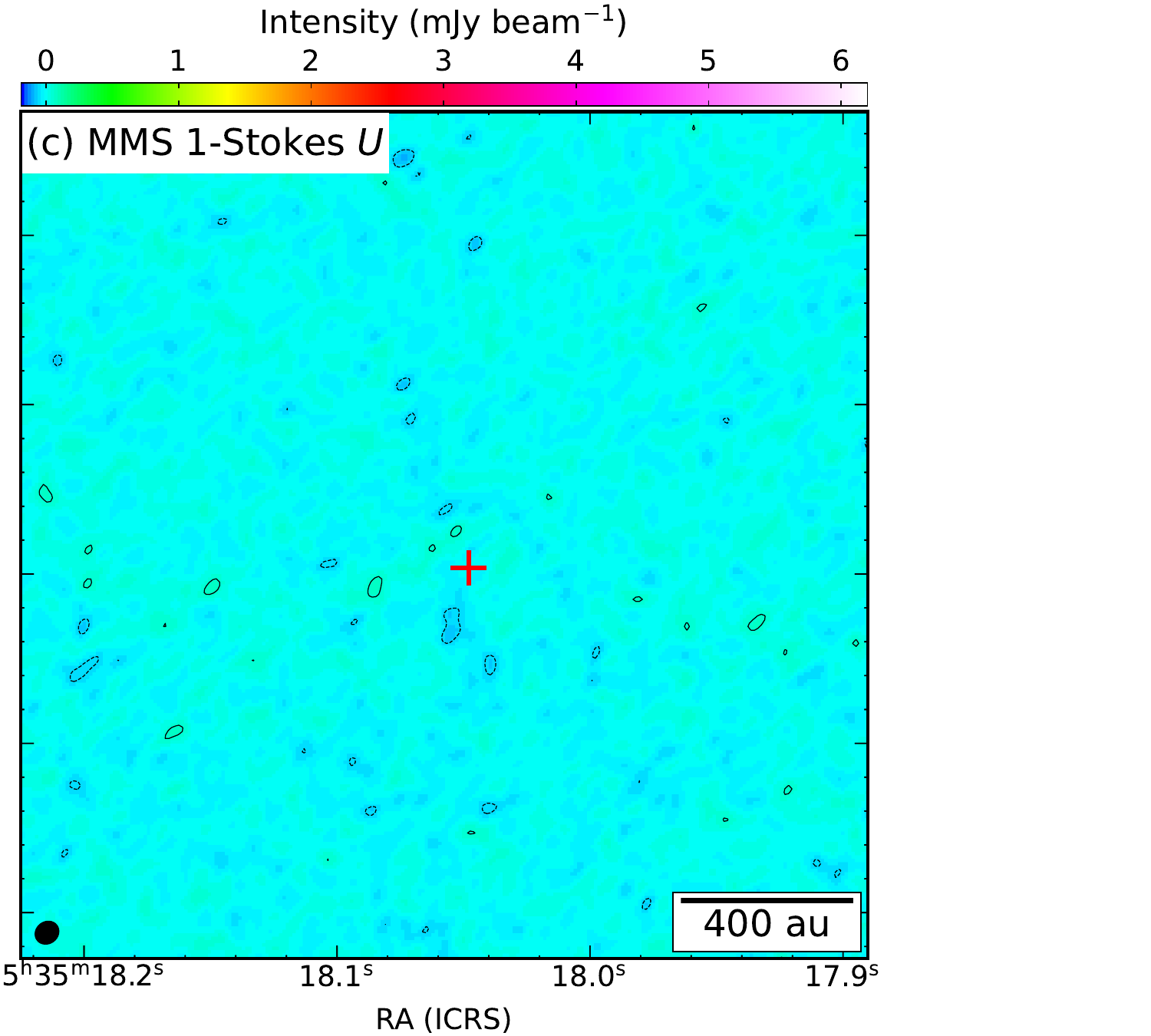}{0.43\textwidth}{}}
\gridline{\hspace{-1.5\baselineskip}
          \fig{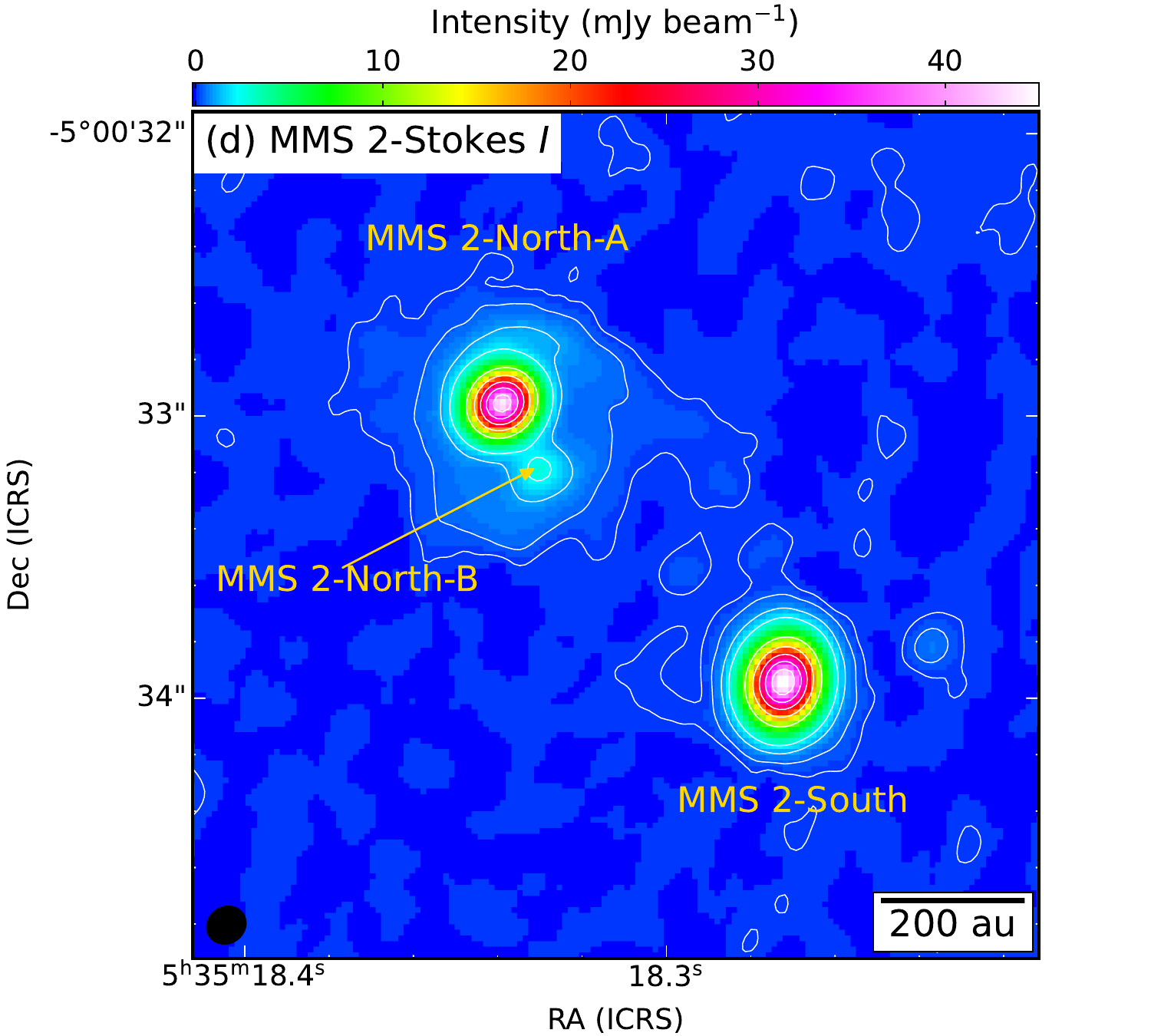}{0.43\textwidth}{}
          {}\hspace{-2.3\baselineskip}
          \fig{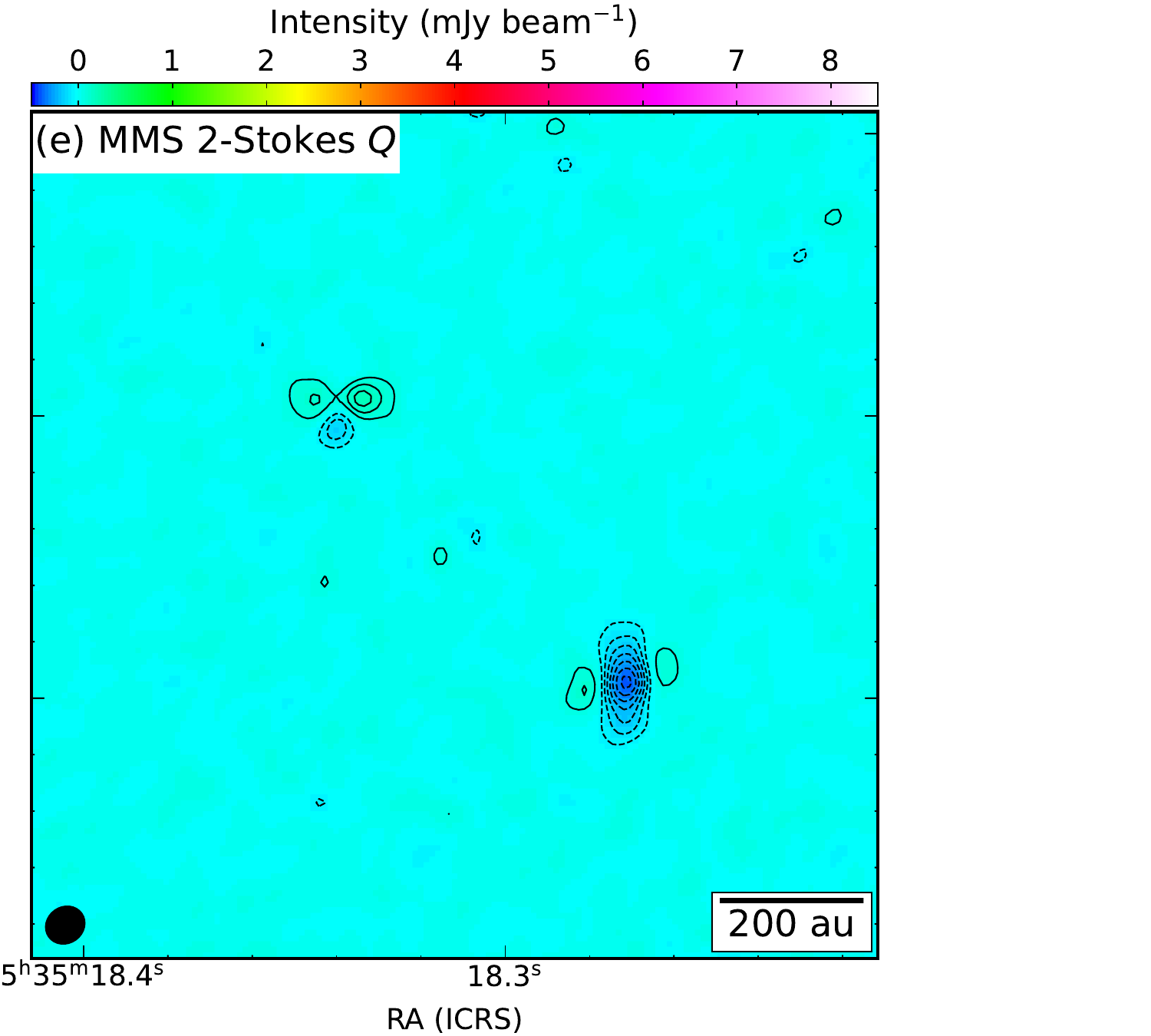}{0.43\textwidth}{}\hspace{-4.9\baselineskip}
          \fig{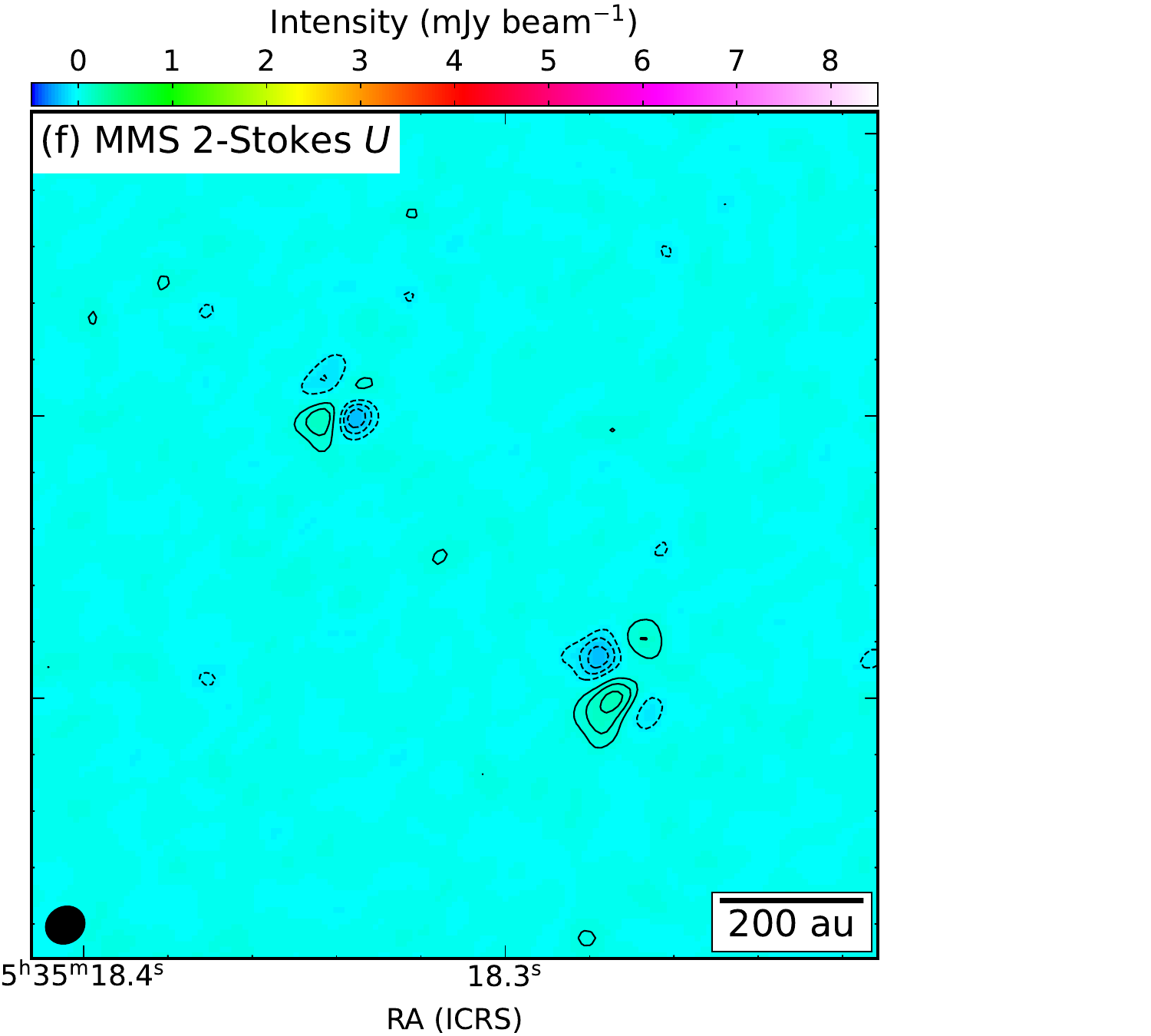}{0.43\textwidth}{}}
\gridline{\hspace{-1.5\baselineskip}
          \fig{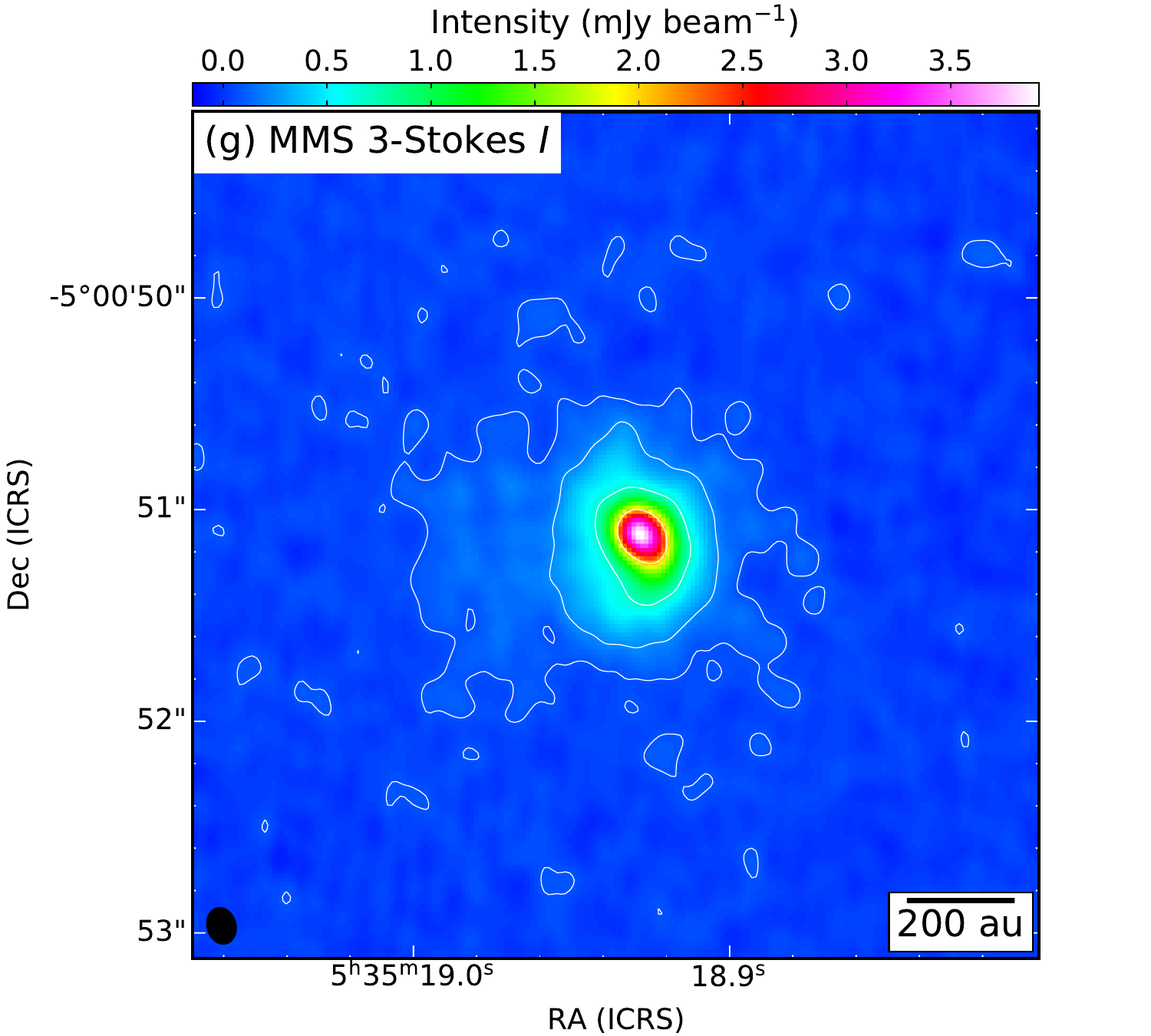}{0.43\textwidth}{}
          {}\hspace{-2.1\baselineskip}
          \fig{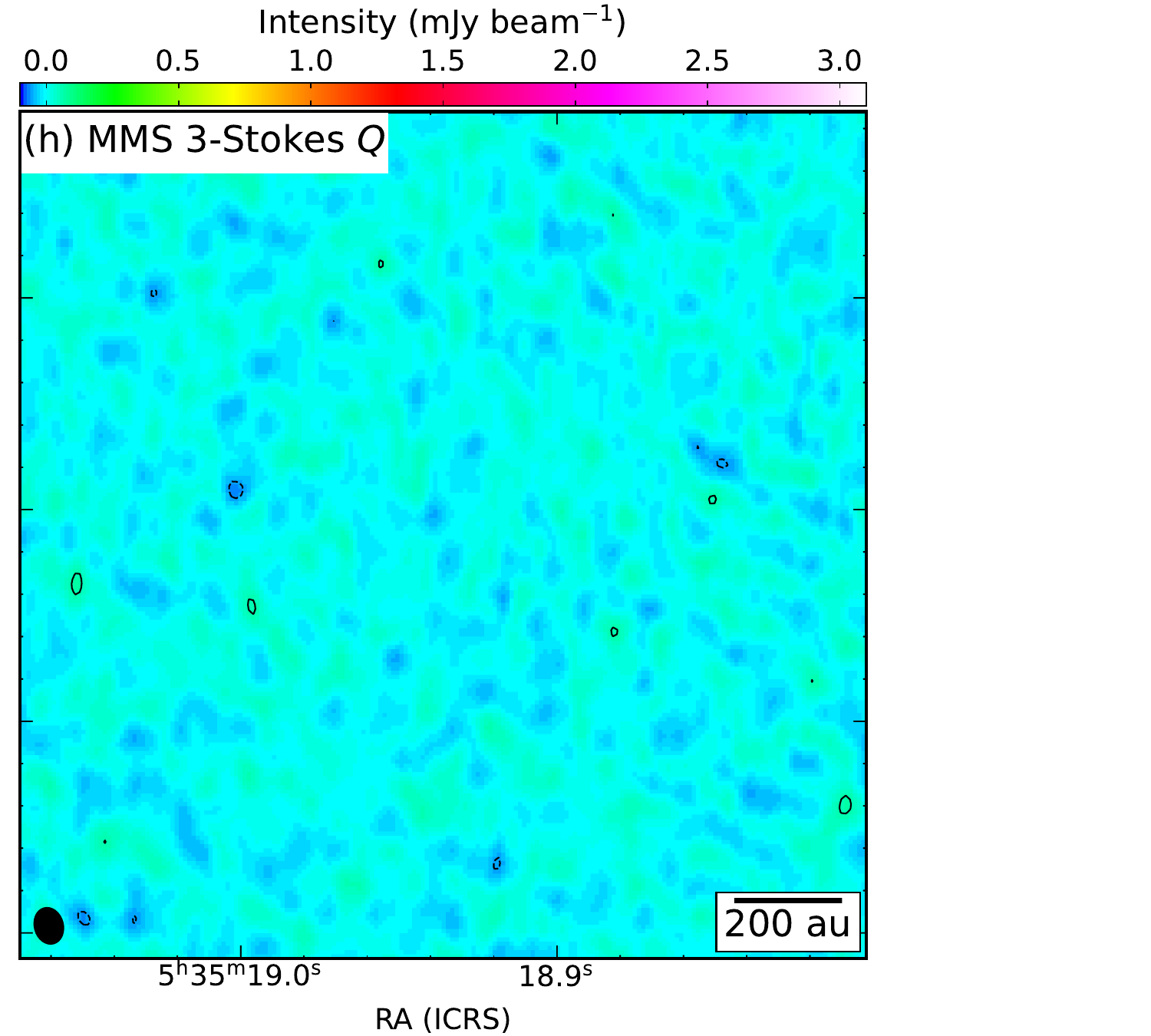}{0.43\textwidth}{}\hspace{-4.9\baselineskip}
          \fig{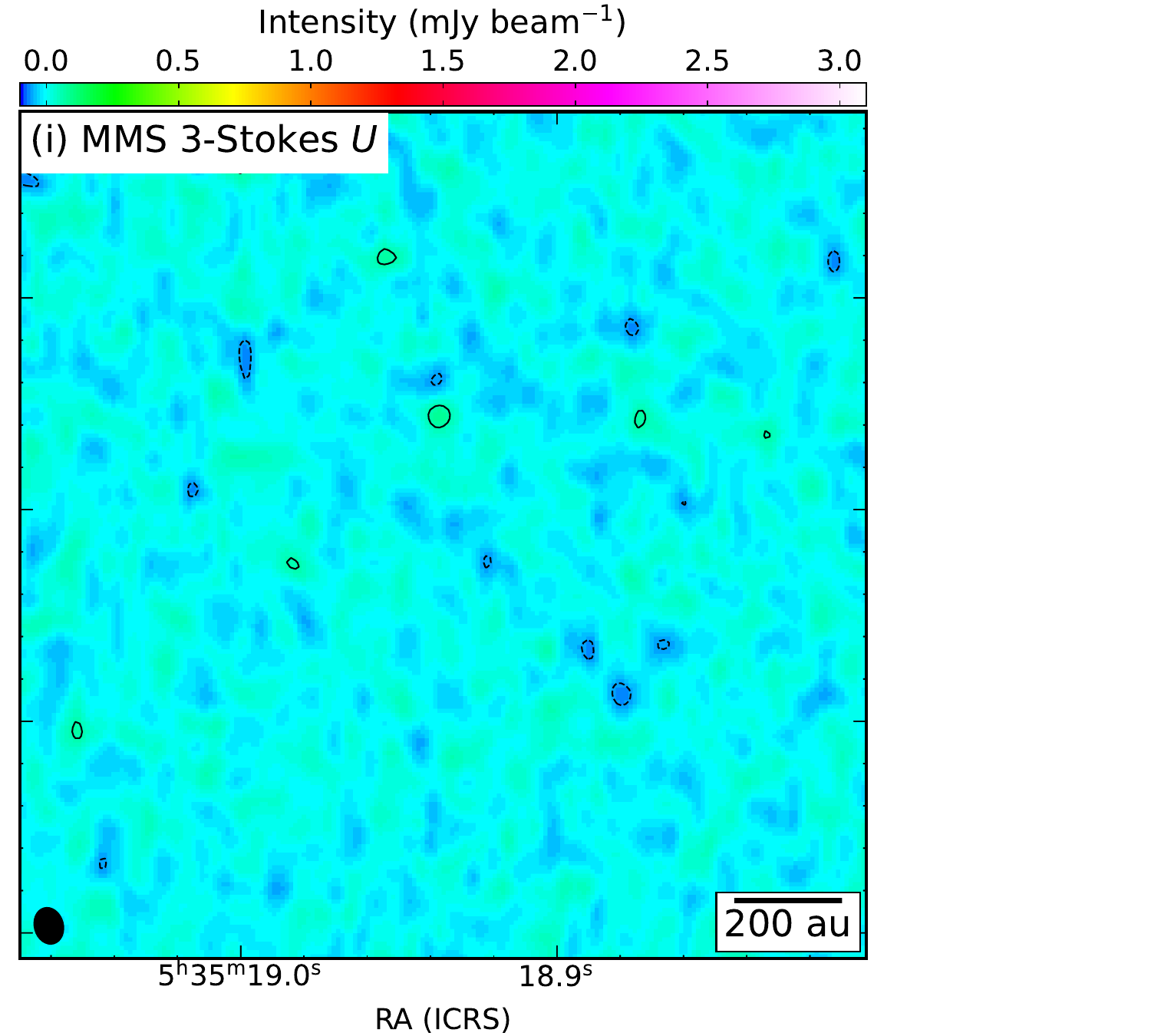}{0.43\textwidth}{}}
\caption{}
\label{fig-stokesIQU-1}
\end{figure*}

\renewcommand{\thefigure}{\arabic{figure} (Continued)}
\addtocounter{figure}{-1}

\begin{figure*}[ht!]
\gridline{\hspace{-1.5\baselineskip}
          \fig{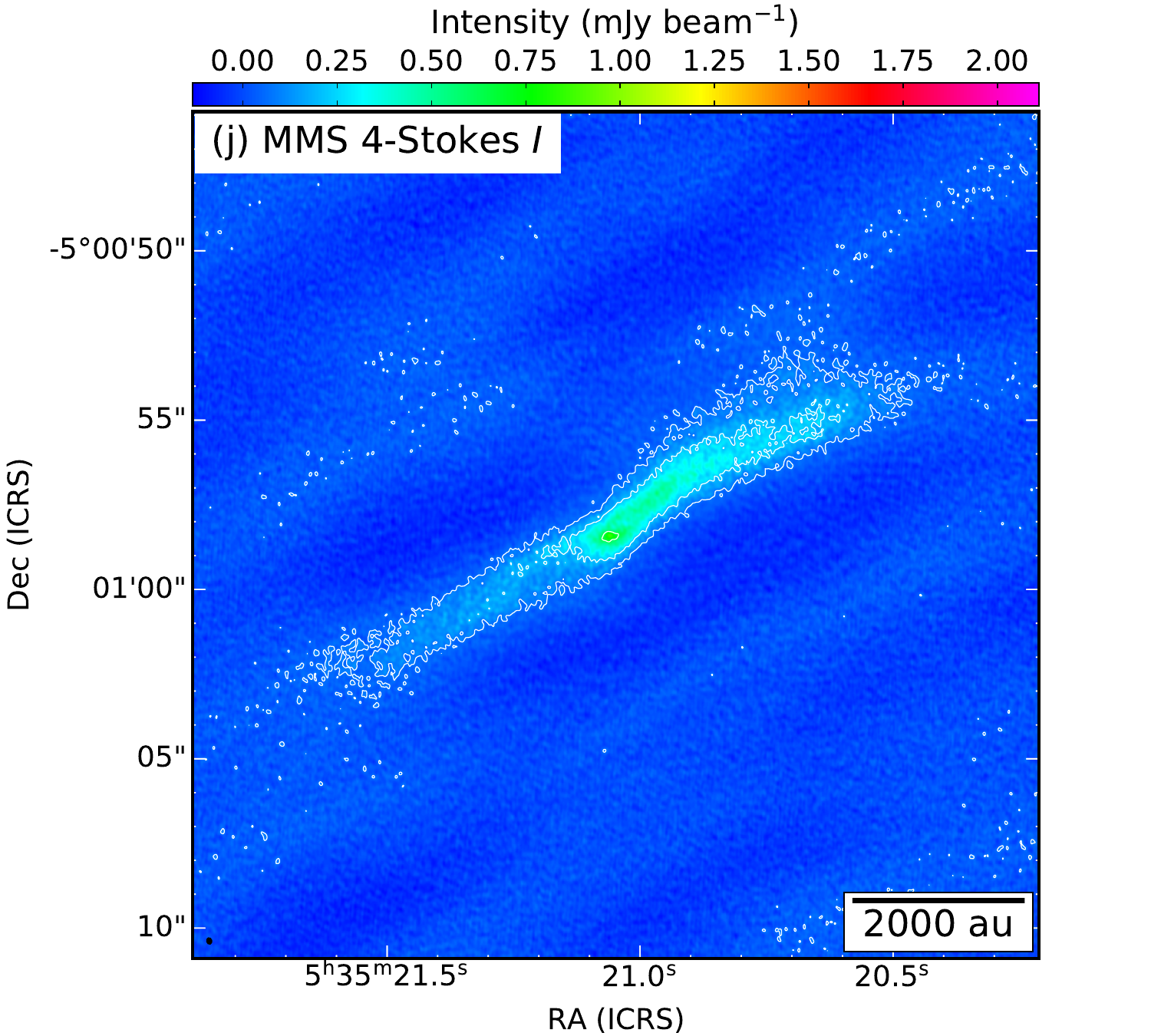}{0.43\textwidth}{}
          {}\hspace{-2.1\baselineskip}
          \fig{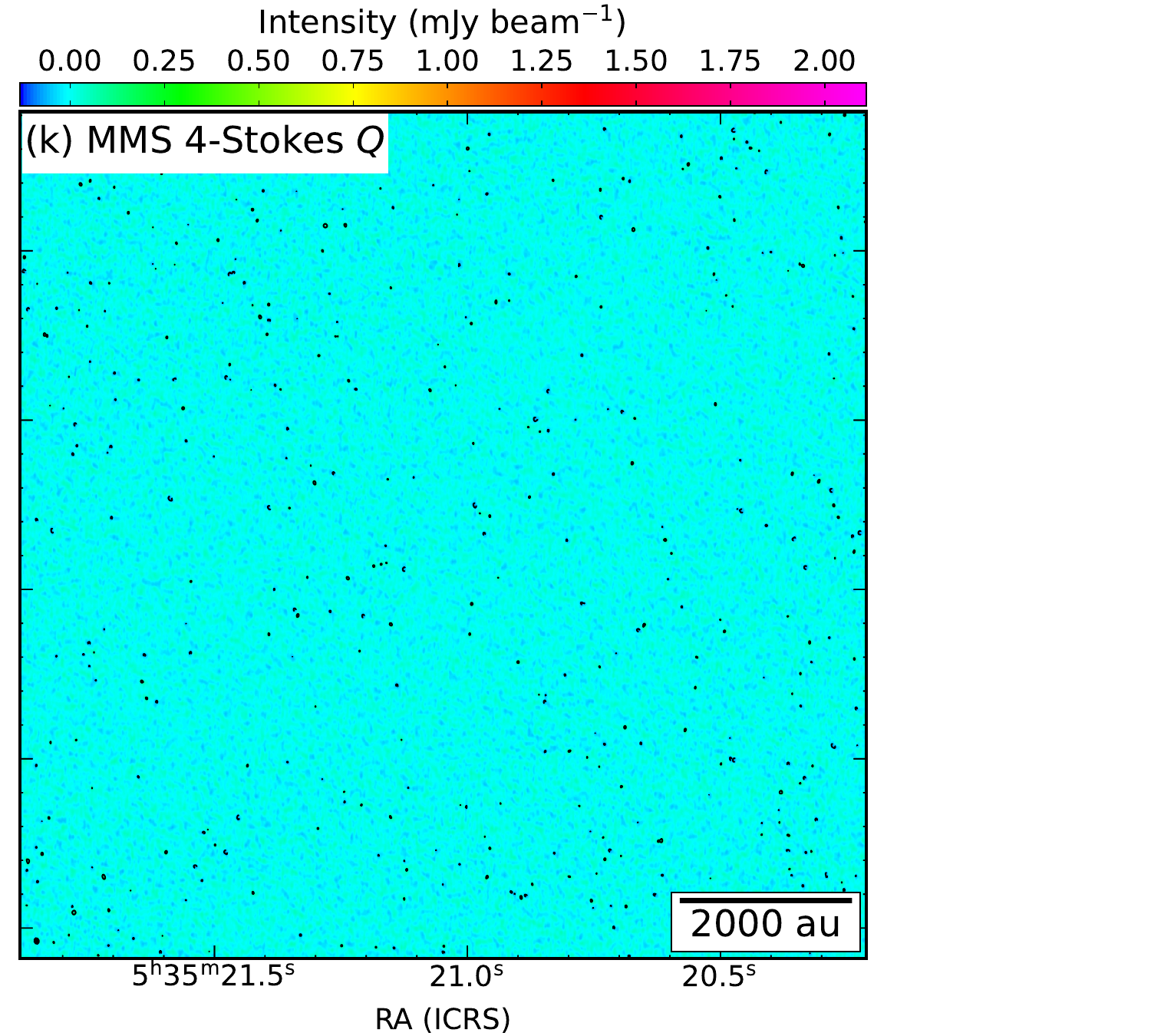}{0.43\textwidth}{}\hspace{-4.9\baselineskip}
          \fig{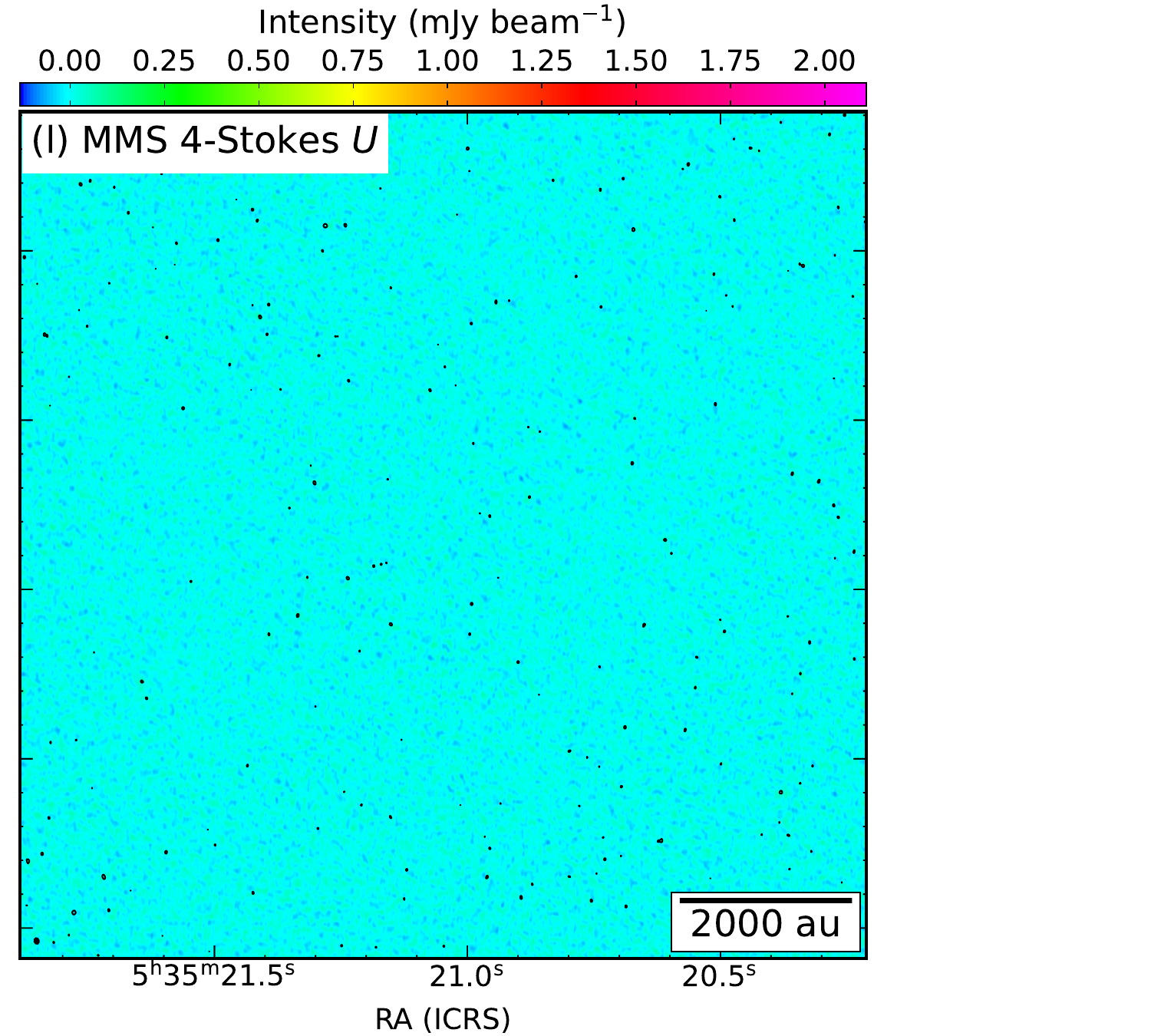}{0.43\textwidth}{}}
\gridline{\hspace{-1.5\baselineskip}
          \fig{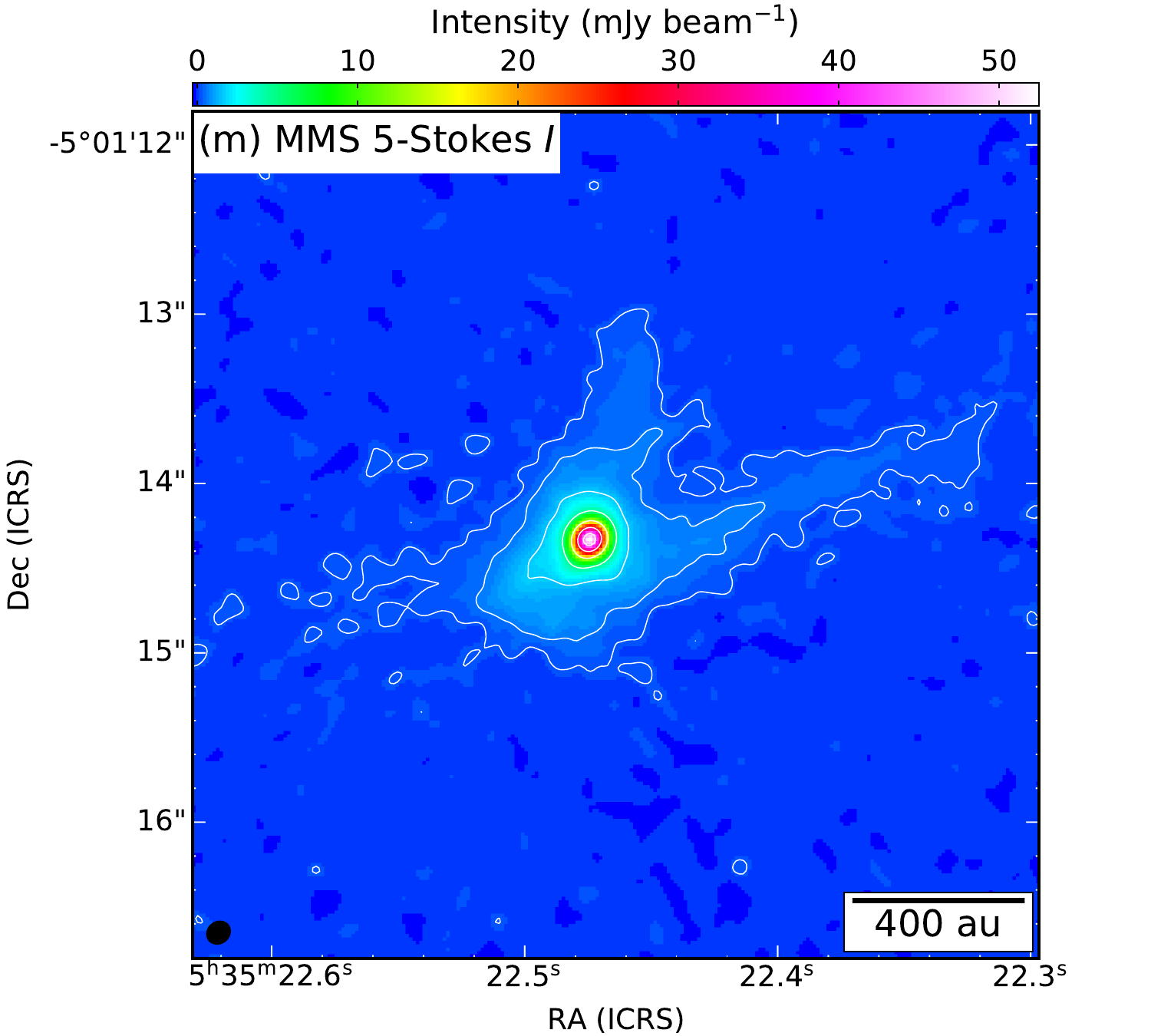}{0.43\textwidth}{}
          {}\hspace{-2.1\baselineskip}
          \fig{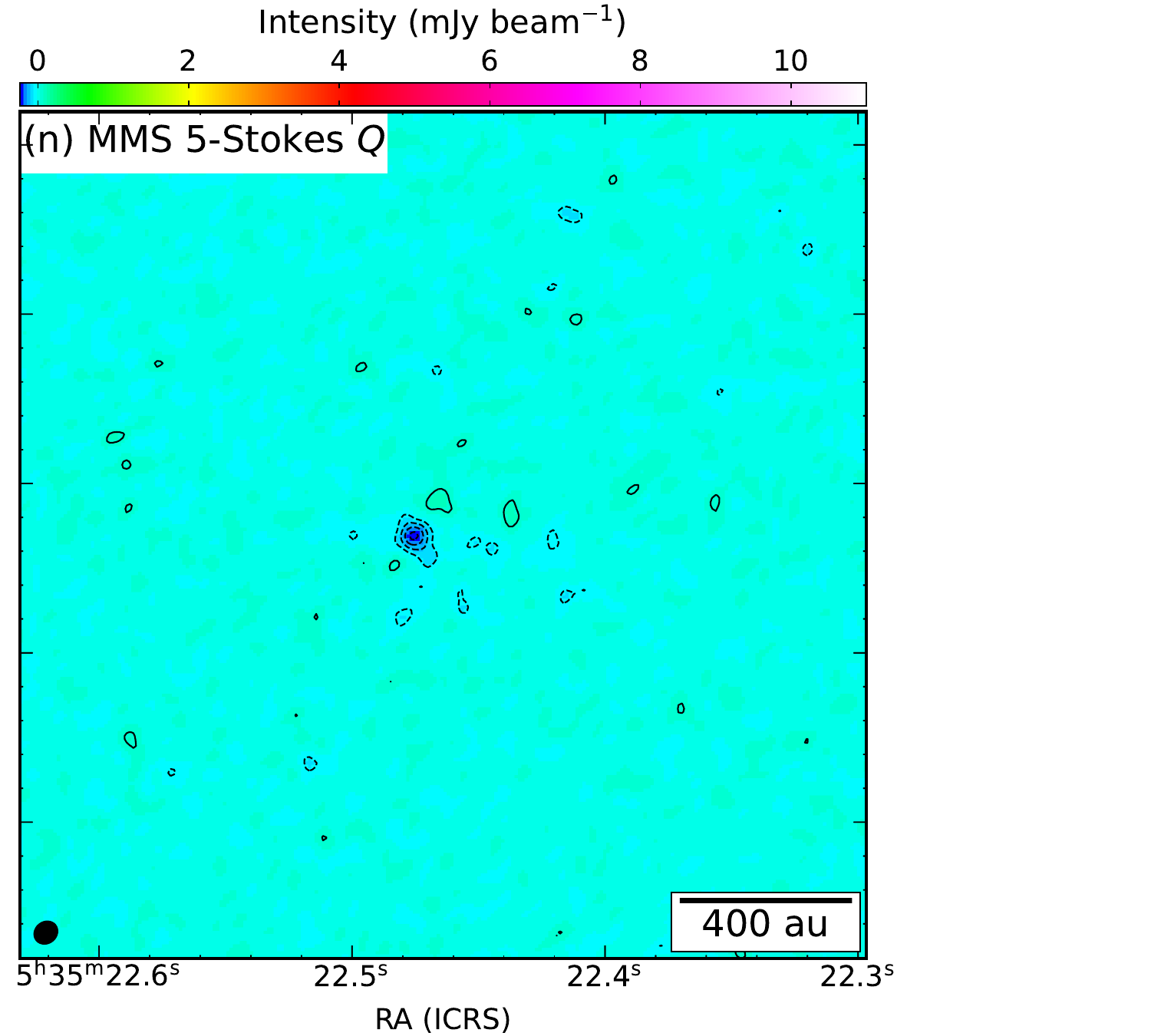}{0.43\textwidth}{}
          {}\hspace{-5.1\baselineskip}
          \fig{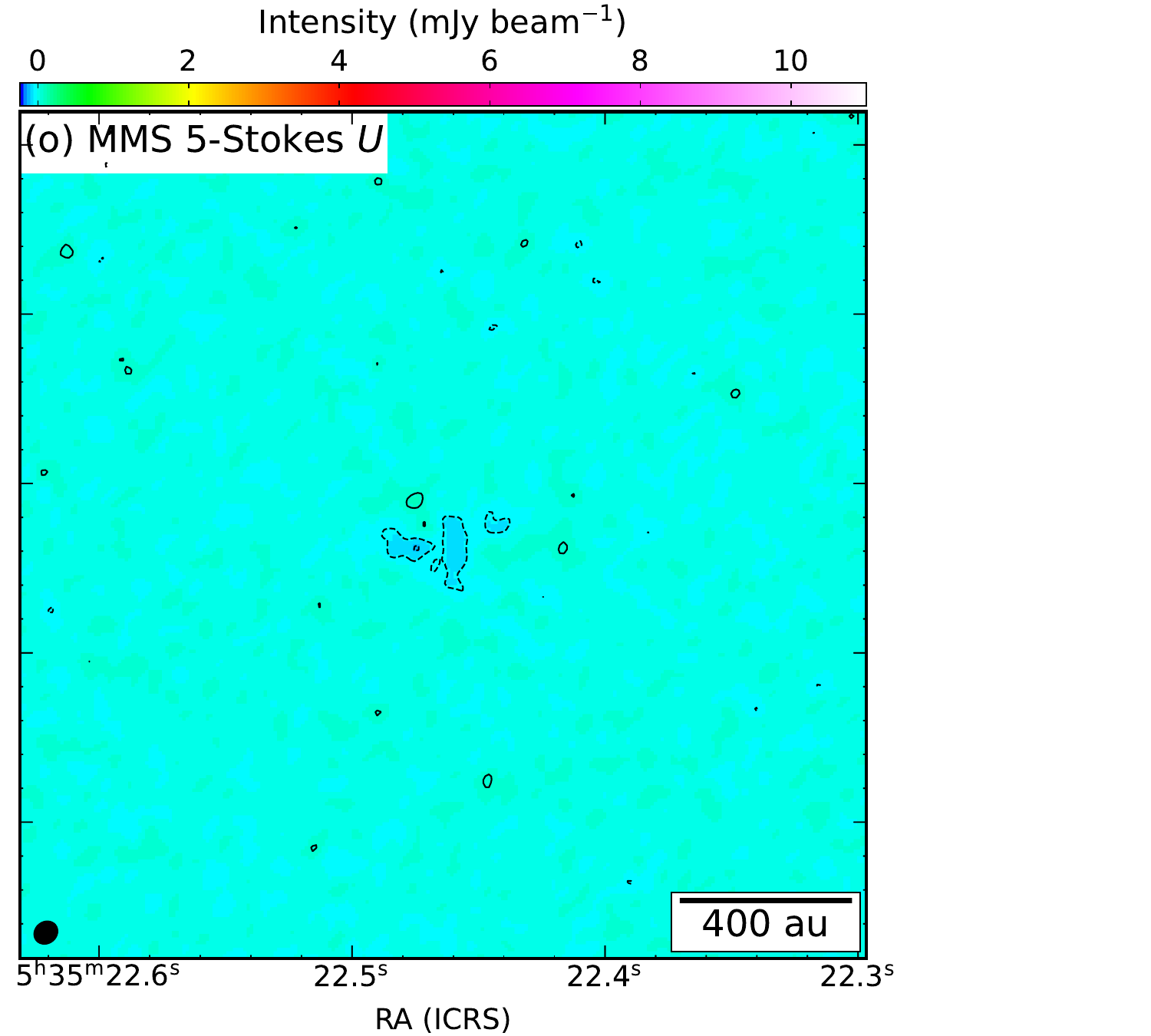}{0.43\textwidth}{}}
\gridline{\hspace{-1.5\baselineskip}
          \fig{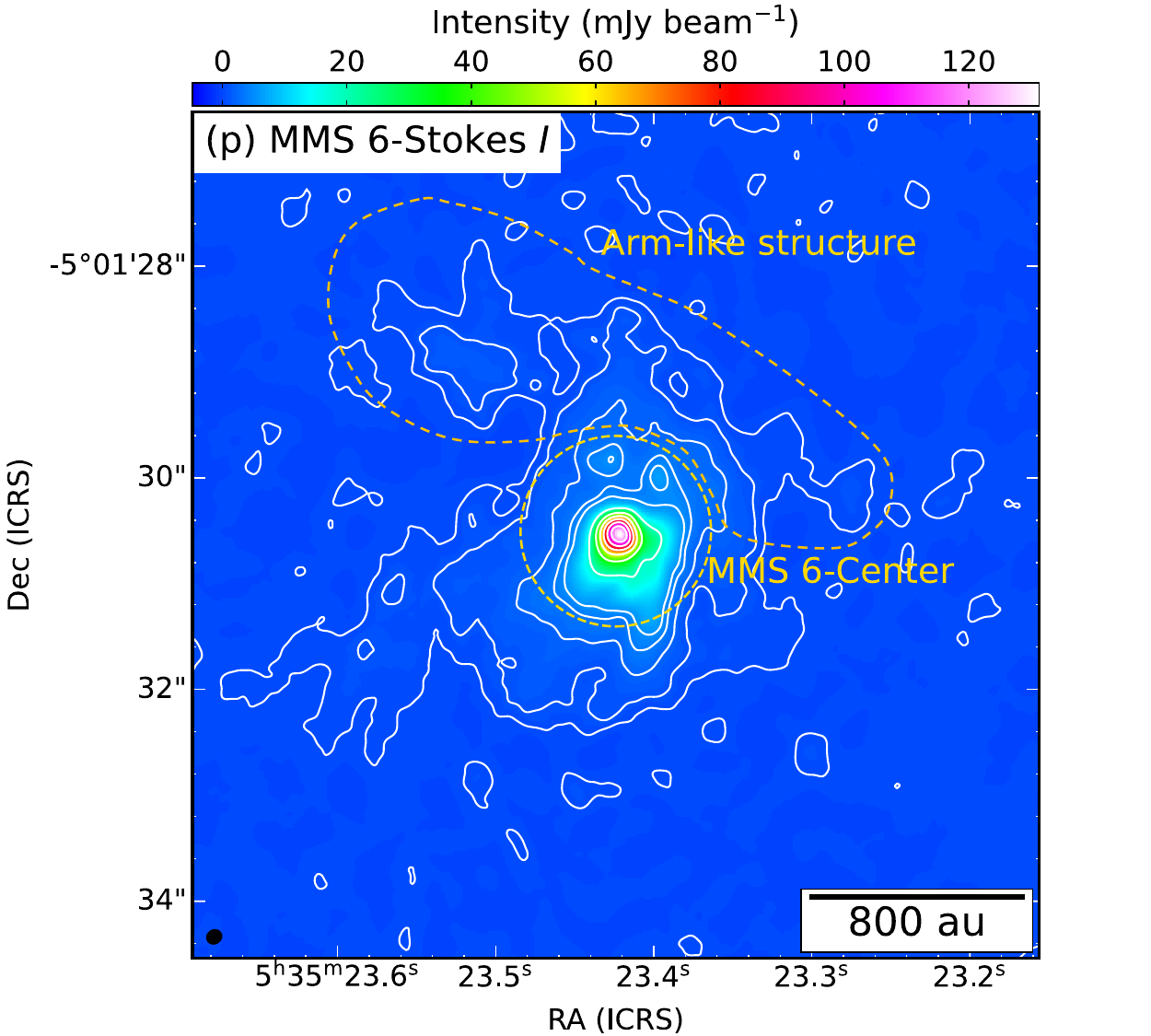}{0.43\textwidth}{}
          {}\hspace{-2.1\baselineskip}
          \fig{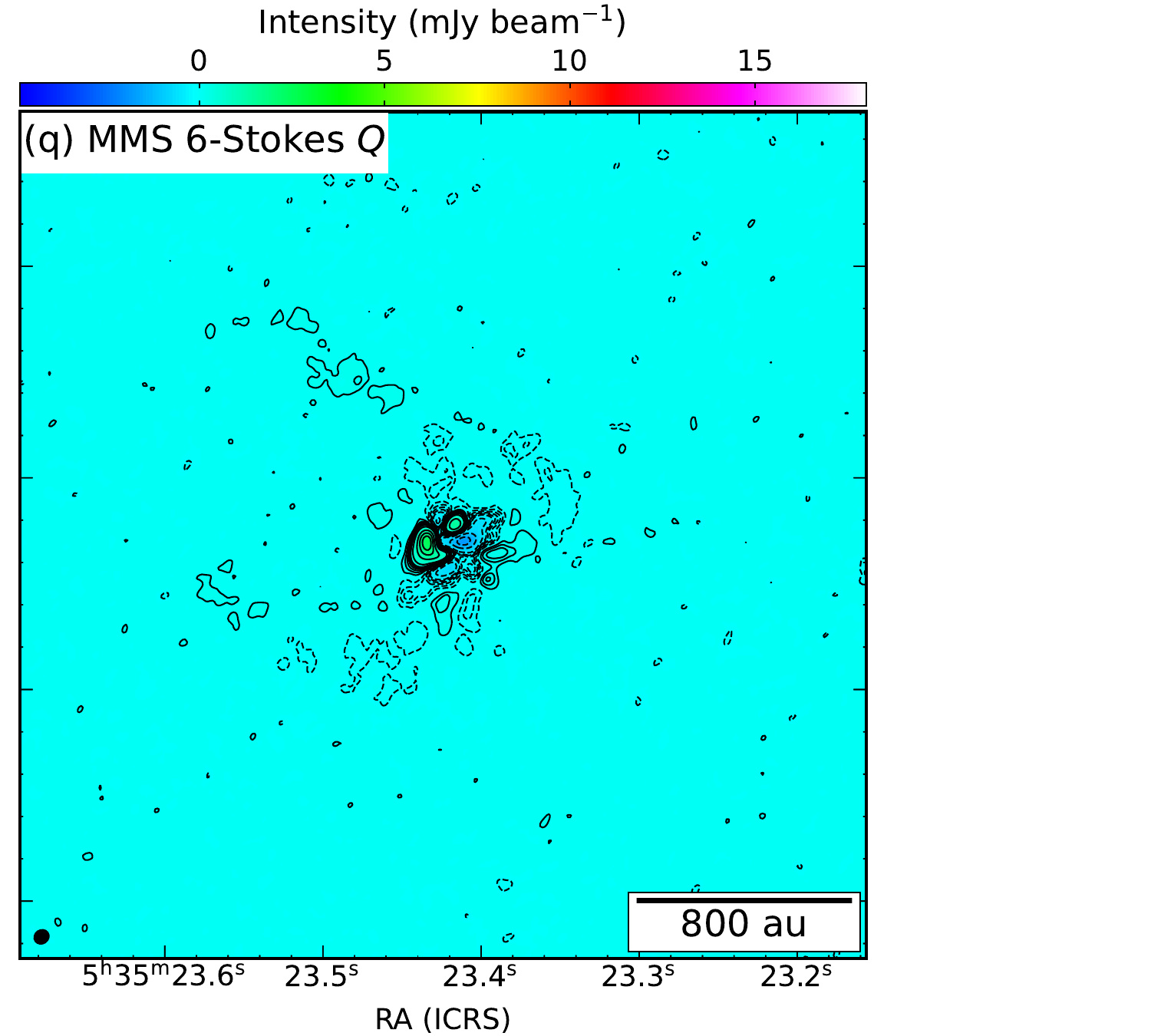}{0.43\textwidth}{}
          {}\hspace{-5.1\baselineskip}
          \fig{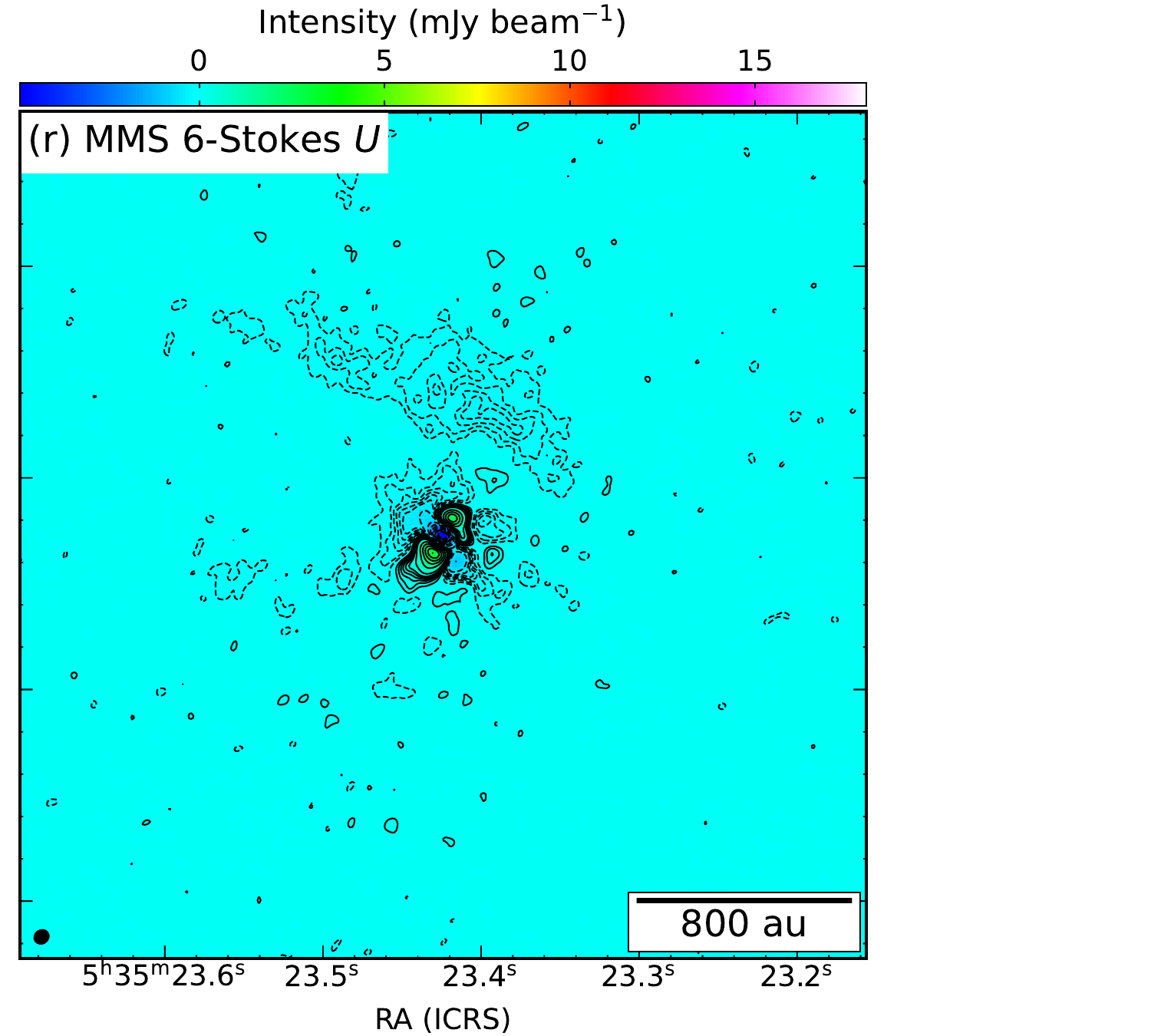}{0.43\textwidth}{}}
\caption{}
\label{fig-stokesIQU-2}
\end{figure*}
\renewcommand{\thefigure
}{\arabic{figure}}
\addtocounter{figure}{-1}

\begin{figure*}[ht!]
\end{figure*}

\begin{figure*}[ht!]
\gridline{\hspace{-1.5\baselineskip}
          \fig{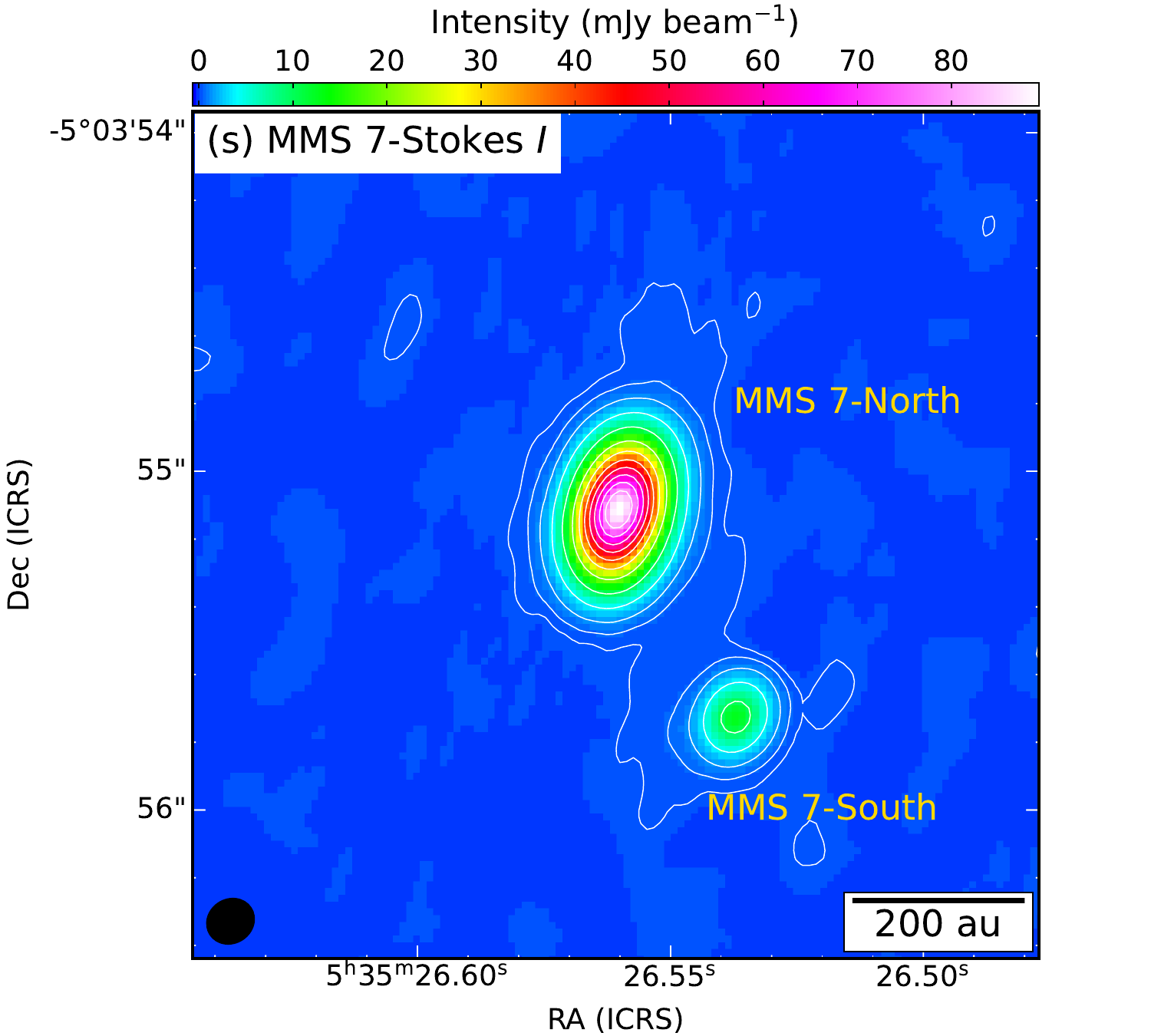}{0.43\textwidth}{}
          {}\hspace{-2.1\baselineskip}
          \fig{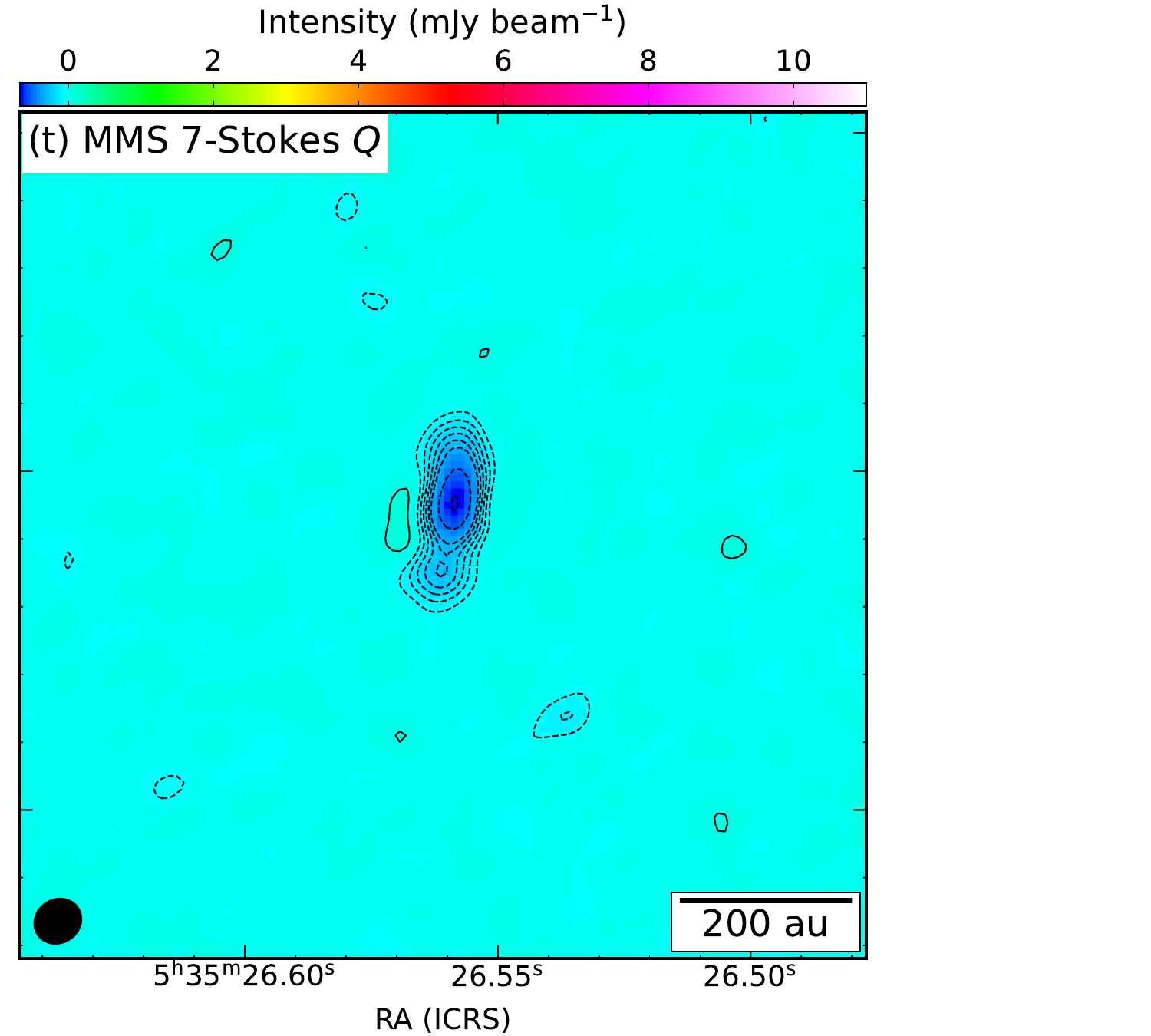}{0.43\textwidth}{}
          {}\hspace{-5.1\baselineskip}
          \fig{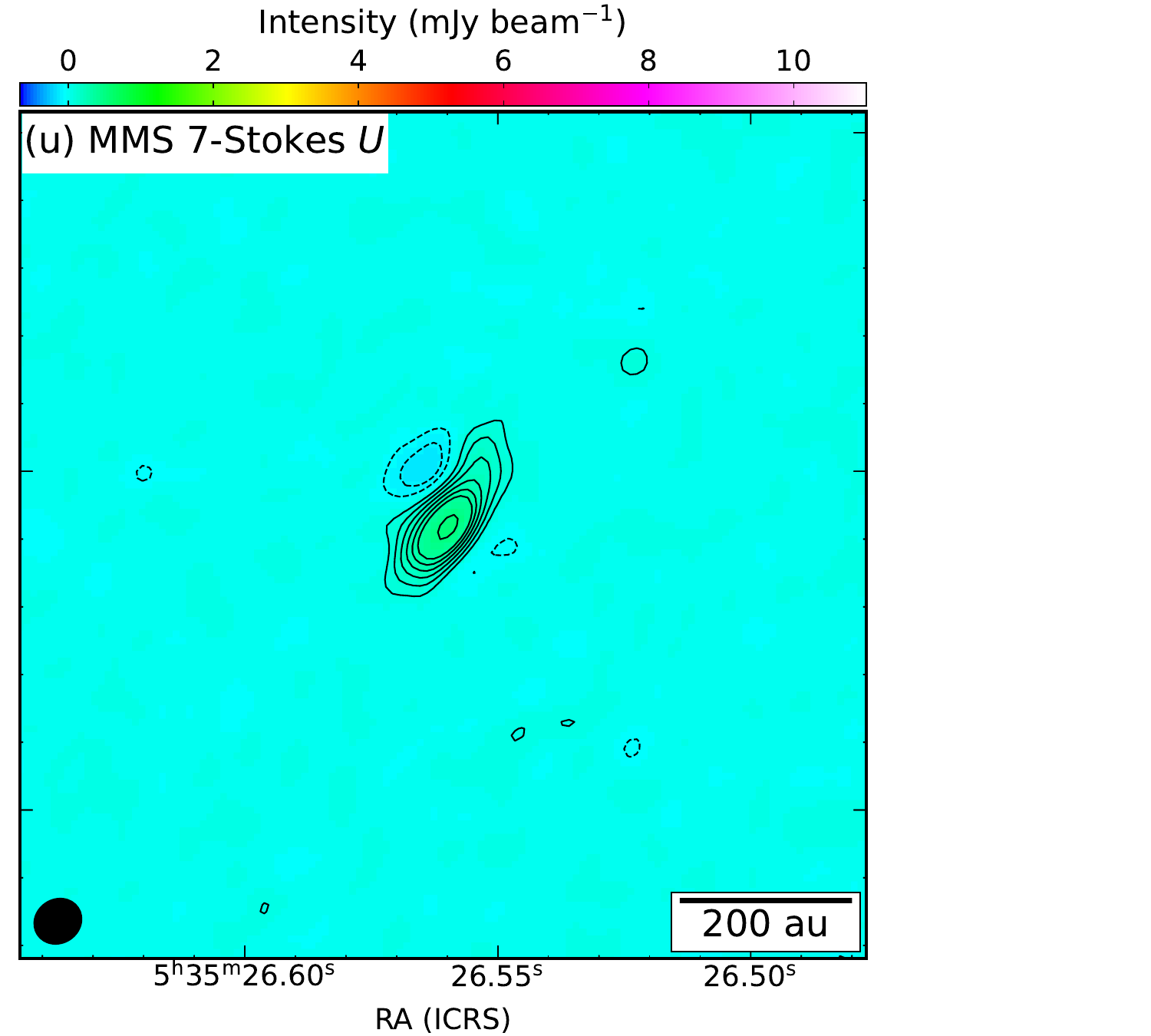}{0.43\textwidth}{}}
\vspace{-1\baselineskip}
\caption{Left panels: The colour and white contours show the Stokes $\textit{I}$ obtained from the ALMA 1.1\,mm dust continuum data. The white contours are (a) [3, 9, 27, 81, 243, 729, 2187]$\times\sigma$ (1$\sigma$ = 35.2 $\mu$Jy\,beam$^{-1}$), (d) [3, 9, 27, 60, 260, 460, 660, 860, 1060]$\times\sigma$ (1$\sigma$ = 36.7 $\mu$Jy\,beam$^{-1}$), (g) [3, 9, 27, 81]$\times\sigma$ (1$\sigma$ = 25.0 $\mu$Jy\,beam$^{-1}$), (j) [3, 9, 27]$\times\sigma$ (1$\sigma$ = 23.6 $\mu$Jy\,beam$^{-1}$), (m) [3, 9, 27, 81, 250, 500]$\times\sigma$ (1$\sigma$ = 60.0 $\mu$Jy\,beam$^{-1}$), (p) [3, 9, 27, 50, 70, 250, 500, 750, 1000, 1250, 1500, 1750]$\times\sigma$ (1$\sigma$ = 75.6 $\mu$Jy\,beam$^{-1}$), (s) [3, 9, 27, 81, 200, 400, 600, 800, 1000, 1200, 1400, 1600]$\times\sigma$ (1$\sigma$ = 49.3 $\mu$Jy\,beam$^{-1}$). Middle panels: The colour and black contours show the Stokes $\textit{Q}$ obtained from the ALMA 1.1\,mm dust continuum data. The black contours are (b) [-3, 3]$\times\sigma$ (1$\sigma$ = 18.8 $\mu$Jy\,beam$^{-1}$), (e) [-21, -18, -15, -12, -9, -6, -3, 3, 6, 9, 12, 15, 18]$\times\sigma$ (1$\sigma$ = 18.4 $\mu$Jy\,beam$^{-1}$), (h) [-3, 3]$\times\sigma$ (1$\sigma$ = 19.2 $\mu$Jy\,beam$^{-1}$), (k) [-3, 3]$\times\sigma$ (1$\sigma$ = 18.8 $\mu$Jy\,beam$^{-1}$), (n) [-15, -12, -9, -6, -3, 3, 6]$\times\sigma$ (1$\sigma$ = 18.5 $\mu$Jy\,beam$^{-1}$), (q) [-216, -189, -162, -135, -108, -81, -54, -27, -18, -15, -12, -9, -6, -3, 3, 6, 9, 12, 15, 18, 27, 54, 81, 108, 135, 162]$\times\sigma$ (1$\sigma$ = 19.0 $\mu$Jy\,beam$^{-1}$), (t) [-45, -36, -27, -18, -15, -12, -9, -6, -3, 3, 6, 9, 12, 15, 18, 27, 54, 81, 108, 135, 162]$\times\sigma$ (1$\sigma$ = 18.1 $\mu$Jy\,beam$^{-1}$). Right panels: The colour and black contours show the Stokes $\textit{U}$ obtained from the ALMA 1.1\,mm dust continuum data. The black contours are (c) [-3, 3]$\times\sigma$ (1$\sigma$ = 18.4 $\mu$Jy\,beam$^{-1}$), (f) [-21, -18, -15, -12, -9, -6, -3, 3, 6, 9, 12, 15, 18]$\times\sigma$ (1$\sigma$ = 18.2 $\mu$Jy\,beam$^{-1}$), (i) [-3, 3]$\times\sigma$ (1$\sigma$ = 19.0$\mu$Jy\,beam$^{-1}$), (l) [-3, 3]$\times\sigma$ (1$\sigma$ = 18.8 $\mu$Jy\,beam$^{-1}$), (o) [-15, -12, -9, -6, -3, 3, 6, 9, 12, 15]$\times\sigma$ (1$\sigma$ = 18.8 $\mu$Jy\,beam$^{-1}$), (r) [-216, -189, -162, -135, -108, -81, -54, -27, -18, -15, -12, -9, -6, -3, 3, 6, 9, 12, 15, 18, 27, 54, 81, 108, 135, 162]$\times\sigma$ (1$\sigma$ = 19.4 $\mu$Jy\,beam$^{-1}$), (u) [-45, -36, -27, -18, -15, -12, -9, -6, -3, 3, 6, 9, 12, 15, 18, 27, 54, 81, 108, 135, 162]$\times\sigma$ (1$\sigma$ = 18.7 $\mu$Jy\,beam$^{-1}$). The red crosses in panel (b) and (c) indicate the peak position of Stokes $I$ for MMS\,1 as shown in panel (a). The synthesized beam size is denoted by a filled ellipse in the bottom left corner of all panels.}
\label{fig-stokesIQU-3}          
\end{figure*}

\subsubsection{MMS 1}
MMS\,1 was first identified by \cite{chini1997dust} in the 1300\,$\mu$m survey and classified as Class 0 source because it has a bolometric luminosity-to-submillimeter luminosity ratio ($L_{\textnormal{bol}}/L_{\textnormal{smm}}$) of $<$ 200 \citep{andre1993submillimeter}. This source was also identified as CSO\,5 in the 350\,$\mu$m observations \citep{lis1998350} and as SMM\,2 in the 850\,$\mu$m SMA observations \citep{takahashi2013hierarchical} as listed in Table~\ref{table:targets-summary} fourth column. Despite the centrally peaked structure being as small as $\sim$1000 au, the number density is as high as 1.5$\times$10$^8$ cm$^{-3}$, as reported by \cite{takahashi2013hierarchical}. No HOPS source associated with MMS\,1 was identified \citep{furlan2016herschel}. Furthermore, there was no evidence to show the existence of the molecular outflow in previous studies \citep{yu1997shock,stanke2002unbiased,aso2000dense,williams2003high,takahashi2008millimeter,feddersen2020carma}. With our recent ALMA study by Takahashi et al.\,(2023a, submitted to ApJ), we have discovered a very compact bipolar CO outflow and EHV jet associated with this source as shown in Figure~\ref{mom0-mms1} (in Appendix~\ref{co-image}). One-sided length of the outflow is $\sim$2400 au and the line-of-sight velocity of the EHV jet is up to $\sim$65 km\,s$^{-1}$. These characteristics suggest that MMS\,1 could be one of the youngest protostars among the observed sources in this region. 

Figure~\ref{fig-stokesIQU-3}\,(a) shows that we have detected both centrally concentrated and extended components associated with MMS\,1 in Stokes $I$. This source is fitted with one compact component and one extended component as listed in Table~\ref{tab:Summary of the fitting results}. The compact component has a linear size of $\sim$19 au $\times$ 13 au with a position angle (P.A.) $= 154^{\circ}$, presumably tracing a very compact dust disk around the protostar. One of the extended components has a linear size of $\sim$684 au $\times$ 614 au (P.A. $=$ 14$^{\circ}$), which likely traces the inner part of an envelope. In addition, an extended tail-like structure is detected at the 9$\sigma$ level in Stokes $I$ pointing towards the southwest direction. This tail-like structure is spatially correlated with the molecular outflow, presumably tracing a swept-up gas by the outflow (Takahashi et al. 2023a, submitted to ApJ).

MMS\,1 is not detected in Stokes $Q$ and $U$ as shown in Figure~\ref{fig-stokesIQU-3}\,(b) and (c). The 3$\sigma$ upper limit of the polarization fraction is $\sim$0.5$\%$. The 3$\sigma$ upper limit of the polarization fraction for MMS\,1 is larger than the minimum detectable polarization fraction for a compact source (0.1$\%$; ALMA Cycle 4 Technical Handbook). Therefore, the sensitivity may be limited in detecting the polarized emission towards the compact dust disk. On the other hand, we notice that the 3$\sigma$ upper limit of the polarization fraction for MMS\,1 is close to the minimum detectable polarization fraction for an extended component (0.3$\%$; ALMA Cycle 4 Technical Handbook). This indicates that the observing sensitivity allows for the detection of the extended polarized emission associated with the envelope and core size scale as presented in Appendix~\ref{mms1poli}. Figure \ref{mms1-poli} shows PI with the magnetic field (rotate the $E$-vectors by 90$^{\circ}$) along the outflow, but more in-depth observations with shorter baselines are necessary to better characterize the magnetic field properties.

\begin{table*}[h]
  \centering
  \rotatebox{90}{
    \begin{minipage}{1.3\textwidth}
      \centering
      \caption{\centering\small The 1.1\,mm Continuum Fitting Results}
      \label{tab:Summary of the fitting results}
      \begin{tabular}{lccccccc}
      \hline\hline \noalign {\smallskip}
        &R.A.$^a$ & Decl.$^a$& Peak Flux & Total Flux Density  & Deconvolved Size$^b$ & P.A. & Inclination Angle ($i$)$^c$  \\
       & (J2000)& (J2000) & (mJy\,beam$^{-1}$) & (mJy)& (milliarcsec $\times$ milliarcsec)&(degree) & (degree)\\
       \hline
        MMS\,1 compact component  & 05 35 18.05 & $-$5 00 17.98 & 12.885$\pm$0.057 & 14.08$\pm$0.1 & 47.5$\pm$3.0$\times$33.8$\pm$3.7& 154$\pm$11 & 44\\
        MMS\,1 extended component 1& 05 35 18.05 & $-$5 00 17.96 & 1.395$\pm$0.039 & 17.34$\pm$0.53 & 616$\pm$19$\times$329$\pm$10& 148.2$\pm$1.8 & 58\\
        MMS\,1 extended component 2 & 05 35 18.06 & $-$5 00 18.07 & 0.4$\pm$0.013 & 65.6$\pm$1.9 & 1740$\pm$52$\times$1563$\pm$47& 14$\pm$11 &26\\
        \hline
        MMS\,2-North-A compact component & 05 35 18.34 & $-$5 00 32.95 & 43.0$\pm$0.085 & 64.30$\pm$0.2 & 100.0$\pm$0.9$\times$90.4$\pm$0.9&149.0$\pm$3.9 &25 \\
        MMS\,2-North-A extended component & 05 35 18.34 & $-$5 00 32.97 & 1.15$\pm$0.05 & 21.5$\pm$1.0 & 622$\pm$28$\times$520$\pm$24& 13$\pm$10 &33 \\
        MMS\,2-North-B compact component$^d$ &05 35 18.33 & $-$5 00 33.21 & 1.92$\pm$0.08 & 2.01$\pm$0.14 & 41$\pm$25$\times$7$\pm$32& 101$\pm$39 &-\\
        MMS\,2-South compact component& 05 35 18.27 & $-$5 00 33.94  & 37.95$\pm$0.086 & 70.9$\pm$0.23 & 148.8$\pm$0.8$\times$102.3$\pm$0.8& 173.0$\pm$1.0 & 47 \\
        MMS\,2-South extended component&05 35 18.27 & $-$5 00 33.94 & 7.91$\pm$0.08 & 27.5$\pm$0.4 & 243.4$\pm$3.7$\times$182.7$\pm$3.2& 5.5$\pm$2.4 &41\\
        \hline
        MMS\,3 compact component &05 35 18.93 & $-$5 00 51.12 & 2.849$\pm$0.037 & 4.912$\pm$0.094 & 162.8$\pm$5.6$\times$95.8$\pm$6.0& 54.0$\pm$3.9 &54  \\
        MMS\,3 extended component 1 &05 35 18.93 & $-$5 00 51.18 & 0.965$\pm$0.026 & 10.09$\pm$0.3 & 542$\pm$17$\times$409$\pm$12& 9.7$\pm$4.4 &41\\
        MMS\,3 extended component 2 &05 35 18.95 & $-$5 00 51.16 & 0.166$\pm$0.01 & 14.5$\pm$0.91 & 1530$\pm$97$\times$1324$\pm$84& 87$\pm$18 &30\\
        \hline
        MMS\,4 relatively compact component & 05 35 21.05 & $-$5 00 58.44 & 0.4$\pm$0.016 & 14.1$\pm$0.63 & 1301$\pm$59$\times$658$\pm$30& 131$\pm$2.5 &60\\
        MMS\,4 extended component &05 35 20.96 & $-$5 00 57.21& 0.327$\pm$0.004 & 226.5$\pm$2.8 & 11662$\pm$142$\times$1388$\pm$17& 120$\pm$0.12 &83\\
        \hline
        MMS\,5 compact component & 05 35 22.47 & $-$5 01 14.33 & 49.24$\pm$0.15 & 67.32$\pm$0.32 & 89.8$\pm$1.4$\times$72.1$\pm$1.7& 1.6$\pm$3.6 &37 \\
        MMS\,5 extended component & 05 35 22.47 & $-$5 01 14.37& 2.30$\pm$0.07 & 72.0$\pm$2.2 & 864$\pm$26$\times$637$\pm$19& 131.5$\pm$3.9 &43\\
        \hline
        MMS\,6 compact component & 05 35 23.42& $-$5 01 30.53& 115.67$\pm$0.47 & 499.3$\pm$2.4 & 
        250.5$\pm$2.2$\times$243.3$\pm$2.2& 7.4$\pm$13.8 &14\\
        MMS\,6 extended component 1 & 05 35 23.42& $-$5 01 30.74 & 22.59$\pm$0.28 & 414.6$\pm$5.5 & 670.5$\pm$9.0$\times$475.7$\pm$6.4& 111.0$\pm$1.5 &45\\
        MMS\,6 extended component 2 & 05 35 23.42& $-$5 01 30.64 & 5.58$\pm$0.11 & 747$\pm$15 & 1945$\pm$40$\times$1256$\pm$26& 168.3$\pm$1.8 &50 \\
        \hline
        MMS\,7-North compact component 1& 05 35 26.56& $-$5 03 55.11 & 82.31$\pm$0.14 & 213.31$\pm$0.47 & 241.5$\pm$0.6$\times$107.2$\pm$0.5& 166.5$\pm$0.1  &64\\
        MMS\,7-North compact component 2& 05 35 26.56& $-$5 03 55.14 & 9.68$\pm$0.13 & 30.65$\pm$0.54 & 298.7$\pm$5.8$\times$116.1$\pm$3.5& 166.0$\pm$0.8  &67\\
        MMS\,7-South compact component & 05 35 26.54 & $-$5 03 55.73  & 12.5$\pm$0.14 & 15.96$\pm$0.28 & 92.9$\pm$4.5$\times$45.9$\pm$7.0& 165.8$\pm$5.1 &60\\
        \hline \noalign {\smallskip}
        \end{tabular}
        \footnotesize $^a${The position of each component refers to the peak position in Stokes $I$, which is determined by the 2D Gaussian fitting.}\\
        \footnotesize $^b${The errors of the deconvolved size are the 2D Gaussian fitting errors.}\\
        \footnotesize $^c${$i$ is estimated using the axis ratio of each fitted component, assuming that they are geometrically thin disk-like structure.}\\
        \footnotesize $^d${This component has a fitting error for the major axis that is approximately half the size of the fitted major axis, and a fitting error for the minor axis three times larger than the fitted minor axis, and thus, we consider the disk associated with this source to be unresolved (see also Appendix~\ref{fitting}).}
    \end{minipage}}
\end{table*}


\begin{figure*}[ht!]
\vspace{-1\baselineskip}
\gridline{
          \fig{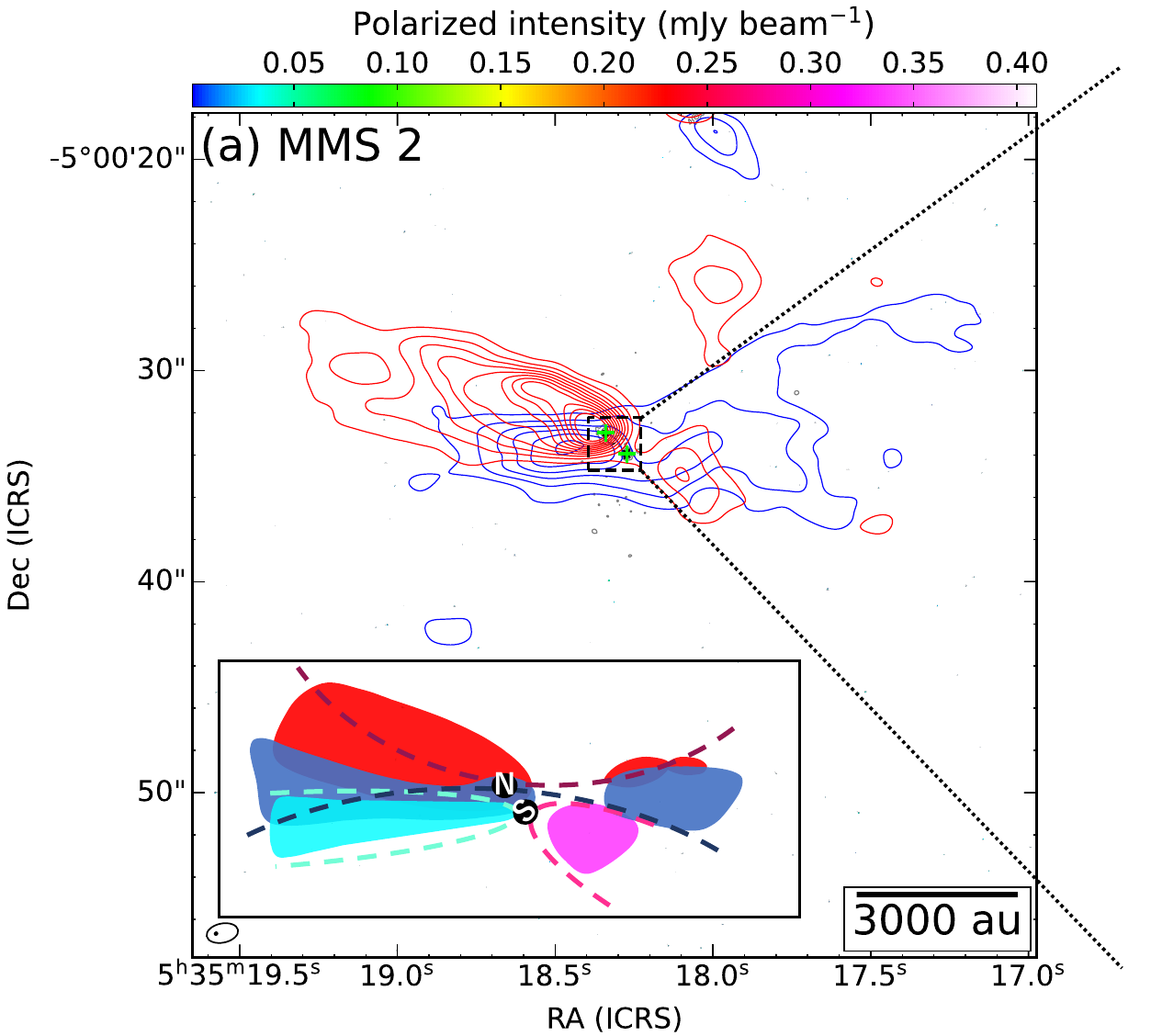}{0.51\textwidth}{}\hspace{-1\baselineskip}
          \fig{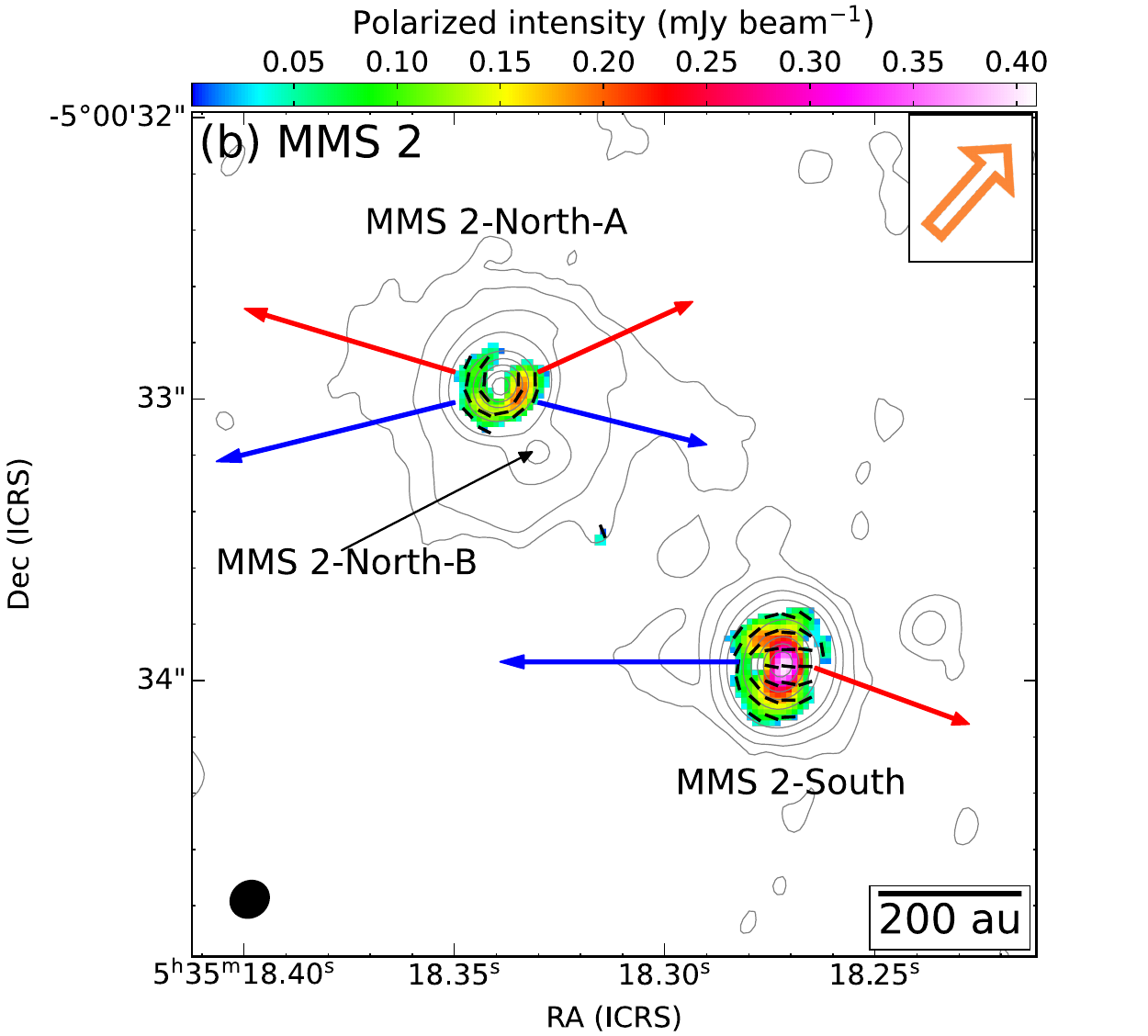}{0.51\textwidth}{}
          }
\vspace{-1.5\baselineskip}
\gridline{
          \fig{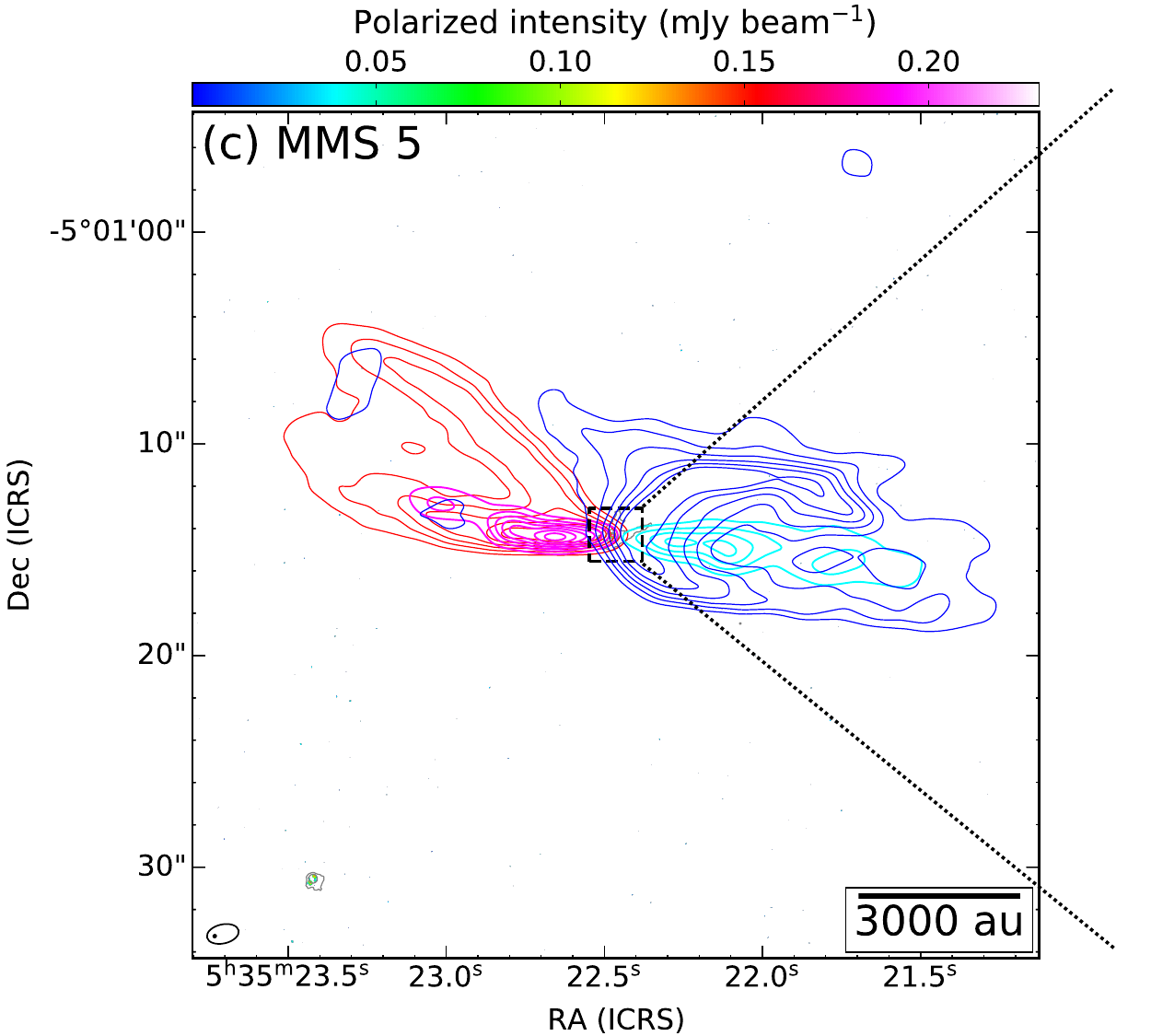}{0.51\textwidth}{}\hspace{-1\baselineskip}
          \fig{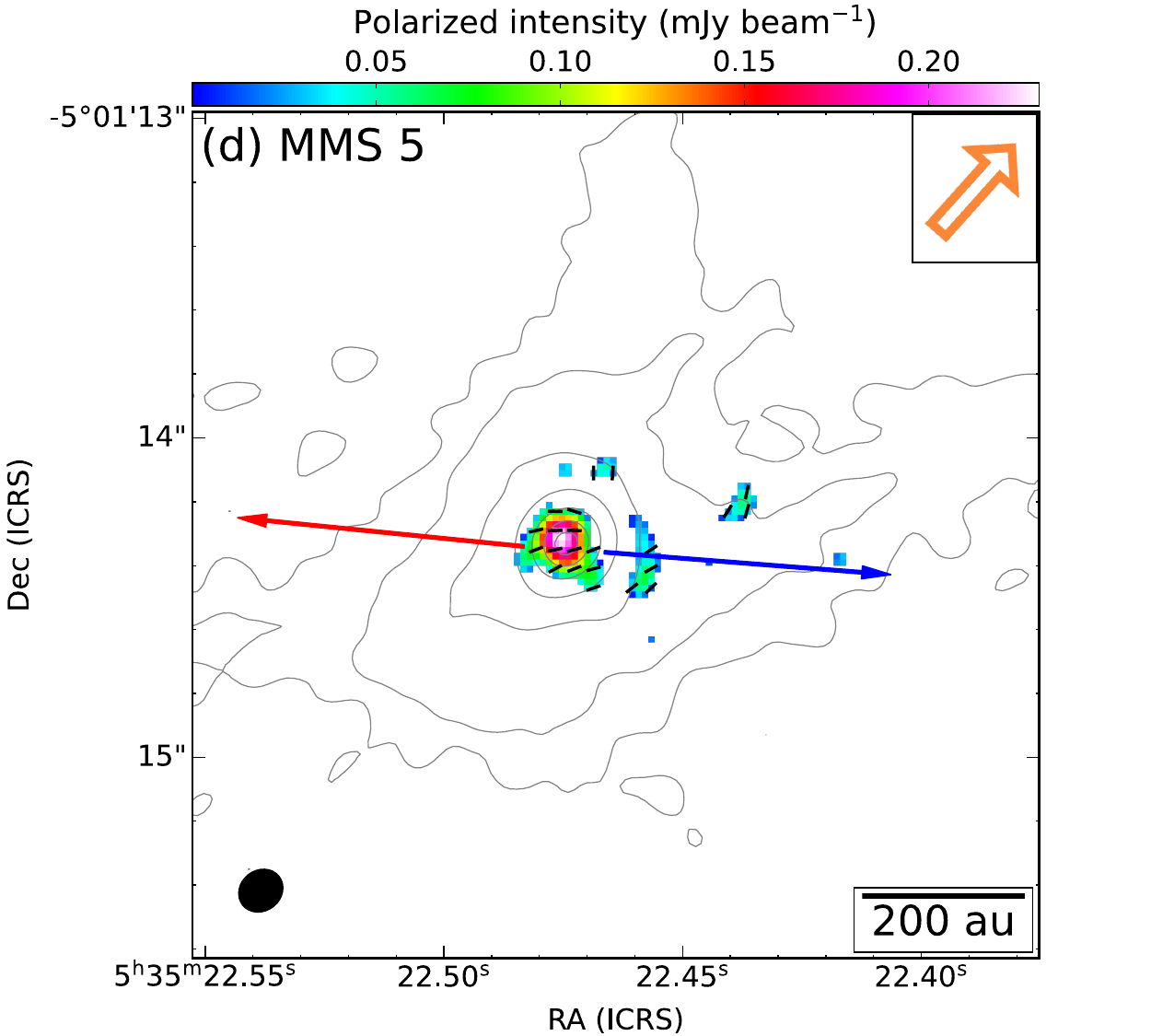}{0.51\textwidth}{}
          }
\vspace{-1.5\baselineskip}
\caption{Left panels: The blue and red contours are the integrated intensity of the CO~($J$ = 2$-$1) emission tracing the outflow. The cyan and magenta contours are the integrated intensity of the CO emission tracing the EHV jet. The CO emission in (a) is integrated from $v_{\textnormal {LSR}}-v_{\textnormal {sys}}$ = $-$1 to $-$17 km\,s$^{-1}$ (blue) and $v_{\textnormal {LSR}}-v_{\textnormal {sys}}$ = 1 to 33 km\,s$^{-1}$ (red), in (c) is integrated from $v_{\textnormal {LSR}}-v_{\textnormal {sys}}$ = $-$5 to $-$50 km\,s$^{-1}$ (blue), $v_{\textnormal {LSR}}-v_{\textnormal {sys}}$ = 5 to 50 km\,s$^{-1}$ (red), $v_{\textnormal {LSR}}-v_{\textnormal {sys}}$ = $-$50 to $-$105 km\,s$^{-1}$ (cyan), and $v_{\textnormal {LSR}}-v_{\textnormal {sys}}$ = 50 to 70 km\,s$^{-1}$ (magenta), respectively. The contour levels in (a) are [5, 10, 15, 20, 25]$\times\sigma$ (1$\sigma$ = 0.32 Jy\,beam$^{-1}$), in (c) are [5, 10, 15, 20, 25, 30, 35, 40, 45, 50, 55]$\times\sigma$ (1$\sigma$ = 0.39 Jy\,beam$^{-1}$ (red and blue)) and [10, 20, 30, 40, 50, 60, 70, 80]$\times\sigma$ (1$\sigma$ = 0.087 Jy\,beam$^{-1}$ (magenta and cyan)). The green crosses in (a) indicate the peak positions of Stokes $I$ associated with MMS\,2-North-A and MMS\,2-South, respectively. The synthesized beam size for continuum is denoted by a filled dot, and for CO, it is denoted by an empty ellipse in the bottom left corner. Illustration in the bottom left of (a) shows the proposed schematic diagram of MMS\,2 CO outflow. The red and blue lobes represent the red- and blue-shifted CO components associated with MMS\,2-North-A. The magenta and cyan lobes represent the red- and blue-shifted CO components associated with MMS\,2-South. Right panels: Zoomed-in images of left panel images overlaid with $E$-vectors (black line segments) and contours of Stokes $I$ (grey). The colour shows the PI. The blue and red arrows show the outflow axes of the blue- and red-shifted components. The contour levels in (b) are [3, 9, 27, 60, 260, 460, 660, 860, 1060]$\times\sigma$ (1$\sigma$ = 37.6 $\mu$Jy\,beam$^{-1}$) and in (d) are [3, 9, 27, 81, 250, 500, 750]$\times\sigma$ (1$\sigma$ = 60.0 $\mu$Jy\,beam$^{-1}$). The $E$-vectors in all panels are set to have a uniform length. The synthesized beam size for continuum is denoted by a filled ellipse in the bottom left corner. The nearside of the disk is on the red-shifted CO emission. The orange arrow in the upper right corner indicates the large-scale $E$-vector orientation presented by \cite{matthews2000magnetic}.}
\label{fig-co-1}
\end{figure*}


\renewcommand{\thefigure}{\arabic{figure} (Continued)}
\addtocounter{figure}{-1}

\begin{figure*}[]
\gridline{
          \fig{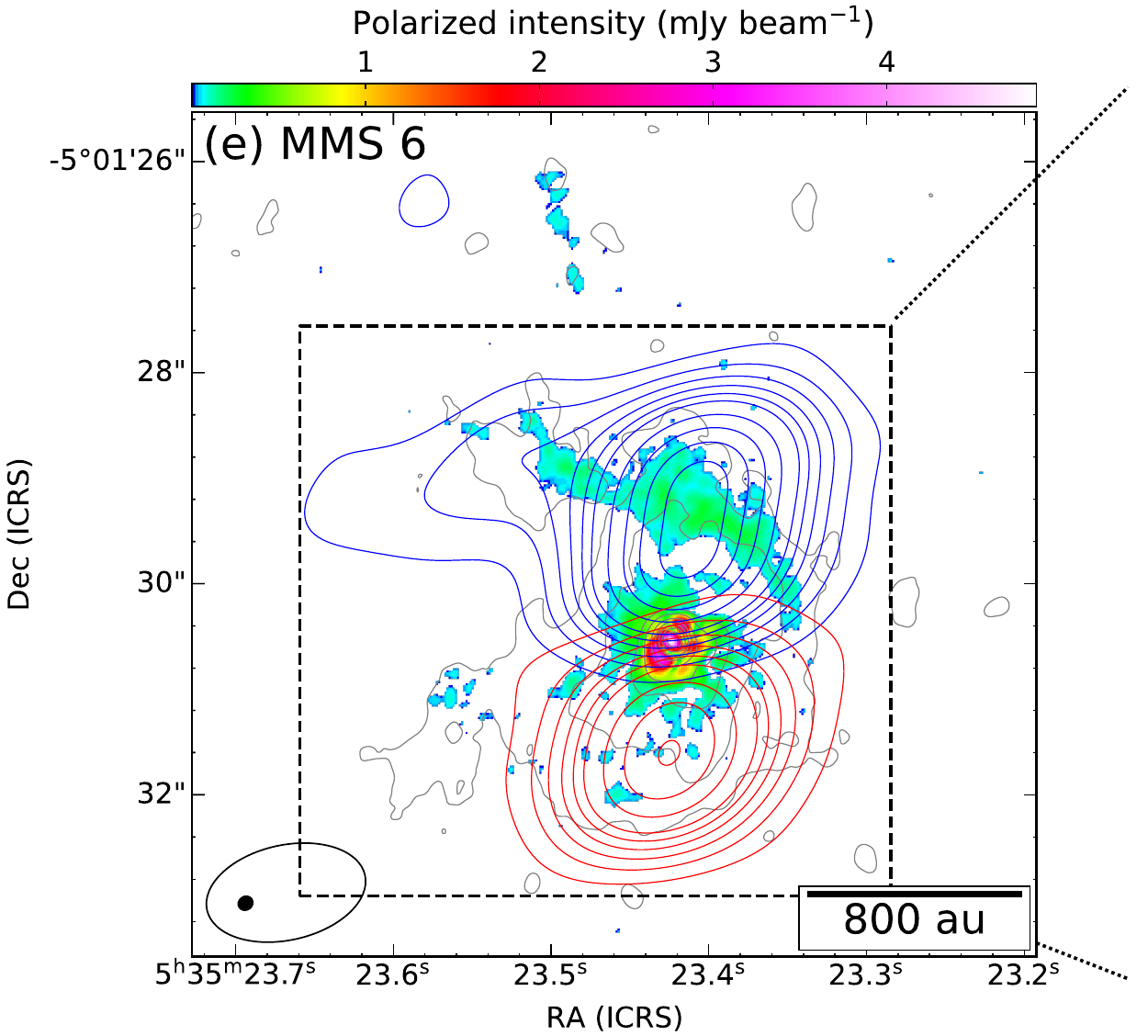}{0.51\textwidth}{}\hspace{-1\baselineskip}
          \fig{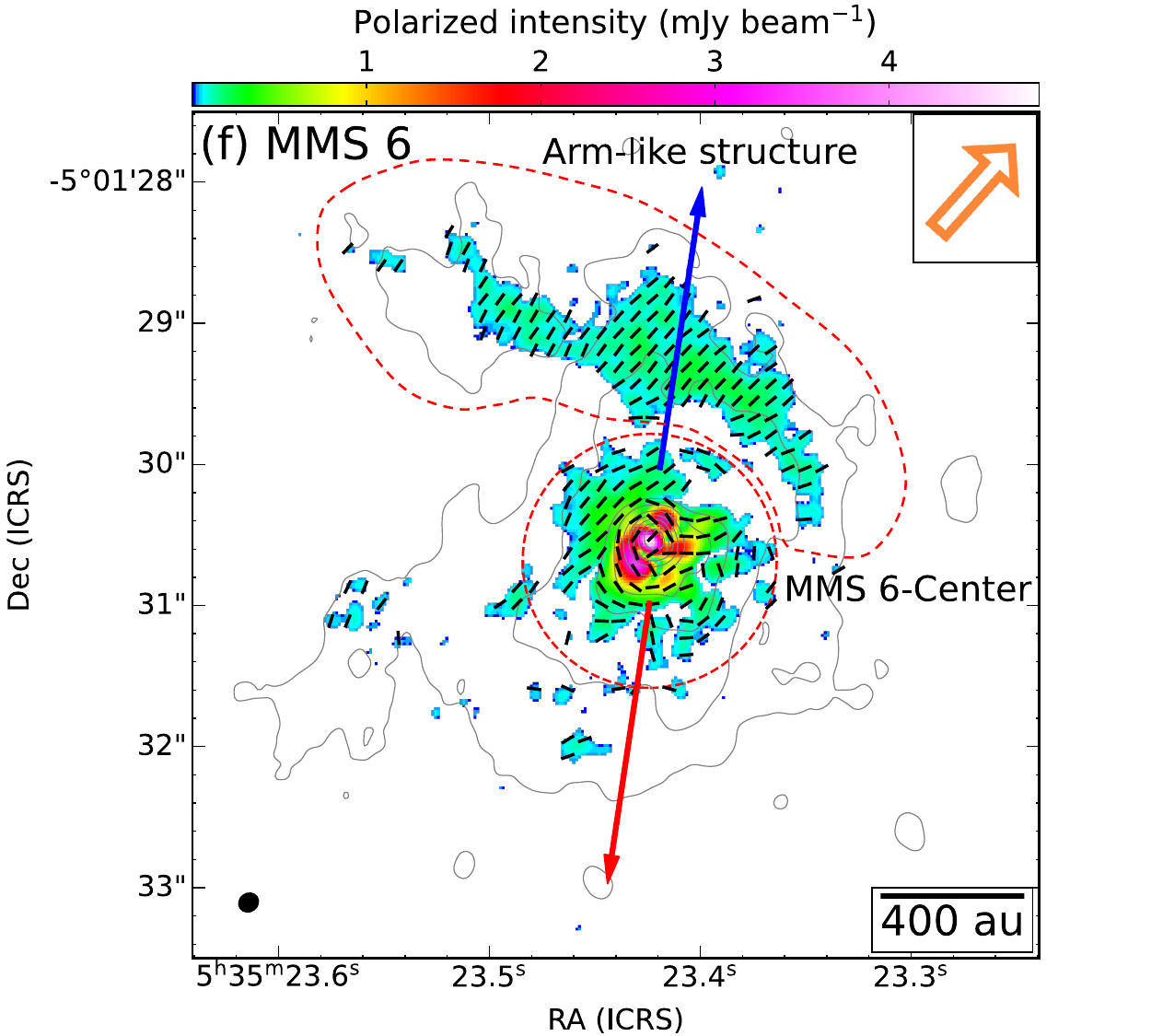}{0.51\textwidth}{}
          }
\vspace{-1.5\baselineskip}
\gridline{
          \fig{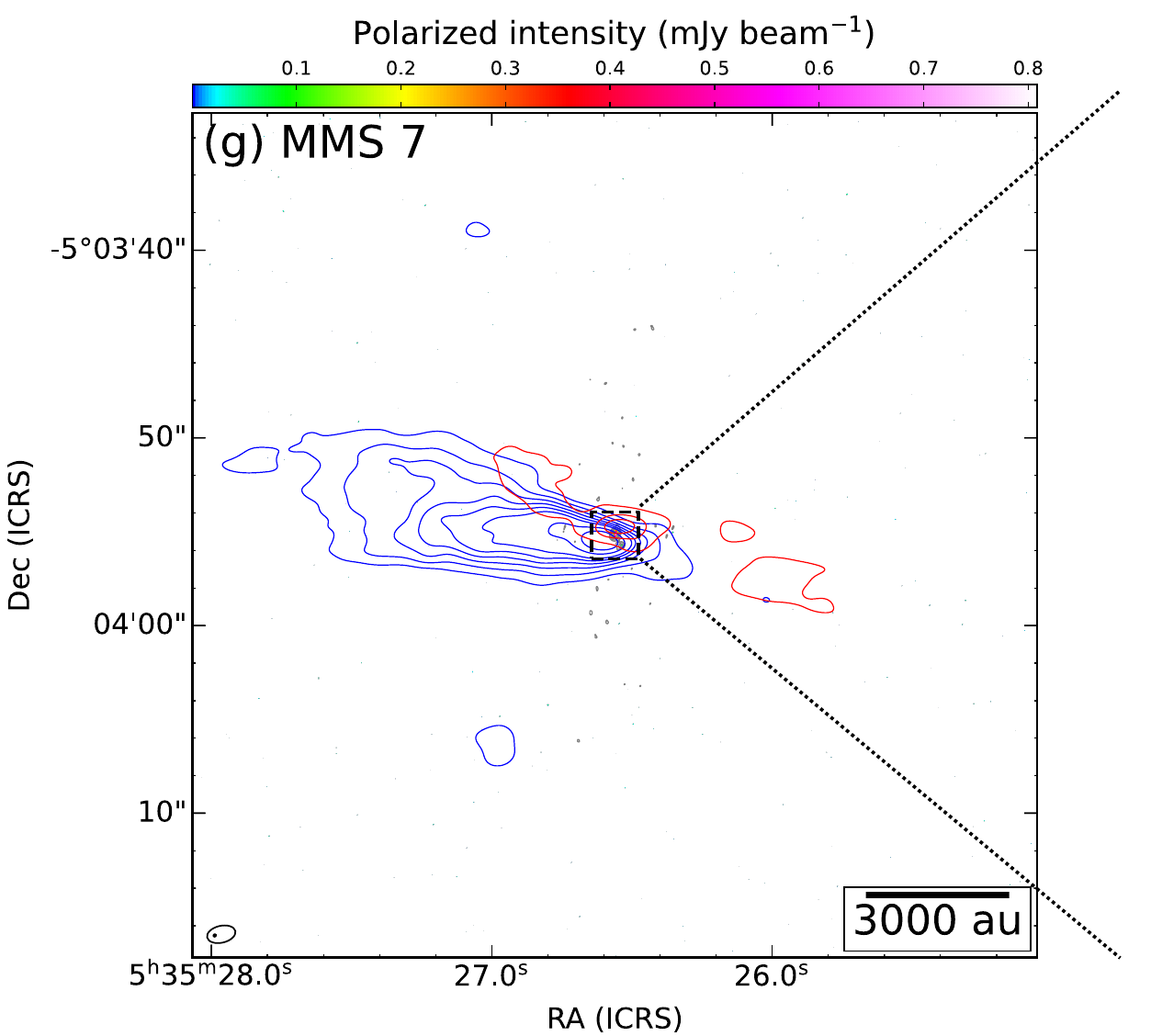}{0.51\textwidth}{}\hspace{-1\baselineskip}
          \fig{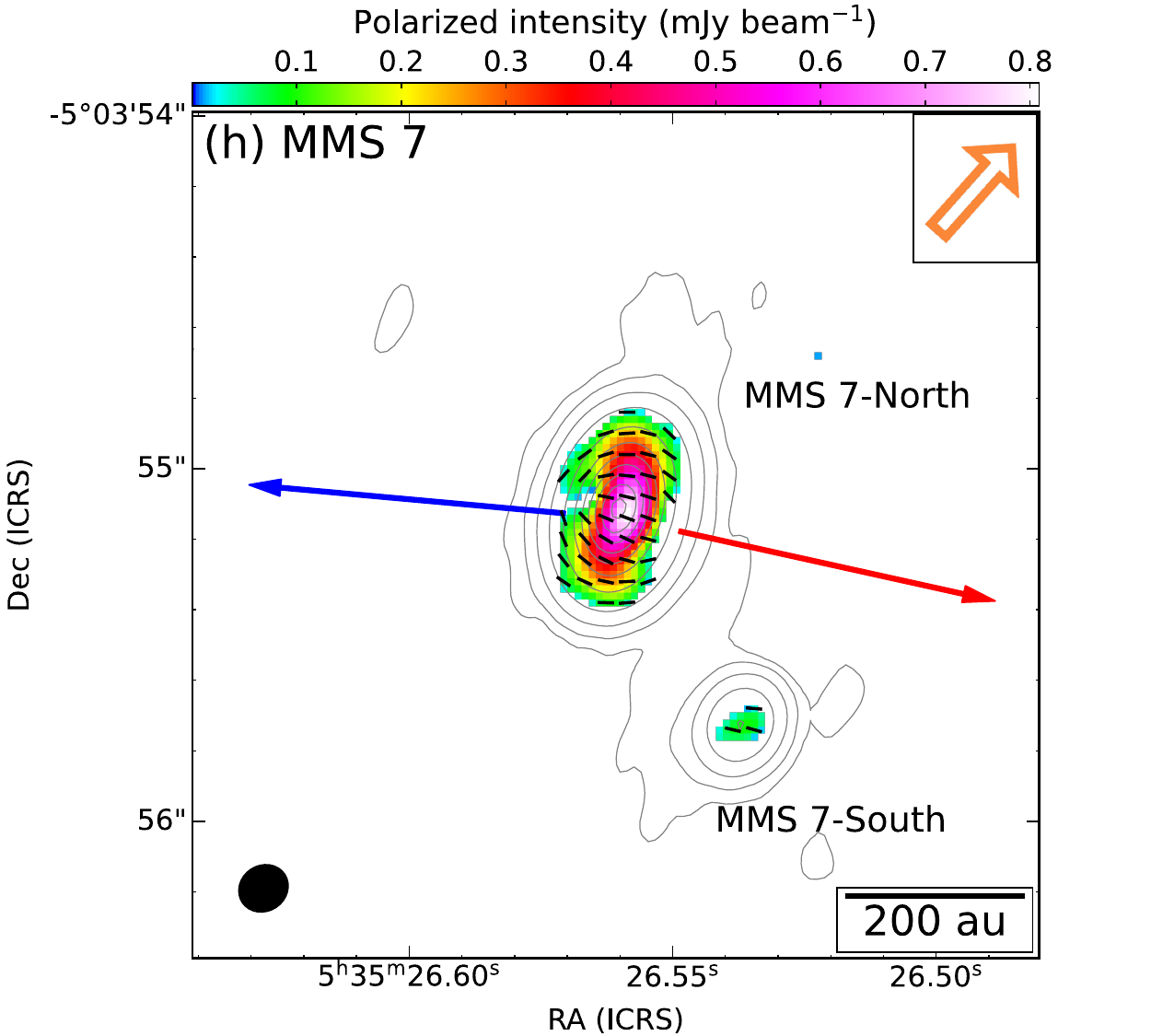}{0.51\textwidth}{}
          }
\vspace{-1.5\baselineskip}
\caption{Left panels: The blue and red contours are the integrated intensity of the CO~($J$ = 2$-$1) emission tracing the outflow. The CO emission in (e) is integrated from $v_{\textnormal {LSR}}-v_{\textnormal {sys}}$ = $-$5 to $-$40 km\,s$^{-1}$ (blue) and $v_{\textnormal {LSR}}-v_{\textnormal {sys}}$ = 5 to 50 km\,s$^{-1}$ (red), in (g) is integrated from $v_{\textnormal {LSR}}-v_{\textnormal {sys}}$ = $-$1 to $-$13 km\,s$^{-1}$ (blue) and $v_{\textnormal {LSR}}-v_{\textnormal {sys}}$ = 1 to 7 km\,s$^{-1}$ (red), respectively. The grey contours in (e) are the Stokes $I$. The contour levels in (e) are [5, 10, 15, 20, 25, 30, 40, 50, 60, 70]$\times\sigma$ (1$\sigma$ = 0.0925 Jy\,beam$^{-1}$ (red and blue)) and [5, 25, 125, 250, 500, 750, 1000, 1250, 1500, 1750]$\times\sigma$ (1$\sigma$ = 75.6 $\mu$Jy\,beam$^{-1}$ (grey)). The contour levels in (g) are [5, 10, 15, 20, 25, 30, 40, 50]$\times\sigma$ (1$\sigma$ = 0.11 Jy\,beam$^{-1}$ (red and blue)). The colour shows the PI. The synthesized beam size for continuum is denoted by a filled dot, and for CO, it is denoted by an empty ellipse in the bottom left corner. Right panels: Zoomed-in images of left panel images overlaid with $E$-vectors (black line segments) and contours of Stokes $I$ (grey). The blue and red arrows show the outflow axes of the blue- and red-shifted CO components. The contour levels in (f) are [5, 25, 125, 250, 500, 750, 1000, 1250, 1500, 1750]$\times\sigma$ (1$\sigma$ = 75.6 $\mu$Jy\,beam$^{-1}$) and in (h) are [3, 9, 27, 81, 250, 500, 750, 1000, 1250, 1500, 1750]$\times\sigma$ (1$\sigma$ = 49.3 $\mu$Jy\,beam$^{-1}$). The $E$-vectors in all panels are set to have a uniform length. The synthesized beam size for continuum is denoted by a filled ellipse in the bottom left corner. The nearside of the disk is on the red-shifted CO emission. The orange arrow in the upper right corner indicates the large-scale $E$-vector orientation presented by \cite{matthews2000magnetic}.}
\label{fig-co-2}
\end{figure*}


\subsubsection{MMS 2}
MMS\,2 was first identified by \cite{chini1997dust} in the 1300\,$\mu$m survey, also known as CSO\,6 \citep{lis1998350} and SMM\,3 \citep{takahashi2013hierarchical} as listed in Table~\ref{table:targets-summary} fourth column. The 10\,$\mu$m observations found that this object is a binary system consisting of two infrared sources showing the Class I-type spectral index\footnote{Spectral index ($\alpha$) describes an exponential factor relating the flux density of a radio source to its frequency, which is used to classify young stellar objects (YSOs) into the evolutionary stages in the near- and mid-infrared regions, given by $\alpha=\frac{d\textnormal{log}(\lambda F_{\lambda})}{d(\textnormal{log($\lambda$)})}$, where $\lambda$ is the wavelength and $F_{\lambda}$ is the flux density \citep[e.g.,][]{lada1987,andre1994,greene1994}.} \citep{nielbock2003stellar}. The separation between these two sources is $1\dotarcsec3$ \citep{nielbock2003stellar}. MMS\,2 was also identified as HOPS-92 by \cite{furlan2016herschel} through the multi-wavelength infrared observations. MMS\,2 was subsequently classified as a flat-spectrum source based on the SED study, in which its mid-infrared (4.5$–$24 $\mu$m) spectral index is between $-\,0.3$ and 0.3 \citep{furlan2016herschel}. The bolometric luminosity of this source is estimated to be $\sim$17.6 $L_{\odot}$ \citep{furlan2016herschel}. In the recent 0.87\,mm ALMA observations, HOPS-92 was further resolved into three sources, HOPS-92-A-A, HOPS-92-A-B, and HOPS-92-B \citep{tobin2020vla}. 
A bipolar CO outflow with one-sided length of $\sim$0.2 pc (P.A. $\sim$ 90$^{\circ}$) elongated along the east-west direction, was observed in previous studies \citep{aso2000dense,williams2003high,takahashi2008millimeter,tanabe2019nobeyama}. Recent CARMA-Nobeyama Radio Observatory Orion (CARMA-NRO Orion) survey reported a more elongated one-sided length of $\sim$0.48 pc associated with this outflow \citep{feddersen2020carma}. Furthermore, the Infrared Imager (IRIM) by \cite{yu1997shock} detected an east-west oriented outflow consisting of a chain of H$_2$ knots associated with MMS\,2.

As shown in Figure~\ref{fig-stokesIQU-3}\,(d), we have detected centrally condensed structures associated with all three HOPS sources in Stokes $I$, in which our observations are consistent with the 0.87\,mm ALMA observations \citep{tobin2020vla}. In our observations, MMS\,2-North-A, MMS\,2-North-B, and MMS\,2-South are associated with HOPS-92-A-A, HOPS-92-A-B, and HOPS-92-B, respectively. MMS\,2-North-B is detected 0$\dotarcsec$3 southeast with respect to MMS\,2-North-A. This source is fitted with five components as listed in Table~\ref{tab:Summary of the fitting results}. Both MMS\,2-North-A and MMS\,2-South are fitted with a single compact component and a single extended component, while MMS\,2-North-B is fitted with only a single compact component. MMS\,2-North-A compact component has a linear size of $\sim$39 au $\times$ 36 au (P.A. $=$ 149$^{\circ}$), and is likely tracing a nearly face-on disk with an inclination angle of $i \sim 26^{\circ}$. The fitted compact component of MMS\,2-South has a linear size of $\sim$59 au $\times$ 40 au (P.A. = 173$^{\circ}$), which likely traces a disk with an inclination angle of $i \sim 46^{\circ}$. The extended emission of MMS\,2-North-A has a fitted linear size of $\sim$244 au $\times$ 204 au (P.A. $=$ 13$^{\circ}$), while the extended emission of MMS\,2-South has a linear size of $\sim$96 au $\times$ 72 au (P.A. $=$ 5.5$^{\circ}$). Both the northern and southern extended emissions show a spiral arm-like structure, which is detected at the 3$\sigma$ level in Stokes $I$. These spiral arm-like structures seem to bridge the two compact millimeter sources, which becomes more clear when we produce the image with the natural weighting (see Figure~\ref{mms2-natural} in Appendix~\ref{mms2natural}). MMS\,2-North-B is one the weakest source and has a very compact structure with a linear size of $\sim$16 au $\times$ 2.8 au (P.A. = 101$^{\circ}$). We consider this component is unresolved because the fitting error is larger than its fitted deconvolved size (see also Appendix~\ref{fitting}).

Figure~\ref{fig-co-1}\,(a) shows the outflows associated with MMS\,2-North-A and MMS\,2-South, which are traced by the CO emissions detected at $\gtrsim 5\sigma$ level in the velocity range of $v_{\textnormal {LSR}}-v_{\textnormal {sys}}$ = $-$17 to 33 km\,s$^{-1}$ (see Figure~\ref{channel-mms2} in Appendix~\ref{co-image}). We detected the outflows elongated mainly along the east-west direction, in which the blue- and red-shifted components are overlapping. These results are consistent with previous single dish observations \citep{aso2000dense,williams2003high,takahashi2008millimeter,tanabe2019nobeyama,feddersen2020carma}. In addition, we resolved the red-shifted components, which are possibly associated with MMS\,2-North-A and MMS\,2-South, respectively. The red-shifted component on the eastern side likely traces the outflow cavity and appears to be associated with MMS\,2-North-A, while the red-shifted component in the southwest seems to be more associated with MMS\,2-South. We could not spatially resolved the blue-shifted components associated with these two sources because the blue-shifted components seem to merge with each other and appears as one along the east-west direction. We propose a scenario for MMS\,2 CO outflow (see the illustration in the bottom left of Figure~\ref{fig-co-1}\,(a)). For MMS\,2-North-A, the red-shifted (red) and blue-shifted (dark blue) components are detected from both eastern and western sides (see Figures~\ref{channel-mms2}\,(d)$-$(g)). The red- and blue-shifted components can be connected by the parabolic lines, respectively. Moreover, MMS\,2-North-A has a nearly face-on disk and the red- and blue-shifted components are overlapping. We think that this could be a nearly pole-on outflow elongated along the north-south direction. MMS\,2-South has a blue-shifted (cyan) component with P.A. $\sim$ 90$^{\circ}$, and a red-shifted (magenta) component in the southwest direction (see the illustration in the bottom left of Figure~\ref{fig-co-1}\,(a)).

As shown in Figure~\ref{fig-co-1}\,(b), PI is detected for both MMS\,2-North-A and MMS\,2-South, while no PI is detected towards MMS\,2-North-B. PI of MMS\,2-North-A shows a ring-like structure and the peak of the PI is located within the ring (P.A. = 13$^{\circ}$). The PI peak associated with MMS\,2-South is offset to the southwest by $\sim$0$\dotarcsec$02 with respect to the Stokes $I$ peak. The PI peak shifts towards the red-shifted outflow, which corresponds to the nearside of the disk (consider that the outflow is ejected in the perpendicular direction of the disk). The PI peak shift is larger than the positional accuracy by a factor of $\sim$5. The orientations of the $E$-vectors are completely different among these two sources (see Figure~\ref{fig-co-1}\,(b)). The $E$-vectors of MMS\,2-North-A trace a ring-like structure in an azimuthal direction, whereas the $E$-vectors of MMS\,2-South are almost parallel to the minor axis of its disk in the central disk region. The P.A. of the $E$-vectors for MMS\,2-South peaked at 66$^{\circ}$$-$87$^{\circ}$ (see Figure~\ref{hist-pa}\,(b)). Additionally, the $E$-vectors of MMS\,2-South are distributed azimuthally in the outer disk region. 

Figure~\ref{fig-fraction}\,(a) shows the polarization fraction detected towards MMS\,2. The polarization fractions are detected in the ranges of 0.4$-$1.6\% and 0.8$-$2.4\% for MMS\,2-North-A and MMS\,2-South, respectively (see Figure~\ref{hist-pfrac}\,(a) and (b)). The differences in the polarization fractions among these two sources could be induced by the inclination effect, which will be discussed in Section~\ref{tendency}.

\subsubsection{MMS 3}
MMS\,3 was first identified by \cite{chini1997dust} in the 1300\,$\mu$m survey, also known as CSO\,7 \citep{lis1998350} and SMM\,4 \citep{takahashi2013hierarchical} as listed in Table~\ref{table:targets-summary} fourth column. This millimeter source was classified as Class 0 source by \cite{chini1997dust} with a $L_{\textnormal{bol}}$/$L_{\textnormal{smm}}$ of $<$ 200 \citep{andre1993submillimeter}. Multi-wavelength infrared observations by \cite{furlan2016herschel} found that HOPS-91 is associated with MMS\,3, and this source is classified as Class 0 source based on the SED study. The bolometric luminosity of MMS\,3 is estimated to be $\sim$3.6 $L_{\odot}$ \citep{furlan2016herschel}. The classification is in agreement with \cite{chini1997dust}. There was no sign of outflows towards this source, as had been reported by previous observations \citep{aso2000dense,williams2003high,takahashi2008millimeter,tanabe2019nobeyama,feddersen2020carma}. Recent ALMA observations by \cite{morii2021revealing} detected a centrally concentrated compact continuum source and also a compact bipolar outflow with one-sided length of 5800 au (P.A. $\sim$ 54$^{\circ}$) in CO~($J$ = 2$-$1) emission for the first time (see Figure~\ref{mom0-mms3} in Appendix~\ref{co-image}). They could confirm that MMS\,3 is a protostellar source not a prestellar source. \cite{morii2021revealing} also suggested that MMS\,3 is a Class 0 source with a lower accretion rate compared to the other protostellar sources in this region. In addition, MMS\,3 also has the lowest Stokes $I$ peak flux among all other protostellar sources observed in this region. 

Figure~\ref{fig-stokesIQU-3}\,(g) shows that we have detected both centrally concentrated and extended components associated with MMS\,3 in Stokes $I$. This source is fitted with one compact component and two extended components as listed in Table~\ref{tab:Summary of the fitting results}. The fitted compact component of MMS\,3 has a linear size of $\sim$64 au $\times$ 38 au (P.A. = 54$^{\circ}$), possibly traces a dust disk. This compact component was also detected in the 1.3\,mm continuum observations with an angular resolution of $\sim0\dotarcsec2$ (78 au), extending along the north-south direction and having a linear size of $\sim$176 au $\times$ 125 au \citep{morii2021revealing}. The size of this compact structure obtained from \cite{morii2021revealing} is about twice as large as ours. The differences in sizes could be attributed to the differences in angular resolutions as well as the fitting methods. First, the angular resolution of the observations by \cite{morii2021revealing} is slightly lower than ours. Second, \cite{morii2021revealing} fitted this source with a single component, which could result in a larger size than fitting with multiple components, this will be discussed in Section~\ref{tendency}. Furthermore, we detected two extended components. These two extended components have the linear sizes of $\sim$213 au $\times$ 161 au (P.A. = 9.7$^{\circ}$) and $\sim$601 au $\times$ 520 au (P.A. = 87$^{\circ}$), respectively. The larger extended component (P.A. = 87$^{\circ}$) presumably traces an envelope elongated along the east-west direction.

MMS\,3 is not detected in Stokes $Q$ and $U$ as shown in Figure~\ref{fig-stokesIQU-3}\,(h) and (i). The 3$\sigma$ upper limit of the polarization fraction is $\sim$2.1$\%$. This value is much higher than the minimum detectable linear polarization for both compact and extended sources, which means that our observations may be limited in sensitivity in order to detect the polarization fraction at a minimum of 3$\sigma$ level towards this source.

\renewcommand{\thefigure}{\arabic{figure}}
\begin{figure*}[ht!]
\gridline{
          \fig{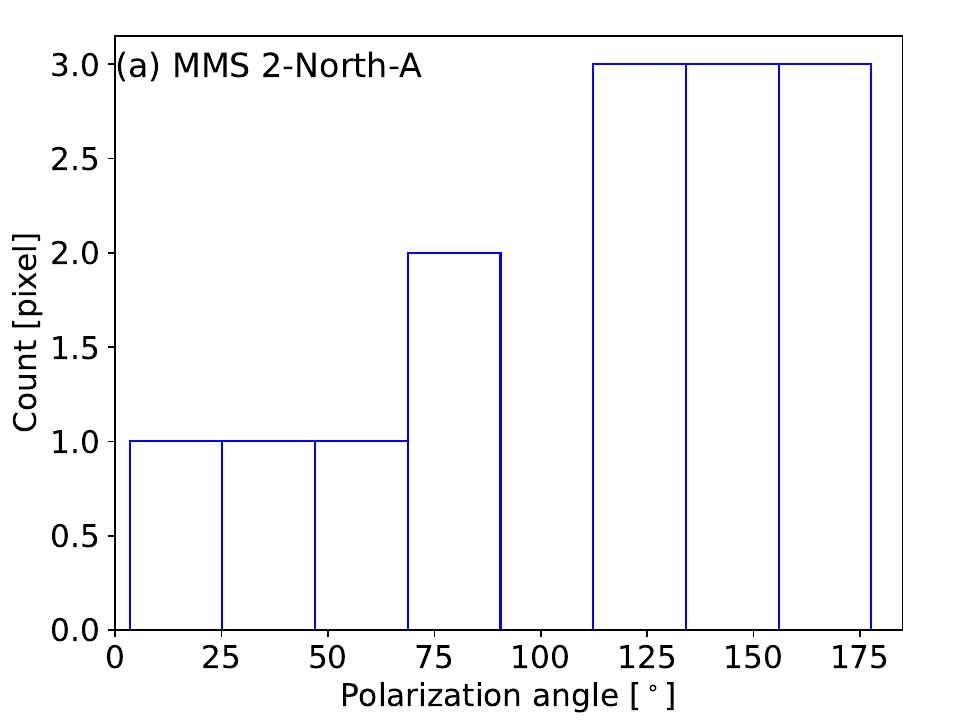}{0.35\textwidth}{}\hspace{-1\baselineskip}
          \fig{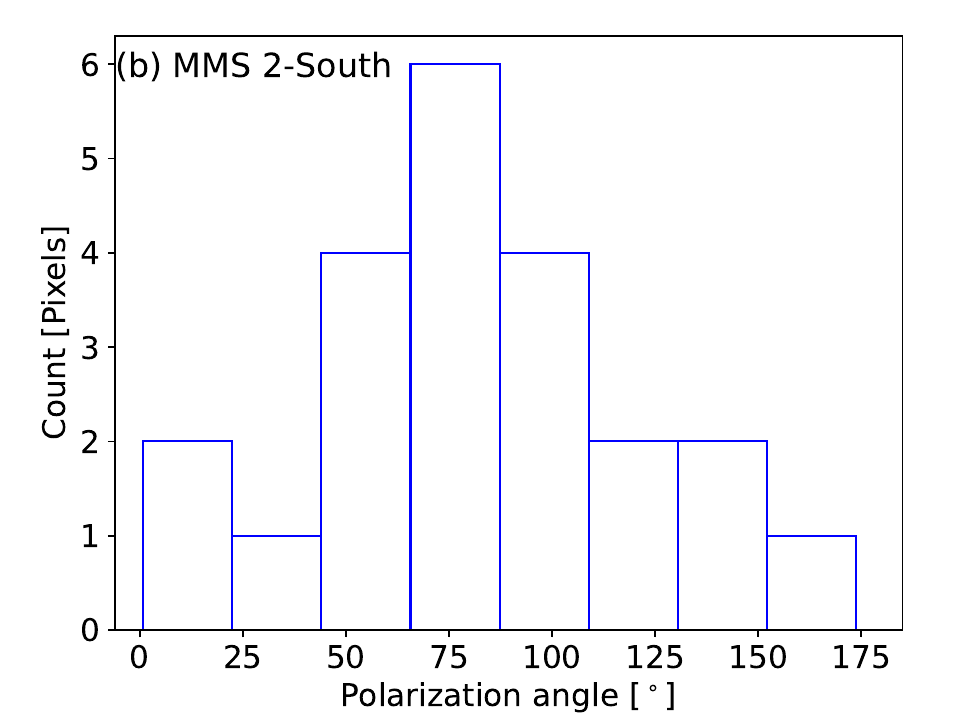}{0.35\textwidth}{}\hspace{-1\baselineskip}
          \fig{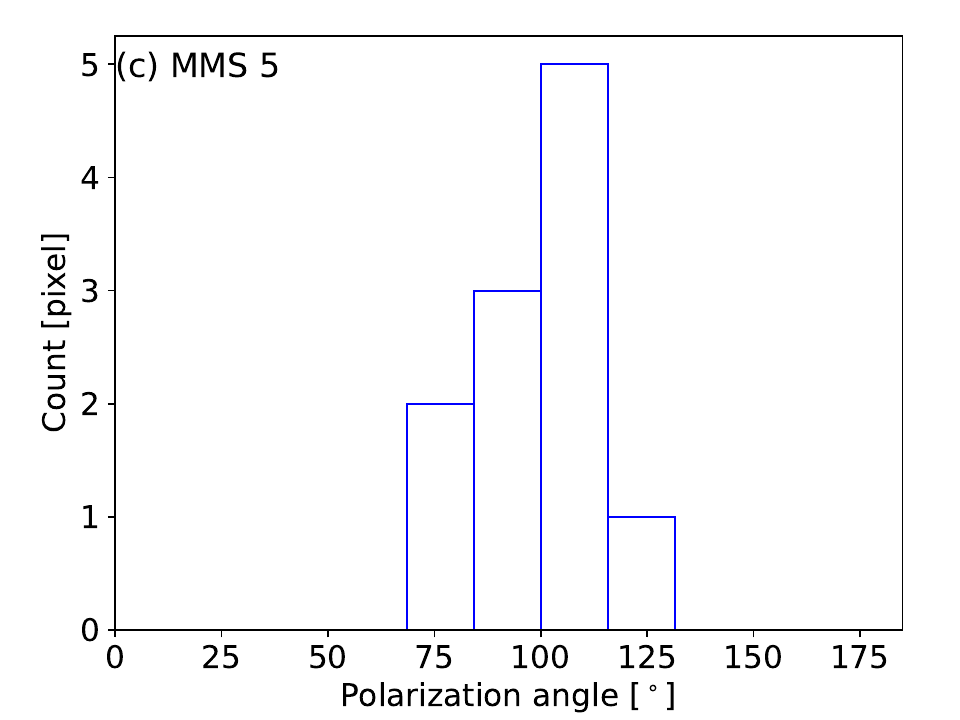}{0.35\textwidth}{}}
\vspace{-1\baselineskip}
\gridline{
          \fig{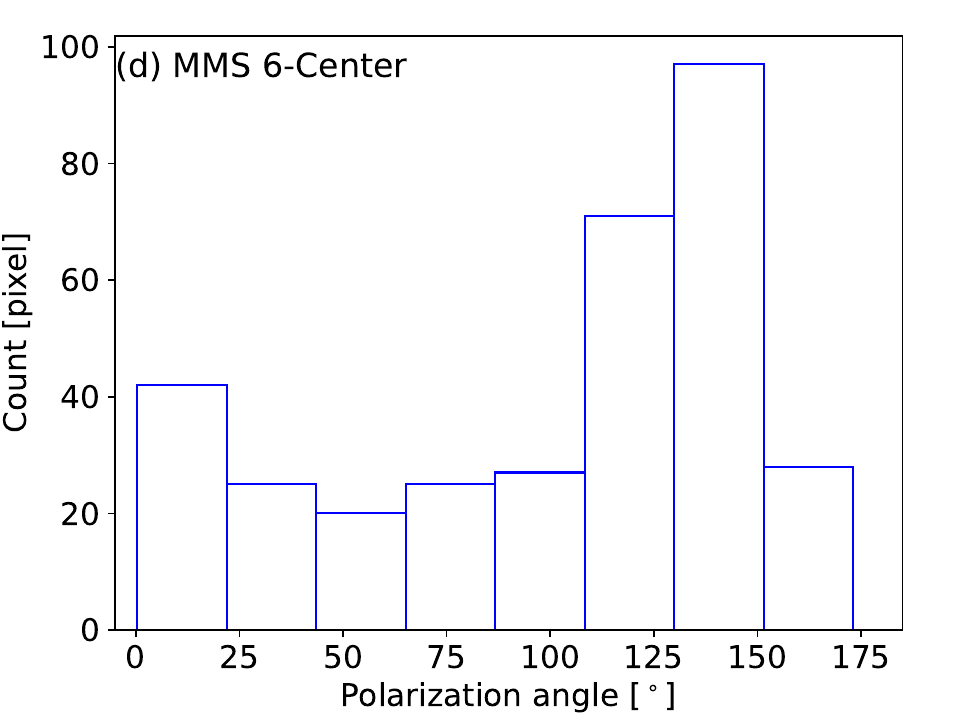}{0.35\textwidth}{}\hspace{-1\baselineskip}
          \fig{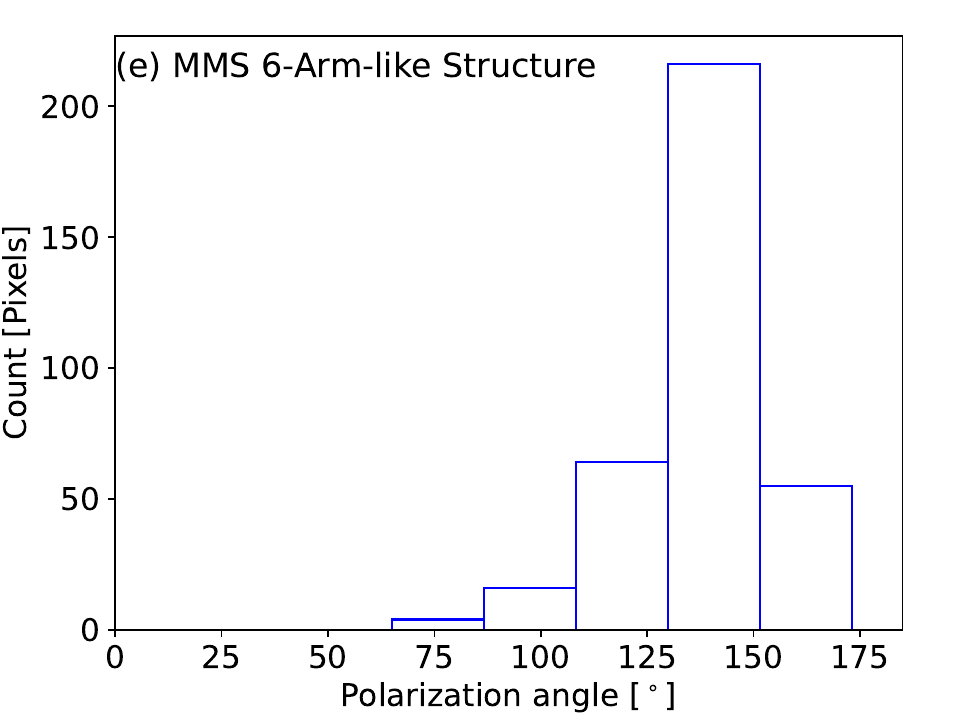}{0.35\textwidth}{}\hspace{-1\baselineskip}
          \fig{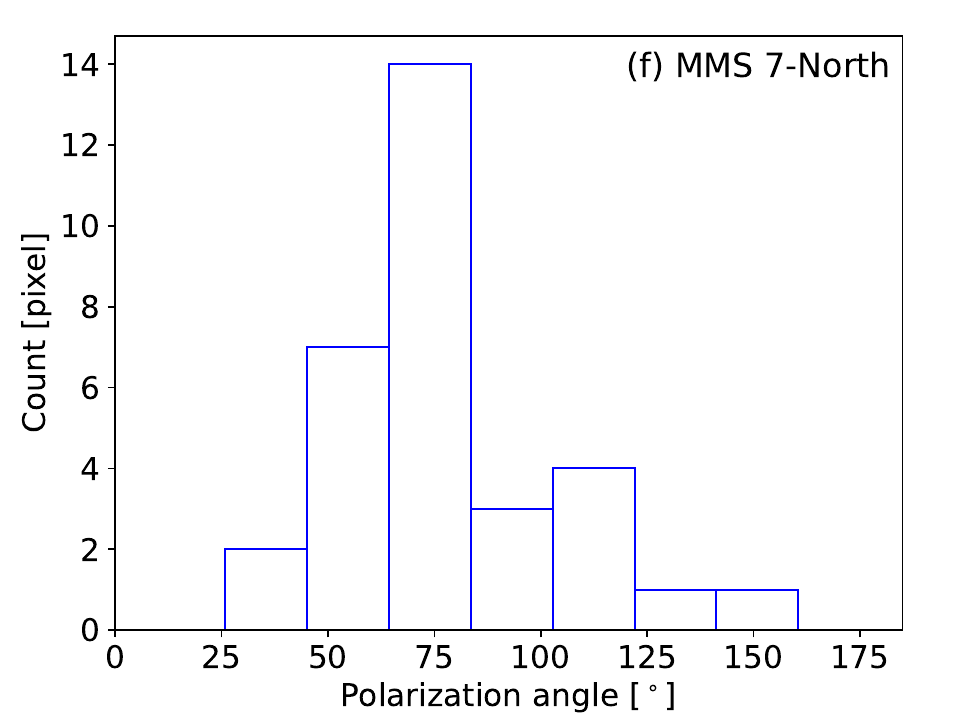}{0.35\textwidth}{}}
\vspace{-1\baselineskip}
\gridline{\hspace{-29.8\baselineskip}
          \fig{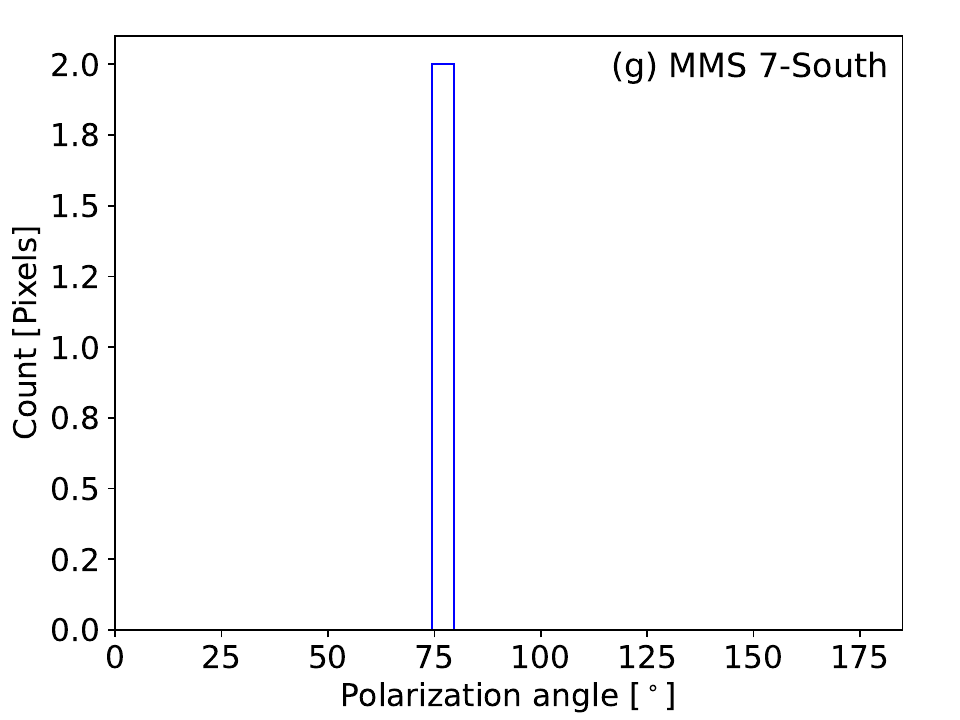}{0.35\textwidth}{}}
\vspace{-1\baselineskip}
\caption{Histograms of the $E$-vector position angles for MMS\,2-North-A, MMS\,2-South, MMS\,5, MMS\,6-Center, MMS\,6-Arm-like structure, MMS\,7-North, and MMS\,7-South. By taking into account of the 180$^{\circ}$ ambiguity in linear polarization position angles, we added 180$^{\circ}$ to any $E$-vector position angle that is $<0^{\circ}$ to better present the distributions of the $E$-vector position angles. The $E$-vector position angles are measured from north towards east.}

\label{hist-pa}          
\end{figure*}

\subsubsection{MMS 4}
MMS\,4 was first identified by \cite{chini1997dust} in the 1300\,$\mu$m survey and classified as Class 0 source based on the $L_{\textnormal{bol}}$/$L_{\textnormal{smm}}$ of $<$ 200 \citep{andre1993submillimeter}. This source was also identified as CSO\,8 \citep{lis1998350} and SMM\,5 \citep{takahashi2013hierarchical} as listed in Table~\ref{table:targets-summary} fourth column. There was no evidence of molecular outflow towards this source and no infrared counter part had been detected for this source \citep{yu1997shock,stanke2002unbiased,aso2000dense,williams2003high,takahashi2008millimeter,feddersen2020carma}. \cite{takahashi2013hierarchical} revealed that MMS\,4 has a triangular-shaped structure with a linear size of 7870 au $\times$ 2150 au. The gas number density of this source is a factor of 10 smaller than most of the sources observed in this region \citep{takahashi2013hierarchical}, which is comparable to the other prestellar cores \citep[e.g., $\sim$1$\times10^6$ cm$^{-3}$;][]{caselli2019}. Therefore, MMS\,4 was considered as a prestellar source and the youngest source among the sources presented in this paper. In addition, \cite{hirano2023} proposed that the nucleus of MMS\,4 is either on the brink of forming the first hydrostatic core or is a potential candidate for the first hydrostatic core. 

Figure~\ref{fig-stokesIQU-3}\,(j) shows an elongated structure with axis ratio of $\sim$0.1 detected towards MMS\,4 in Stokes $I$. This source is fitted with a relatively compact component and a largely extended component as listed in Table~\ref{tab:Summary of the fitting results}. We refer to the compact component detected in MMS\,4 as a ``$relatively$ compact component'' because it is larger than the compact components of the other sources by a factor of $\sim$4$-$50. The relatively compact component has a linear size of $\sim$511 au $\times$ 259 au (P.A. = 131$^{\circ}$), which is extended to the northwest with respect to the Stokes $I$ peak. In addition, there is a narrow and extended emission with a linear size of $\sim$4583 au $\times$ 546 au (P.A. = 120$^{\circ}$), which is elongated along the southeast-northwest direction. This elongated structure is consistent with the structure of SMM\,5 detected by \cite{takahashi2013hierarchical}, which likely traces either the outer part of the envelope or the inner part of a core. 

MMS\,4 is not detected in Stokes $Q$ and $U$ as shown in Figure~\ref{fig-stokesIQU-3}\,(k) and (l). This source is the youngest and the emission is also the weakest among all other sources. The 3$\sigma$ upper limit of the polarization fraction is $\sim$3.7$\%$, which is much higher than the minimum detectable linear polarization. As with MMS\,3, our observations may be limited in sensitivity.

Although MMS\,4 is identified as a prestellar source, it is possible that a compact core has not yet developed. Furthermore, the number density of $\sim$2 $\times 10^{9}$ cm$^{-3}$ is quite high, of which the value is comparable to that of the compact component of the Class 0 source MMS\,3 (number densities can be found in the last column of Table~\ref{tab:Summary of the Physical properties}). This value is also approaching the theoretical value expected for the first hydrostatics core \citep[$\gtrsim 10^{10}$ cm$^{-3}$; e.g.,][]{larson1969,matsunaga2000,machida2008,joos2012}. In addition, we have detected a low-velocity blue-shifted CO emission associated with the relatively compact Stokes $I$ emission of MMS\,4 (see Figure~\ref{mom0-mms4} in Appendix~\ref{co-image}). This blue-shifted CO emission is elongated towards the western direction with respect the Stokes $I$ peak, with a velocity ranging from $v_{\textnormal {LSR}}=6$ to $7$ km\,s$^{-1}$. The high number density and marginal detection of CO emission indicate that MMS\,4 might be in the first hydrostatics core stage, which is in agreement with \cite{hirano2023}.


\begin{figure*}[]
\gridline{\hspace{-1.5\baselineskip}
          \fig{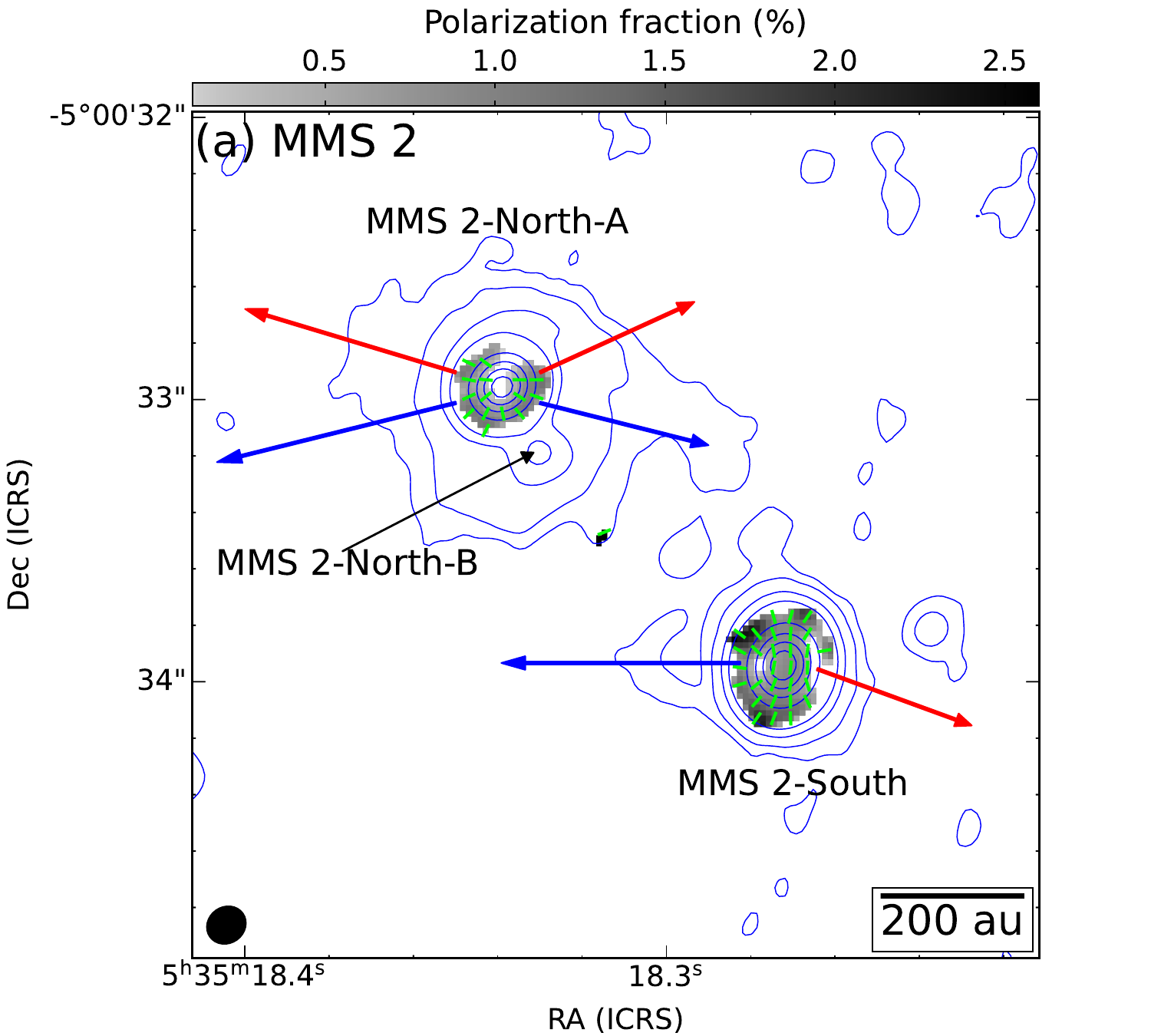}{0.55\textwidth}{}
          \hspace{-2\baselineskip}
          \fig{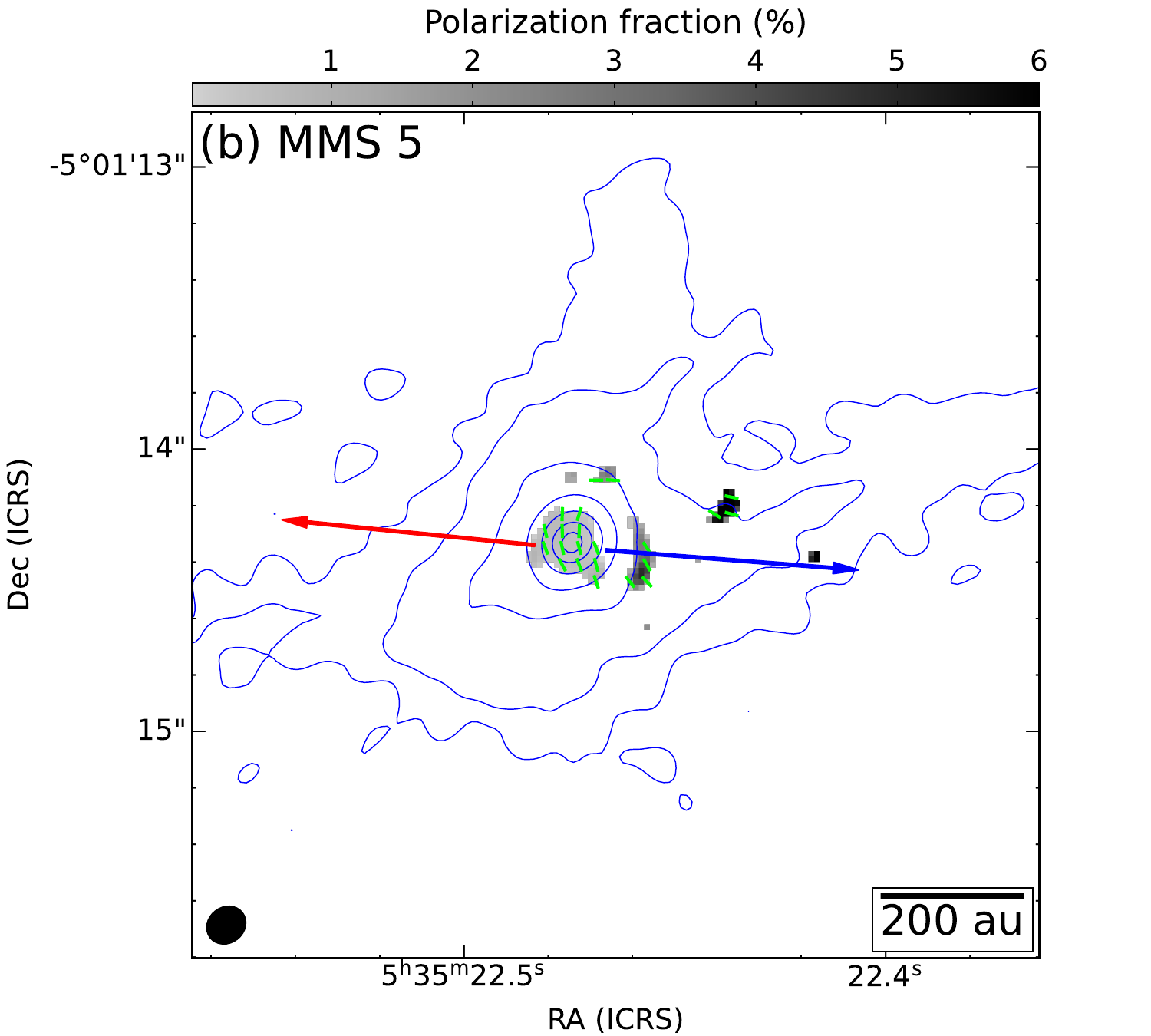}{0.55\textwidth}{}
          }
\vspace{-1\baselineskip}
\gridline{\hspace{-1.5\baselineskip}
          \fig{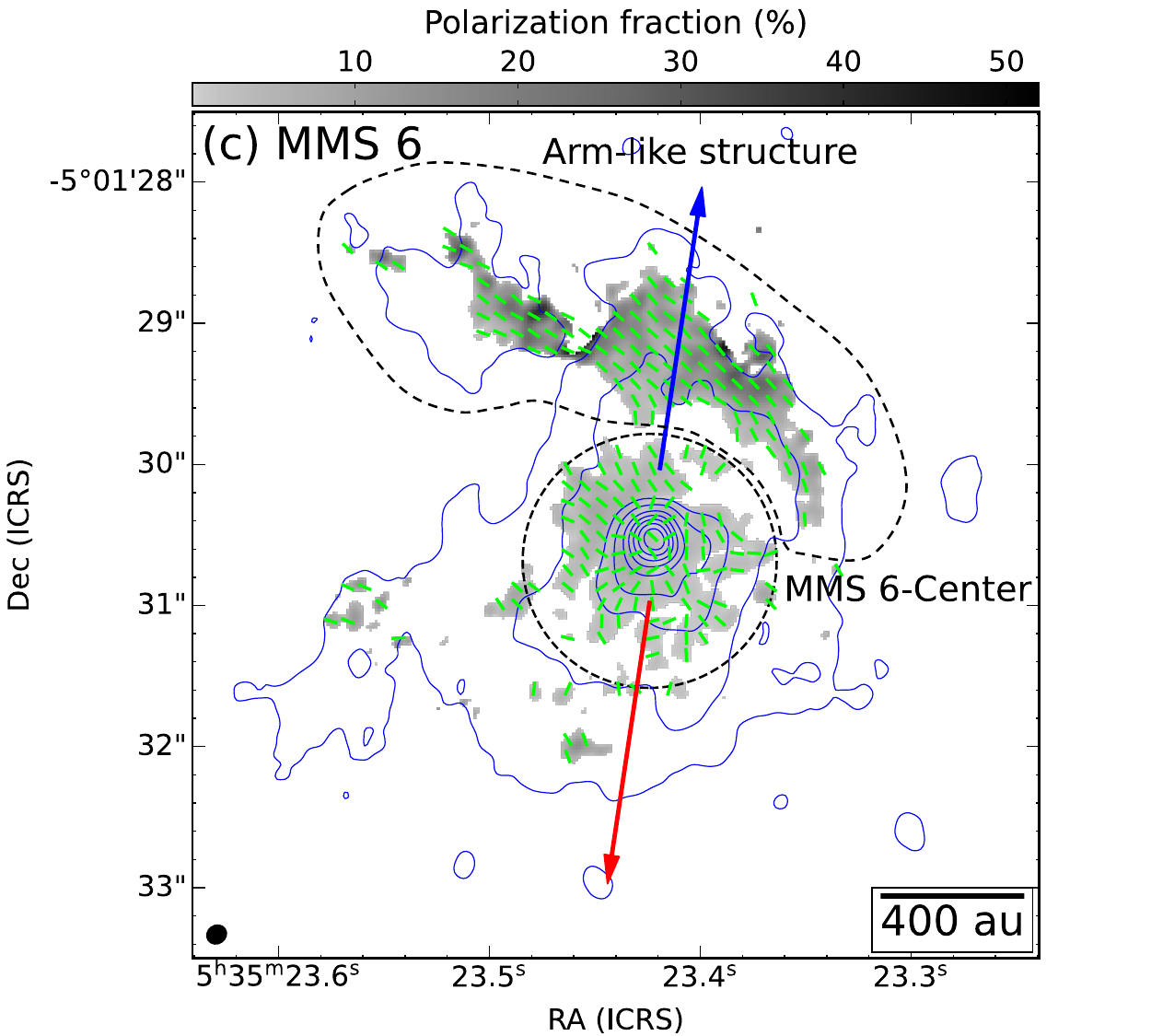}{0.55\textwidth}{}\hspace{-2\baselineskip}
          \fig{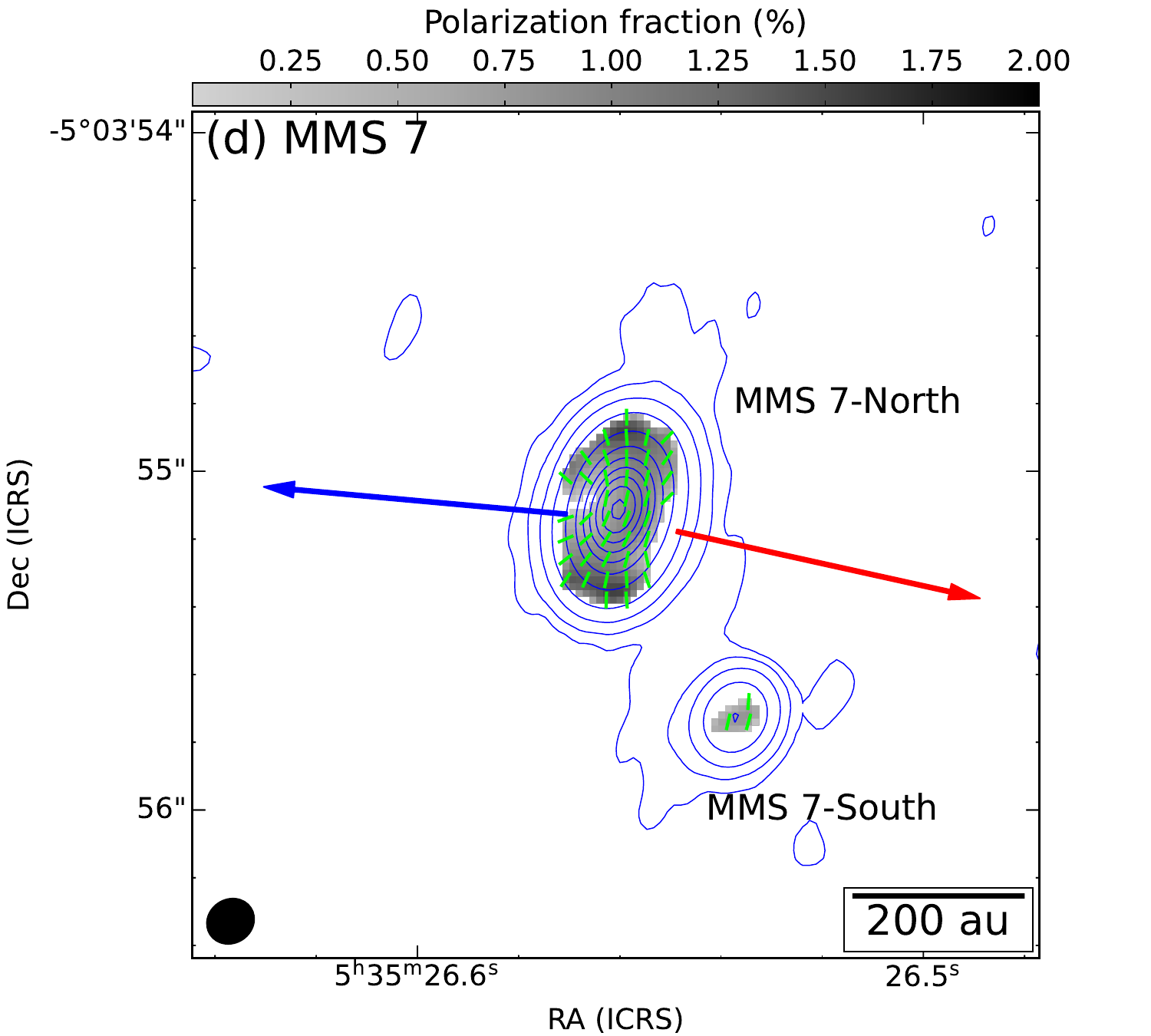}{0.55\textwidth}{}
          }
\vspace{-1\baselineskip}

\caption{Polarization fraction (greyscale) overlaid with the contours of Stokes $I$ (blue) and the inferred $B$-vectors (green line segments) for (a) MMS\,2, (b) MMS\,5, (c) MMS\,6, and (d) MMS\,7. The blue contours correspond to [3, 9, 27, 60, 260, 460, 660, 860, 1060]$\times\sigma$ (1$\sigma$ = 36.7 $\mu$Jy\,beam$^{-1}$) in panel (a), [3, 9, 27, 81, 250, 500, 750]$\times\sigma$ (1$\sigma$ = 60.0 $\mu$Jy\,beam$^{-1}$) in panel (b), [5, 25, 125, 250, 500, 750, 1000, 1250, 1500, 1750]$\times\sigma$ (1$\sigma$ = 75.6 $\mu$Jy\,beam$^{-1}$) in panel (c), and [3, 9, 27, 81, 200, 400, 600, 800, 1000, 1200, 1400, 1600]$\times\sigma$ (1$\sigma$ = 49.3 $\mu$Jy\,beam$^{-1}$) in panel (d). The inferred $\textit{B}$-vectors are obtained by rotating the $E$-vectors (in Figure~\ref{fig-co-1}) by 90$^{\circ}$, assuming that magnetic field is the polarization origin. The $B$-vectors in all panels are set to have a uniform length. The $B$-vectors and the polarization fraction are shown where the PI is $\ge$ 3$\sigma$ level. The red and blue arrows show the outflow axes for the red- and blue-shifted CO components as presented in Figure~\ref{fig-co-1}.}
\label{fig-fraction}          
\end{figure*}

\subsubsection{MMS 5}
MMS\,5 was identified by \cite{chini1997dust} in the 1300\,$\mu$m survey. This source is also known as CSO\,9 \citep{lis1998350} and SMM\,6 \citep{takahashi2013hierarchical} as listed in Table~\ref{table:targets-summary} fourth column. Multi-wavelength infrared observations by \cite{furlan2016herschel} found that HOPS-88 is associated with MMS\,5. The bolometric luminosity of this source is estimated to be $\sim$13.8 $L_{\odot}$ \citep{furlan2016herschel}. MMS\,5 is classified as a Class 0 source \citep{chini1997dust,furlan2016herschel}. A compact bipolar CO outflow with one-sided length of $\sim$0.1 pc (P.A. $\sim$ $-$90$^{\circ}$), which is elongated along the east-west direction, as reported in the CO outflow surveys by \cite{aso2000dense}, \cite{williams2003high}, \cite{takahashi2008millimeter}, \cite{tanabe2019nobeyama}, and \cite{feddersen2020carma}. Moreover, \cite{yu1997shock} and \cite{stanke2002unbiased} reported a chain of H$_2$ emission knots along the east-west direction, among which three knots associated with the blue-shifted lobe of the CO outflow were also reported by \cite{takahashi2008millimeter}. Recent ALMA CO~(2$-$1) observations by \cite{matsushita2019very} spatially resolved the CO outflow showing the cavity-like structure, which extends up to $\sim$14000 au (P.A. $\sim$ 79$^{\circ}$). Additionally, \cite{matsushita2019very} also detected an EHV jet, which extends up to $\sim10000$ au (P.A. $\sim$ 96$^{\circ}$) with a line-of-sight velocity of 100 km\,s$^{-1}$. The EHV jet was also detected in CO~(6$-$5) observations by \cite{gomez2019warm}. Furthermore, \cite{matsushita2019very} imaged an envelope on a $\sim$2000 au scale in the C$^{18}$O~(2$-$1) emission, revealing a rotational motion of the envelope. The rotation could be fitted by either Keplerian rotation or rotation due to angular momentum conservation \cite{matsushita2019very}.

As shown in Figure~\ref{fig-stokesIQU-3}\,(m), we have detected the Stokes $I$ emission associated with MMS\,5. The detected emission shows a centrally concentrated compact component with the V-shaped extended emission opening to the northwest direction. This source is fitted with a single compact component and a single extended component as listed in Table~\ref{tab:Summary of the fitting results}. The compact component has a linear size of $\sim$35 au $\times$ 28 au (P.A. = 1.6$^{\circ}$), which traces a dust disk around the protostar. The western arm of the extended V-shape structure seems to be partially correlated with the blue-shifted component of the EHV jet, while the northern arm of the extended V-shape structure is more or less correlated with the wide-opening blue-shifted component of the outflow. This implies that the extended emission traces the warm dust caused by the interaction between the outflow and the surrounding dense gas.

The molecular line observations for MMS\,5 were reported by \cite{matsushita2019very} as described earlier. Figure~\ref{fig-co-1}\,(c) shows the outflow and EHV jet associated with MMS\,5, which are traced by the CO emissions detected at $\gtrsim 5\sigma$ in the velocity ranges of $v_{\textnormal {LSR}}-v_{\textnormal {sys}}$ = $\pm$5 to $\pm$50 km\,s$^{-1}$ and $v_{\textnormal {LSR}}-v_{\textnormal {sys}}$ = $-$105 to 70 km\,s$^{-1}$, respectively (see Figure~\ref{channel-mms5}\, in Appendix~\ref{co-image}). Both blue- and red-shifted components of the outflow show a cavity-like structure as seen in Figure~\ref{fig-co-1}\,(c). The red-shifted component of the outflow has a projected length of $\sim$7000 au with P.A. $\sim$ 69$^{\circ}$ extends along the northeast-southwest direction, while the blue-shifted component of the outflow has a projected length of $\sim$7900 au with P.A. $\sim$ $-95^{\circ}$ extends along the east-west direction. The red-shifted component of the EHV jet has a length of $\sim$4300 au with P.A. $\sim$ 82$^{\circ}$, while the blue-shifted component of the EHV jet has a length of $\sim$6200 au with P.A. $\sim$ $-$98$^{\circ}$. 

As shown in Figure~\ref{fig-co-1}\,(d), PI is mainly detected around the Stokes $I$ peak in the central $\sim$0$\dotarcsec$25 (100 au) region, whereas a small portion of the outlying component is detected at the $\sim$0$\dotarcsec$3 western side with respect to the Stokes $I$ peak. We found that the PI exhibits a peak offset of $\sim$0$\dotarcsec$03 towards the nearside of the disk with respect to the Stokes $I$ peak (see Figure~\ref{fig-co-1}\,(d)). The peak shift is larger than the positional accuracy by a factor of $\sim$4. The $E$-vectors seem to peak at P.A. = 100$^{\circ}$$-$116$^{\circ}$ (see Figure~\ref{hist-pa}\,(c)), which are almost parallel to the minor axis of the centrally compact component traced by the Stokes $I$ emission. 

Figure~\ref{fig-fraction}\,(b) shows the polarization fraction detected towards MMS\,5. The high polarization fractions of $\gtrsim$\,4$-$6\% are mainly originated from the outlying PI component rather than from the central region of the disk. We plot the histogram with a focus only on the central 100 au region of MMS\,5 to avoid the interference of noise from those outlying regions. The centrally compact component has a polarization fraction of 0.4$-$1.3\% (see Figure~\ref{hist-pfrac}\,(c)).

\subsubsection{MMS 6}
MMS\,6 was first identified in the 1300\,$\mu$m survey by \cite{chini1997dust}. This millimeter source is also known as CSO\,12 \citep{lis1998350} and SMM\,8 \citep{takahashi2013hierarchical} as listed in Table~\ref{table:targets-summary} fourth column. Furthermore, this source was identified as HOPS-87 in the multi-wavelength infrared observations \citep{furlan2016herschel}, and was subsequently classified as a Class 0 source due to its positive mid-infrared (4.5–24 $\mu$m) spectral index and a bolometric temperature ($T_{\textnormal{bol}}$) of $<$ 70 K \citep{furlan2016herschel}. The H$_{2}$ knots possibly associated with MMS\,6 were reported by \cite{yu1997shock}, \cite{stanke2002unbiased}, and \cite{williams2003high}. The CO outflow was not clearly observed towards this source \citep{aso2000dense,williams2003high,matthews2005multiscale,takahashi2008millimeter}. MMS\,6 was resolved as MMS\,6-main and MMS\,6-NE in SMA 0.9\,mm continuum observations by \cite{takahashi2009}, for which MMS\,6-main was identified as a bright and the most compact source in the OMC-3 region. \cite{takahashi2012molecular} detected a very compact bipolar outflow with one-sided length of $\sim$1000 au in CO~(3–2) and HCN~(4–3) line observations towards MMS\,6-main. All above evidences suggest that MMS\,6 could be a protostellar core at the earliest stage \citep{takahashi2012molecular,takahashi2012spatially}.

As shown in Figure~\ref{fig-stokesIQU-3}\,(p), we have detected a centrally condensed Stokes $I$ emission associated with MMS\,6. In the northern region, we have also detected an extended emissions showing an arm-like structure, which is mostly elongated along the northeast-southwest direction. Here, we refer to the centrally condensed emission as MMS\,6-Center (within central $\sim$500 au region), and the extended emission as arm-like structure (see Figure~\ref{fig-stokesIQU-3}\,(p)). MMS\,6 is one of the youngest sources, while it has the highest Stokes $I$ peak flux among all other sources.

As shown in Figure~\ref{fig-co-1}\,(e), MMS\,6 is associated with the most compact CO outflow among MMS\,2, MMS\,5, and MMS\,7. The outflow traced by the CO emission is detected at $\gtrsim 5\sigma$ in the velocity range of $v_{\textnormal {LSR}}-v_{\textnormal {sys}}$ = $-$40 to 50 km\,s$^{-1}$ (see Figure~\ref{channel-mms6} in Appendix~\ref{co-image}). At $v_{\textnormal {LSR}}-v_{\textnormal {sys}}$ = 50 km\,s$^{-1}$, we have detected a red-shifted component that may trace an EHV jet (see Figure~\ref{channel-mms6}\,(j)). The projected lengths of the red- and blue-shifted CO components are comparable with each other, which the one-sided length is $\sim$1200 au with P.A. $\sim$ 6$^{\circ}$ (see Figure~\ref{fig-co-1}\,(e)). The high angular resolution images of the outflow and EHV jet traced by CO~(2$-$1) and SiO~(5$-$4) emissions obtained through the same project, will be presented in Takahashi et al.\,(2023b, in preparation).

In Figure~\ref{fig-co-1}\,(f), the PI associated with MMS\,6-Center exhibits a peak that is offset by $\sim$0$\dotarcsec$04 towards the nearside of the disk with respect to the Stokes $I$ peak. The PI peak shift is larger than the positional accuracy by a factor of $\sim$110. Such a high value results from its extremely large S/N ($\sim$175). $E$-vectors are peaked at P.A. = 130$^{\circ}$$-$150$^{\circ}$ (see Figure~\ref{hist-pa}\,(d)). We have also detected the arm-like structure in the PI from the northern extended emission. The PI associated with this arm-like structure extends up to $\sim$1400 au along the northeast-southwest direction. The $E$-vectors are mainly parallel to the minor axis of this extended arm-like structure \citep{takahashi2019alma}. 

\begin{figure*}[ht!]
\gridline{
          \fig{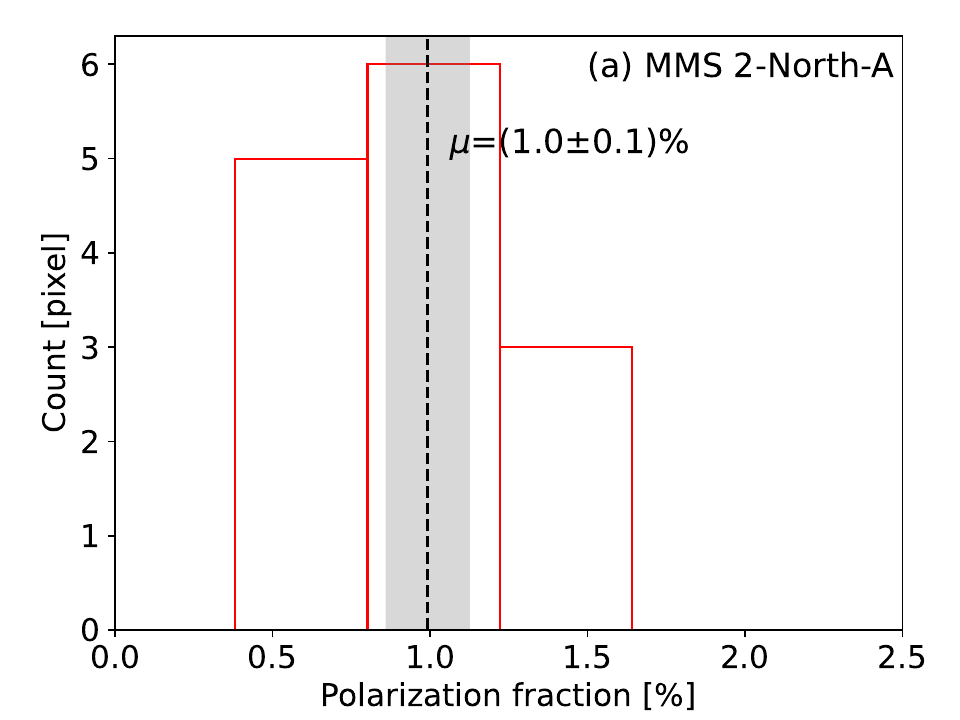}{0.35\textwidth}{}\vspace{-0.5\baselineskip}
          \fig{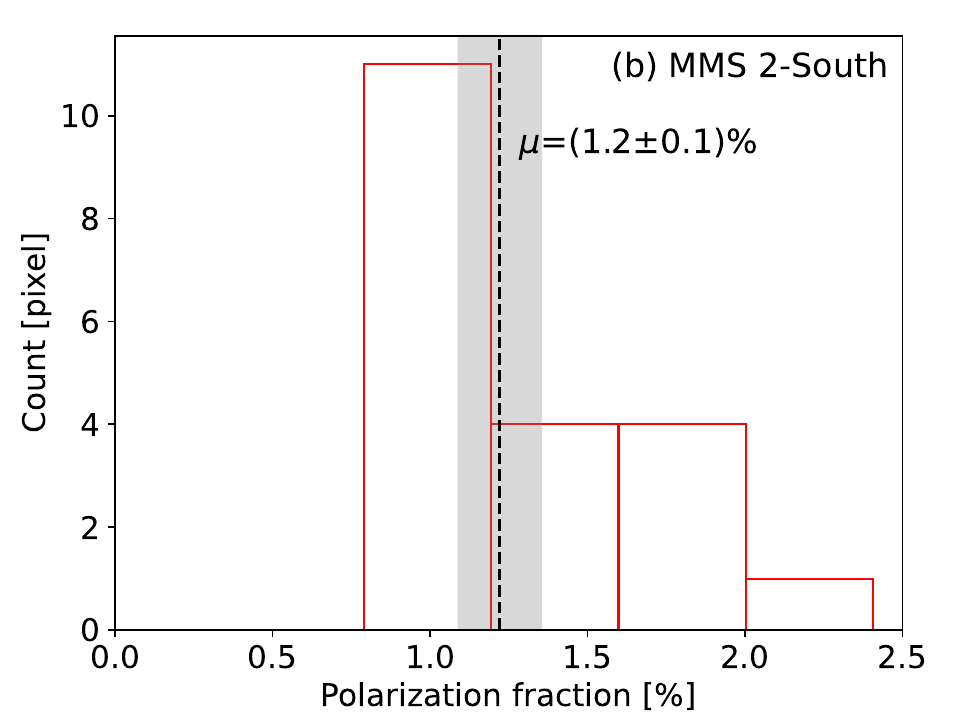}{0.35\textwidth}{}\vspace{-0.5\baselineskip}
          \fig{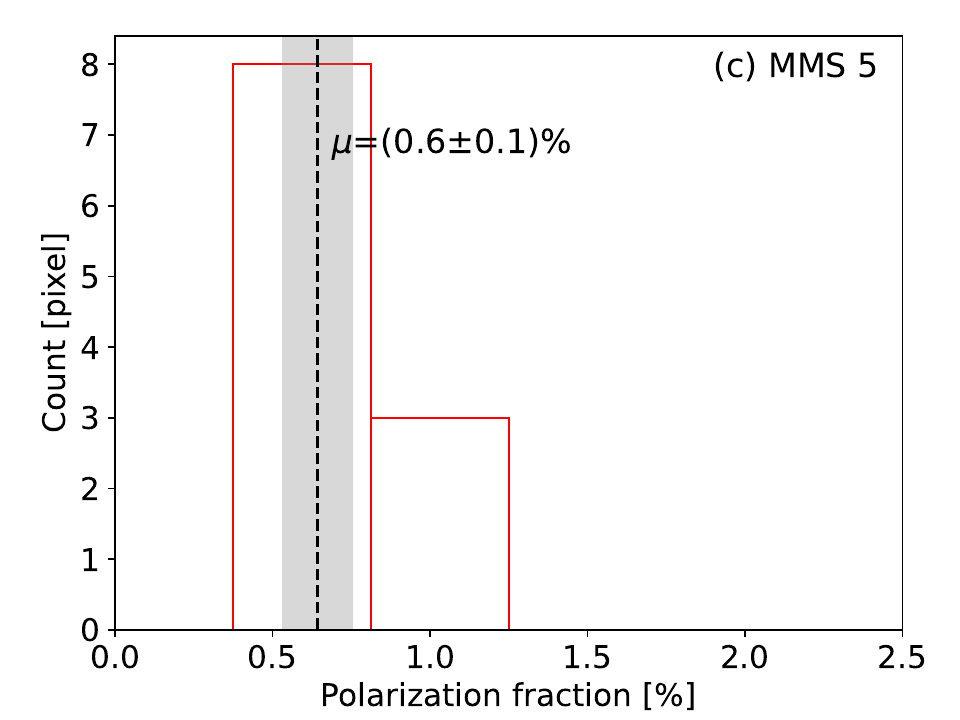}{0.35\textwidth}{}\vspace{-0.5\baselineskip}}
\gridline{
          \fig{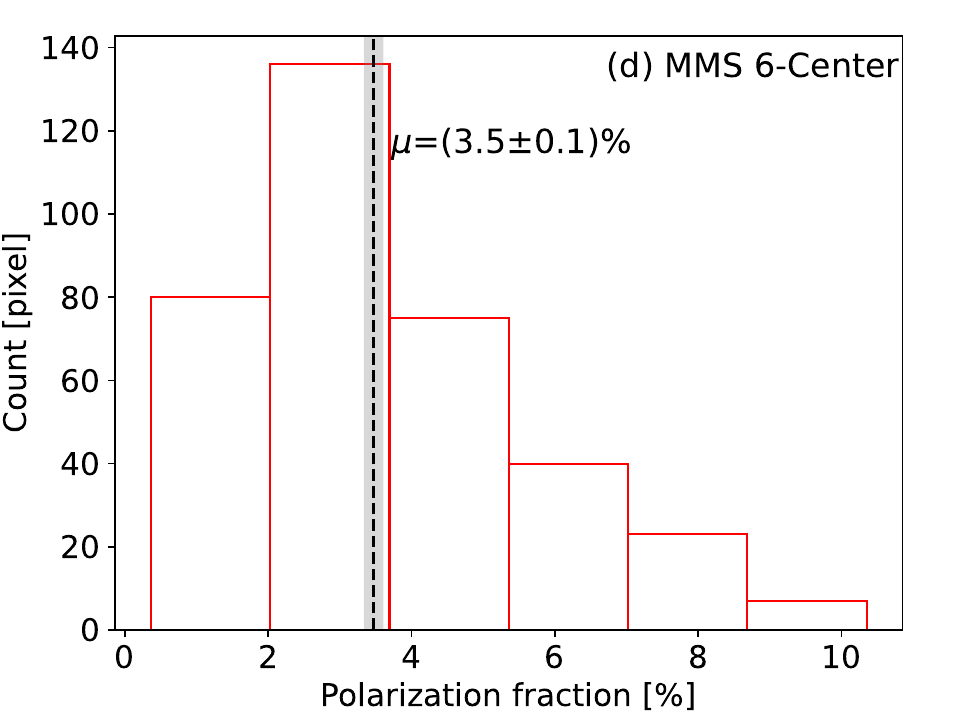}{0.35\textwidth}{}\vspace{-0.5\baselineskip}
          \fig{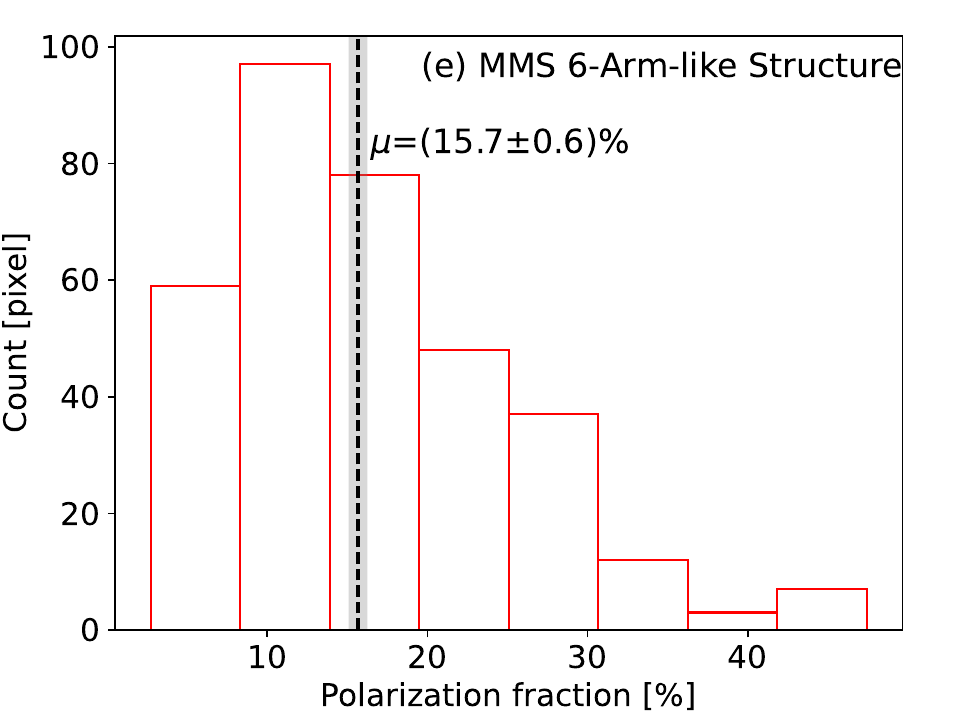}{0.35\textwidth}{}\vspace{-0.5\baselineskip}
          \fig{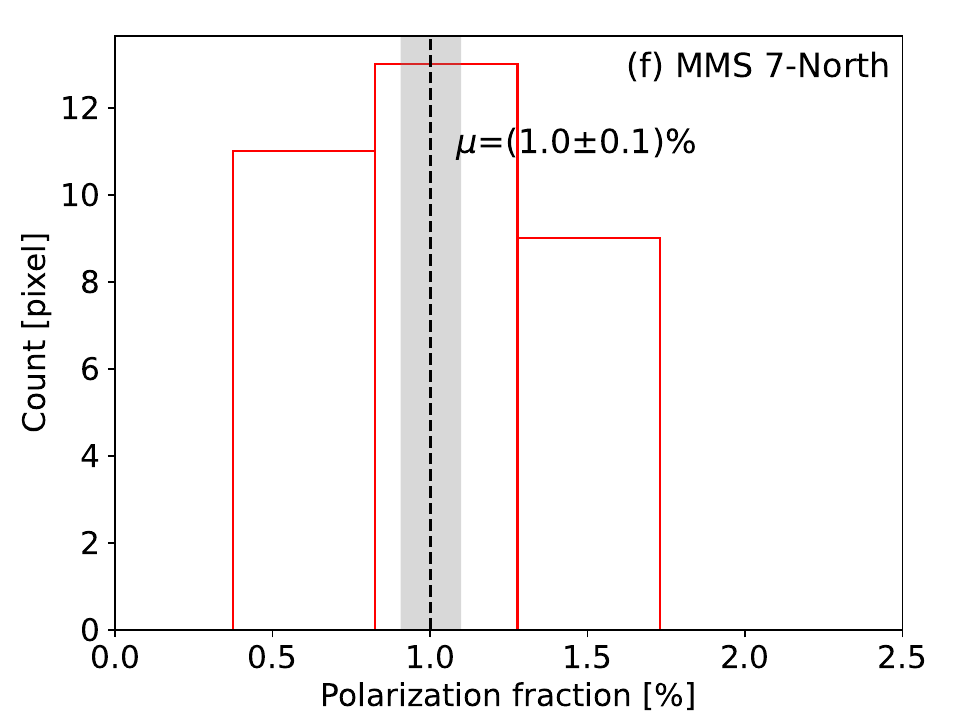}{0.35\textwidth}{}\vspace{-0.5\baselineskip}}
\vspace{-1\baselineskip}
\gridline{\hspace{-30\baselineskip}
          \fig{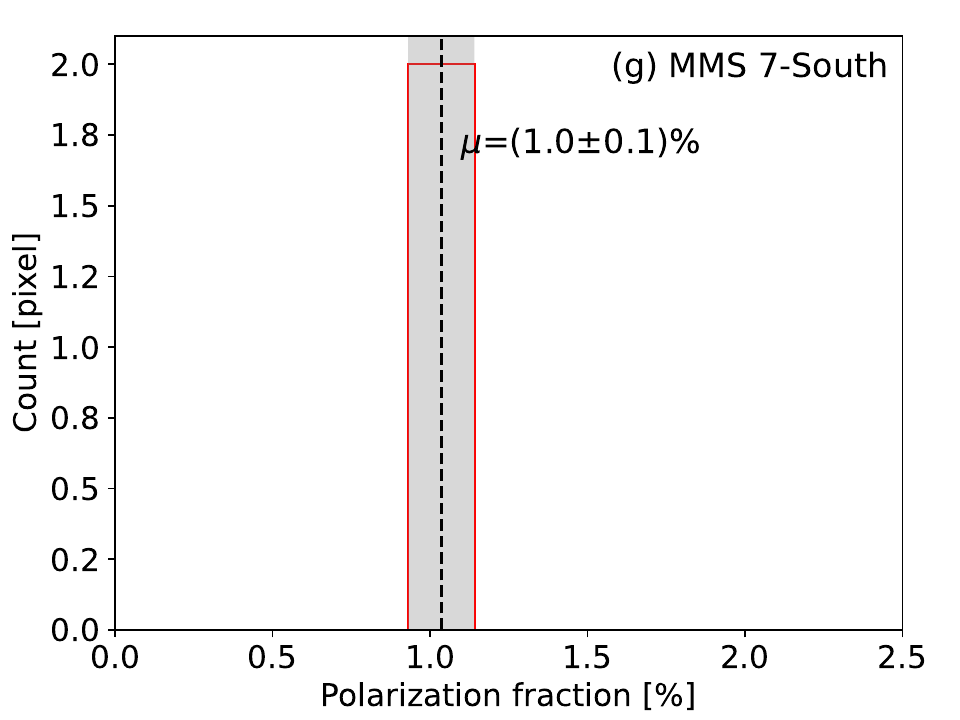}{0.35\textwidth}{}}
\vspace{-1\baselineskip}
\caption{Histograms of the polarization fractions for MMS\,2-North-A, MMS\,2-South, MMS\,5, MMS\,6-Center, MMS\,6-Arm-like structure, MMS\,7-North, and MMS\,7-South. The vertical dashed line and $\mu$ indicate the mean polarization fraction. The errors of the mean, which includes both statistical and systematic errors, is shown in each histogram and is denoted by the shaded region. The method for calculating the mean and its errors is explained in the beginning of Section~\ref{results}.}

\label{hist-pfrac}          
\end{figure*}

As shown in Figure~\ref{fig-fraction}\,(c), MMS\,6-Center exhibits a high polarization fraction of 0.4$-$10.4\%, and the arm-like structure displays a significantly high polarization fraction of 2.7$-$47.4\% (See Figure~\ref{hist-pfrac}\,(d) and (e)). MMS\,6 polarization analysis was thoroughly presented by \cite{takahashi2019alma}.

\subsubsection{MMS 7}
MMS\,7 was first identified by \cite{chini1997dust} in the 1300\,$\mu$m survey, and was classified as Class 0 source based on the $L_{\textnormal{bol}}$/$L_{\textnormal{smm}}$ of $<$ 200 \citep{andre1993submillimeter}. This source is also known as CSO\,12 \citep{lis1998350} and SMM\,11 \citep{takahashi2013hierarchical} as listed in Table~\ref{table:targets-summary} fourth column. The multi-wavelength infrared observations by \cite{furlan2016herschel} found that HOPS-84 is associated with MMS\,7. This source was subsequently classified as Class I source due to its Class I-type SED \citep{furlan2016herschel}. The bolometric luminosity of MMS\,7 is estimated to be $\sim$42.9 $L_{\odot}$ \citep{furlan2016herschel}. This source was detected as a single source in previous observations   \citep{chini1997dust,lis1998350,takahashi2006millimeter,takahashi2008millimeter,takahashi2013hierarchical,furlan2016herschel}. Recent ALMA observations by \cite{tobin2020vla} resolved HOPS-84 into two sources, HOPS-84-A and HOPS-84-B in the 0.87\,mm continuum image. In this paper, MMS\,7-North and MMS\,7-South are associated with HOPS-84-A and HOPS-84-B, respectively. A giant bipolar molecular outflow elongated along the east-west direction was observed in the CO observations \citep{aso2000dense,williams2003high,takahashi2006millimeter}, and the red-shifted lobe of the CO outflow is significantly more elongated than the blue-shifted lobe \citep{takahashi2008millimeter, tanabe2019nobeyama, feddersen2020carma}. The size of the blue-shifted lobe is $\sim$0.12 pc and the red-shifted lobe extends up to $\sim$0.83 pc \citep{feddersen2020carma}. Furthermore, MMS\,7 was also identified as VLA\,4 in the 3.6\,cm VLA detection. The observed flux associated with VLA\,4 might be due to free-free emission from the radio jets \citep{reipurth1999vla,reipurth2004radio}. The faint knots along the CO outflow were also detected by \cite{takahashi2008millimeter}. \cite{takahashi2006millimeter} detected a disk-like envelope with a fan-shaped structure around MMS\,7 by Nobeyama Millimeter Array (NMA) in H$^{13}$CO$^{+}$(1$-$0) emission. Moreover, \cite{zhiyuan2021} identified the filamentary infall motion towards MMS\,7 region using the ALMA and SMA. 

As shown in Figure~\ref{fig-stokesIQU-3}\,(s), we have detected centrally condensed Stokes $I$ emissions associated with both MMS\,7-North and MMS\,7-South. MMS\,7-South is detected $\sim$0$\dotarcsec$7 (275 au) southwest with respect to MMS\,7-North. MMS\,7 is fitted with two components, in which both MMS\,7-North and MMS\,7-South are fitted with a single compact component as listed in Table~\ref{tab:Summary of the fitting results}. The compact component of MMS\,7-North has a linear size of $\sim$95 au $\times$ 42 au (P.A. = 166.5$^{\circ}$), which possibly traces a compact dust disk. MMS\,7-South has a linear size of $\sim$37 au $\times$ 17 au (P.A. = 165.8$^{\circ}$), which is about twice more compact than MMS\,7-North and also shows a disk-like structure. The disk of MMS\,7-North is elongated almost perpendicular to its associated outflow. The disks of both MMS\,7-North and MMS\,7-South are nearly edge-on with an inclination angle of $i \sim 64^{\circ}$ and $i \sim 60^{\circ}$, respectively. 

Figure~\ref{fig-co-1}\,(g) shows the outflow associated with MMS\,7, which is traced by the CO emission detected at $\gtrsim 5\sigma$ in the velocity range of $v_{\textnormal {LSR}}-v_{\textnormal {sys}}=-13$ to 7 km\,s$^{-1}$ (see Figure~\ref{channel-mms7} in Appendix~\ref{co-image}). We detected a blue-shifted component of the CO emission showing a cavity-like structure extended to $\sim$10000 au along the northeast-southwest direction with P.A. $\sim$ 80$^\circ$. The red-shifted component of the CO emission with P.A. $\sim$ 106$^\circ$ is detached into two parts along the northeast-southwest direction.

In Figure~\ref{fig-co-1}\,(h), PI is detected for both MMS\,7-North and MMS\,7-South. A strong PI of MMS\,7-North is detected, and its peak exhibits an offset of $\sim$0$\dotarcsec$02 towards the nearside of the disk with respect to Stokes $I$ peak. The peak shift is larger than the positional accuracy by a factor of $\sim$9. On the other hand, only a small and weak patch of PI is detected within the central 30 au region of MMS\,7-South. The $E$-vectors of both MMS\,7-North and MMS\,7-South are mainly parallel to the minor axes of their disks with P.A. = 64$^{\circ}$$-$84$^{\circ}$ and P.A. = 75$^{\circ}-$$80^{\circ}$, respectively (see Figure~\ref{hist-pa}\,(f) and (g)). Near the edge of the disk of MMS\,7-North, the orientation of the $E$-vectors is more or less in an azimuthal direction (see Figure~\ref{fig-co-1}\,(h)).

Figure~\ref{fig-fraction}\,(d) shows the polarization fraction detected towards MMS\,7. MMS\,7-North exhibits a polarization fraction of 0.4$-$1.7\%, and MMS\,7-South exhibits a polarization fraction of 0.9$-$1.1\% (see Figure~\ref{hist-pfrac}\,(f) and (g)). 



\subsection{Multi-scale E-vector orientations and outflow direction}\label{relation}

\subsubsection{Comparison of E-vectors at 100$-$1000 au and 10000 au}

In the right panels of Figure~\ref{fig-co-1}, we compare the $E$-vector orientations on the disk-scale ($\sim$55 au) for MMS\,2, MMS\,5, MMS\,6, and MMS\,7 with the $E$-vector orientation associated with the filamentary cloud-scale ($\sim$7500 au), as reported by the 850\,$\mu$m SCUMBA/JCMT observations towards OMC-3 region \citep{matthews2000magnetic}. We found that there is no clear correlation in the $E$-vectors between the cloud-scale and the disk-scale for MMS\,2, MMS\,5, MMS\,7, and MMS\,6-Center. Note that MMS\,6 image also shows the $E$-vector orientation of the extended arm-like structure coincides with the $E$-vector orientation observed in the filamentary cloud-scale (see Figure~\ref{fig-co-1}\,(f)). The non-correlation in the $E$-vector orientations between the filamentary cloud-scale and the disk-scale could potentially suggest different polarization origins between the filament and the individual sources. Alternatively, assuming that polarization on disk-scale is originated from the magnetic field, there might be no $B$-vector correlation between the cloud-scale and the disk-scale.

\subsubsection{Comparison of E-vector orientations at 100 au and outflow directions}
The right panels of Figure~\ref{fig-co-1} also present the relation between the $E$-vector orientation and the outflow axis. The $E$-vector orientations seem to correlate with the outflow axes for MMS\,2-South, MMS\,5, MMS\,6, and MMS\,7-North within a deviation of $\sim$$10^{\circ}$$-$$40^{\circ}$. If we assume the polarization origin is magnetic field\footnote{The dust polarization towards these sources can be explained by multiple mechanisms and the details of the polarization origins will be discussed in \ref{discuss-origin}. Here, we assume the magnetic field is the polarization origin to show the inferred magnetic field orientation.}, the orientation of the inferred $B$-vectors for each source is obtained by rotating the $E$-vectors by 90$^\circ$ as shown in Figure~\ref{fig-fraction}. The outflow axes are perpendicular to the inferred $B$-vectors for most of sources, which can be explained by the toroidal magnetic field configuration, as discussed in \cite{takahashi2019alma} for MMS\,6.


\subsection{Physical properties} \label{mass}
The dust mass $M_{\textnormal{dust}}$ of the each component is estimated as \\
\begin{equation}
M_{\textnormal{dust}} = \frac{F_{\lambda}d^2}{\kappa_{\lambda}B_{\lambda}(T_{\textnormal{dust}})},
\end{equation}

\noindent where $F_{\lambda}$ is the total flux density, $d$ = 393 pc is the distance to the source \citep{tobin2020vla}, $B_{\lambda}(T_{\textnormal{dust}})$ is the Planck function as a function of dust temperature $T_{\textnormal{dust}}$, $\kappa_{\lambda}$ is the dust opacity and we adopted $\kappa_{\lambda}= 0.00144$ cm$^{2}$g$^{-1}$ at $\lambda = 1.1$\,mm, which is interpolated from the coagulation model of dust grains coated with thin ice mantles at a gas density of 10$^{8}$ cm$^{-3}$ \citep{ossenkopf1994dust}. We assume a gas-to-dust mass ratio of 100. We estimate the dust temperature from previous studies with two different scenarios. One is from \cite{tobin2020vla}, where the dust is heated by stellar radiation. The dust temperature $T_{\textnormal{dust}}$ adopted from \cite{tobin2020vla} is

\begin{equation}
T_{\textnormal{dust}} = T_0 \left(\frac{L_\textnormal{bol}}{L_{\odot}}\right)^{0.25}\left(\frac{r}{50\textnormal{au}}\right)^{-0.46},
\end{equation}

\noindent where $T_{0}$ = 43 K, $L_\textnormal{bol}$ is the bolometric luminosity, and $r$ is the radius (we use the geometric mean of the semi-major and semi-minor axes of each deconvolved component). $L_\textnormal{bol}$ is obtained from the HOPS catalog. MMS\,1 is not detected in HOPS sources and given that the lowest $L_\textnormal{bol}$ among the detected HOPS sources in OMC-3 region is $<0.05$\,$L_{\odot}$, we assume that MMS\,1 has a bolometric luminosity of $L_\textnormal{bol}\simeq0.05$ $L_{\odot}$. The other scenario is that the dust is heated by disk accretion heating \citep{xu2021formation}. The dust temperatures adopted from disk accretion heating model are interpolated from Figure~16 (leftmost panel: the case of 0.1 $M_{\odot}$) in \cite{xu2021formation}. We notice that for most of the compact components, the temperatures derived from these two models are comparable. However, for disk accretion heating model, temperature is mostly affected by the inner region ($r$ $\lesssim$ 30 au) of the disk. Especially for MMS\,1 compact component, the temperature derived by disk accretion heating is about 10 times higher than that by stellar radiation. The extended components were not significantly affected by disk accretion heating. The derived dust temperatures for both models are listed in Table~\ref{tab:dust temperature}. Since many components have the radii larger than 30 au, we use the temperature derived from stellar radiation \citep{tobin2020vla} to estimate the dust mass. For the prestellar source MMS\,4, we consider that the dust is being heated by the external sources \citep{Zucconi2001}.

\begin{table*}[ht!]
{\small
\begin{center}
\caption{\small Dust Temperature ($T_{\textnormal{dust}}$)}
\label{tab:dust temperature}

\begin{tabular}{lcc}
\hline\hline \noalign {\smallskip}
 & Stellar Radiation Heating$^c$ & Disk Accretion Heating \\
& $T_{\textnormal{dust}}$ & $T_{\textnormal{dust}}$ \\
& [K] & [K] \\
\hline
MMS\,1 compact component  & 47.5$\pm$1.4 & 400\\
MMS\,1 extended component 1  & 15.6$\pm$0.2 & 19\\
MMS\,1 extended component 2 & 8.6$\pm$0.1 & 14\\
\hline
MMS\,2-North-A compact component & 138.5$\pm$0.4 & 140\\
MMS\,2-North-A extended component  & 60.8$\pm$0.9 & 18\\
MMS\,2-North-B compact component$^a$ & $\le$305.0 & 550\\
MMS\,2-South compact component & 122.9$\pm$0.3 & 95\\
MMS\,2-South extended component & 96.0$\pm$0.5 & 25\\ 
\hline
MMS\,3 compact component  & 82.2$\pm$1.4 & 90\\
MMS\,3 extended component 1  & 44.6$\pm$0.4 & 19 \\
MMS\,3 extended component 2 & 26.8$\pm$0.6 & 14\\
\hline
MMS\,4 relatively compact component$^b$  & - & -\\
MMS\,4 extended component$^b$  & - & -\\
\hline
MMS\,5 compact component  & 140.8$\pm$1.0 & 150\\
MMS\,5 extended component  & 50.7$\pm$0.5 & 17\\
\hline
MMS\,6 compact component  & 103.4$\pm$0.3 & 23\\
MMS\,6 extended component 1  & 70.6$\pm$0.3 & 18\\
MMS\,6 extended component 2 & 44.2$\pm$0.3 & 14\\
\hline
MMS\,7-North compact component 1  & 135.9$\pm$0.2 & 39\\
MMS\,7-North compact component 2  & 127.1$\pm$1.1 & 27\\
MMS\,7-South compact component & 205.8$\pm$7.8 & 220\\
\hline \noalign {\smallskip}
\end{tabular}
\end{center}
\footnotesize $^a${MMS\,2-North-B is unresolved, with only the upper limit on $T_{\textnormal{dust}}$ provided.}\\
\footnotesize $^b${MMS\,4 is a prestellar core and $T_{\textnormal{dust}}$ for MMS\,4 is extrapolated from \cite{Zucconi2001}}, in which the\\temperature for compact and extended components are 5 and 7 K, respectively. The dust is considered to be heated by the external sources such as cosmic rays, therefore the outer part has higher temperature than the inner part.\\
\footnotesize $^c${The errors for $T_{\textnormal{dust}}$ are obtained by propagating only the errors for deconvolved size.}  
}
\end{table*}

\begin{table*}[ht!]
{\small
\begin{center}
\caption{\small Physical Properties}
\label{tab:Summary of the Physical properties}

\begin{tabular}{lcccc}
\hline\hline \noalign {\smallskip}
 & Radius ($r$)$^a$ & Dust Mass ($\textit{M}_{\textnormal{dust}}$)$^b$ & Column Density ($N_{\textnormal{H}_2}$)$^c$& Number Density ($n_{\textnormal{H}_2}$)$^d$\\
& (au)  & (10$^{-4}$ \(M_\odot\)) & (10$^{25}$ cm$^{-2}$)& (10$^{10}$ cm$^{-3}$)\\
\hline
MMS\,1 compact component	&7.9$\pm$0.5	&	0.812$\pm$0.027	&	7.901$\pm$1.059	&	66.976$\pm$9.974\\ 

MMS\,1 extended component 1	&88.5$\pm$1.9	&	4.098$\pm$0.077	&	0.317$\pm$0.015	&	0.239$\pm$0.012\\ 

MMS\,1 extended component 2	&324.1$\pm$6.9	&	41.703$\pm$0.929	&	0.240$\pm$0.012	&	0.050$\pm$0.003\\ 
\hline
MMS\,2-North-A compact component	&18.7$\pm$0.1	&	1.160$\pm$0.004	&	2.012$\pm$0.024	&	7.201$\pm$0.094\\ 

MMS\,2-North-A extended component	&111.8$\pm$3.6	&	0.938$\pm$0.015	&	0.045$\pm$0.003	&	0.027$\pm$0.002\\ 


MMS\,2-North-B compact component$^e$	&$\le$3.4	&	$\le$0.016	&	$\le$0.857	&	$\le$17.073\\

MMS\,2-South compact component	&24.2$\pm$0.1	&	1.450$\pm$0.004	&	1.492$\pm$0.014	&	4.113$\pm$0.043\\ 

MMS\,2-South extended component	&41.4$\pm$0.5	&	0.731$\pm$0.004	&	0.257$\pm$0.006	&	0.415$\pm$0.011\\ 
\hline
MMS\,3 compact component	&24.5$\pm$0.9	&	0.154$\pm$0.003	&	0.155$\pm$0.012	&	0.421$\pm$0.035\\ 

MMS\,3 extended component 1	&92.5$\pm$2.0	&	0.625$\pm$0.006	&	0.044$\pm$0.002	&	0.032$\pm$0.002\\ 

MMS\,3 extended component 2	&279.7$\pm$12.5	&	1.652$\pm$0.046	&	0.013$\pm$0.001	&	0.003$\pm$0.000\\ 
\hline
MMS\,4 relatively compact component &181.8$\pm$5.9  &31.036$\pm$1.387 & 0.568$\pm$0.045 & 0.209$\pm$0.018\\
MMS\,4 extended component &790.6$\pm$6.8  &219.018$\pm$2.708 & 0.212$\pm$0.004 & 0.018$\pm$0.000\\
\hline
MMS\,5 compact component	&15.8$\pm$0.2	&	1.193$\pm$0.009	&	2.891$\pm$0.089	&	12.227$\pm$0.420\\ 

MMS\,5 extended component	&145.8$\pm$3.1	&	3.851$\pm$0.043	&	0.110$\pm$0.005	&	0.050$\pm$0.002\\ 
\hline
MMS\,6 compact component	&48.5$\pm$0.3	&	12.258$\pm$0.038	&	3.153$\pm$0.043	&	4.346$\pm$0.065\\ 

MMS\,6 extended component 1	&111.0$\pm$1.0	&	15.351$\pm$0.071	&	0.754$\pm$0.015	&	0.454$\pm$0.010\\ 

MMS\,6 extended component 2	&307.1$\pm$4.5	&	46.718$\pm$0.365	&	0.300$\pm$0.009	&	0.065$\pm$0.002\\ 
\hline
MMS\,7-North compact component 1	&31.6$\pm$0.1	&	3.925$\pm$0.006	&	2.377$\pm$0.013	&	5.028$\pm$0.030\\ 

MMS\,7-North compact component 2	&36.6$\pm$0.7	&	0.606$\pm$0.006	&	0.274$\pm$0.010	&	0.501$\pm$0.021\\ 

MMS\,7-South compact component	&12.8$\pm$1.0	&	0.191$\pm$0.007	&	0.703$\pm$0.118	&	3.666$\pm$0.684\\ 
\hline \noalign {\smallskip}
\end{tabular}
\end{center}
}
\footnotesize $^a${$r$ is the geometric mean of the semi-major and semi-minor axes of each deconvolved component. The errors for $r$ are obtained by propagating the errors for deconvolved size.}\\
\footnotesize $^b${The errors for $\textit{M}_{\textnormal{dust}}$ are obtained by propagating the errors for $\textit{T}_{\textnormal{dust}}$ and total flux density. For MMS\,4, the errors for $\textit{M}_{\textnormal{dust}}$ are obtained by propagating only the errors for total flux density.}\\
\footnotesize $^c${The errors for $N_{\textnormal{H}_2}$ are obtained by propagating the errors for $\textit{T}_{\textnormal{dust}}$ and $r$.}\\
\footnotesize $^d${The errors for $n_{\textnormal{H}_2}$ are obtained by propagating the errors for $N_{\textnormal{H}_2}$ and $r$.}\\
\footnotesize $^e${MMS\,2-North-B is unresolved, with only the upper limits on $r$, $M_{\textnormal{dust}}$,  $N_{\textnormal{H}_2}$, and $n_{\textnormal{H}_2}$ provided.}
\end{table*}

The column density $N_{\textnormal{H}_2}$ is estimated as\\
\begin{equation}
N_{\textnormal{H}_2} = \frac{M_{\textnormal{H}_2}}{\pi r^2\mu_{\textnormal{H}_2}m_{\textnormal{H}}},
\end{equation}

\noindent where $M_{\textnormal{H}_2}$ is the gas mass obtained by multiplying the dust mass $M_{\textnormal{dust}}$ by 100, $r$ is the geometric mean radius of each deconvolved component, $\mu_{_{\textnormal{H}_2}} = 2.8$ is the mean molecular weight per hydrogen molecule \citep{kauffmann2008mambo}, and $m_{\textnormal{H}}=1.67\times10^{-24}$ \textnormal{g} is the mass of hydrogen atom. The number density is estimated as $n_{\textnormal{H}_2}\sim N_{\textnormal{H}_2}/{r}$ by assuming each source is spherical. Table~\ref{tab:Summary of the Physical properties} summarizes the physical properties of all components, where all the estimates are based on the temperature derived by stellar radiation \citep{tobin2020vla}. For the extremely compact disk such as MMS\,1 compact component, the dust mass estimated from the temperature derived by accretion heating is $\sim$$0.085\times10^{-4}$ $M_{\odot}$, which is about 10 times smaller than the dust mass estimated from the temperature derived by stellar radiation.

The compact components of MMS\,1, MMS\,2-North-B, MMS\,5 and MMS\,7-South have diameters smaller than the beam size by a factor of $\sim$2$-$8. The compact sizes could result in their high column densities and number densities compared to the components of other sources.

\section{Discussion}\label{discussion}

\begin{figure*}[ht!]
\end{figure*}

\begin{figure*}[ht!]
\centering
\includegraphics[scale=0.37]{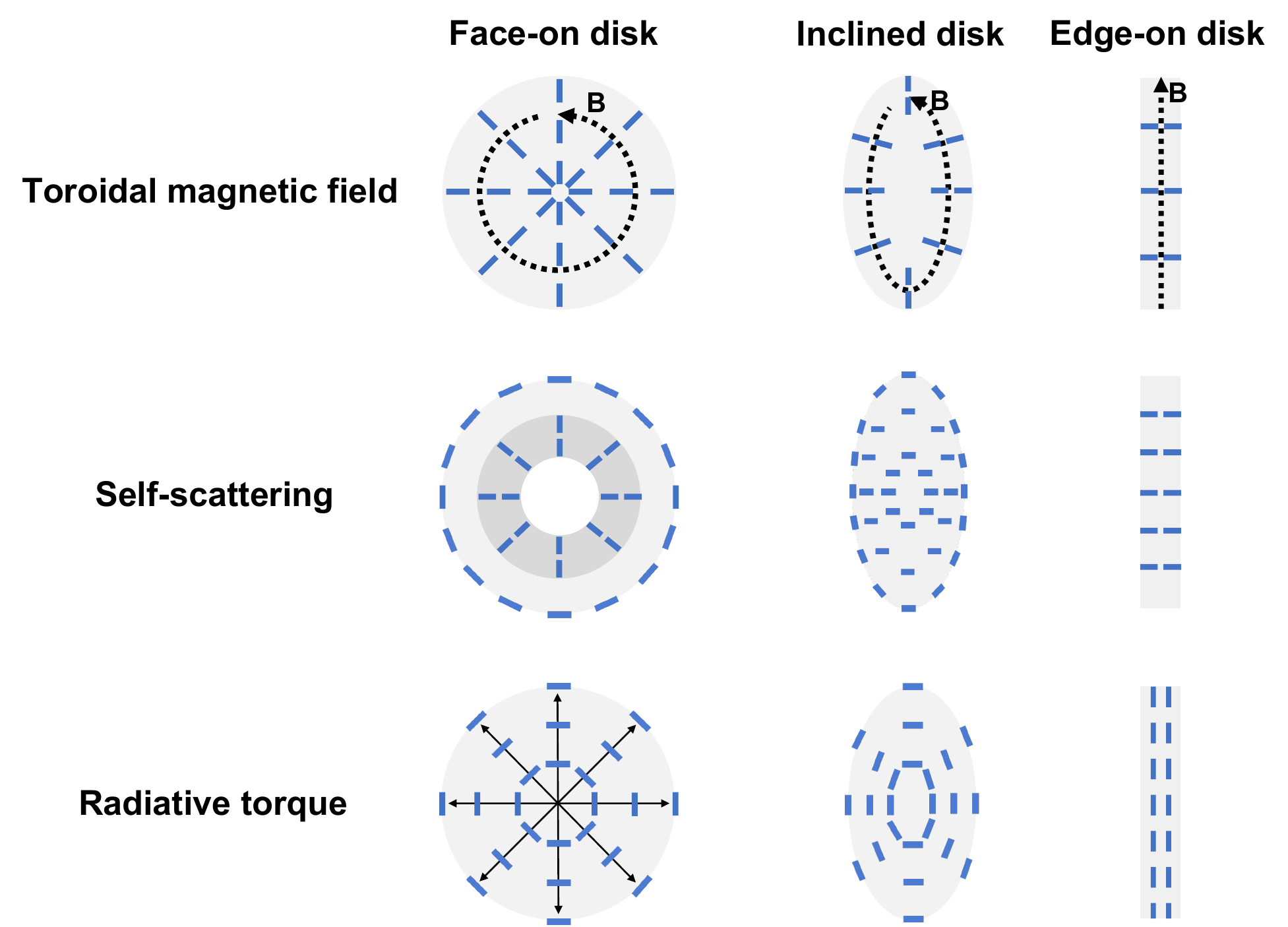}
\caption{Illustration of the $E$-vector orientations. The $E$-vectors are denoted by the blue solid line segments. The major axis of the disk is in the vertical (north-south) direction. Top panel: $E$-vector orientation predicted by dust alignment due to toroidal magnetic field. The black dashed line denotes the toroidal magnetic field. Middle panel: $E$-vector orientation predicted by self-scattering. The two intensity rings for face-on configuration are represented by different greyscales in the left panel. Bottom panel: $E$-vector orientation predicted by radiative torque. The black solid lines denote the radiation field.}
\label{origin}
\end{figure*}

\subsection{Origin of the polarization}\label{discuss-origin}

In this section, we discuss the polarization origin of each source. As explained in Section~\ref{sec:intro}, there are six mechanisms producing polarized emission, which will be explained in detail in the following paragraphs. Among these mechanisms, the polarization fraction of the reversal of self-scattering \citep{yang2020reversal} is expected to be $\sim$0.1\%, which is equivalent to the current ALMA systematic calibration uncertainty. Thus, we do not discuss this mechanism further with our sources.

\begin{enumerate}
{\bf \item Dust alignment by (toroidal) magnetic field:} As the radiative torque rotates the non-spherical dust grains \citep{draine1996}, the spinning dust grains are expected to align their major axes perpendicular to the local magnetic field and emit polarized emission \citep[e.g.,][]{hildebrand1988magnetic,lazarian2007tracing,andersson2015,hull2019interferometric}. Polarization due to the dust alignment by magnetic field has been mostly studied on large-scale \citep[e.g.,][]{rao1998bima,matthews2000magnetic,girart2006sma,koch2022multiscale}. Some observations showed that the polarization fraction could be as high as $\gtrsim$ 10\% \citep[e.g.,][]{rao1998bima,matthews2000magnetic}, in which the high polarization fraction may be attributed to the limited sensitivity of both the SMA and CARMA. It is possible that a weak magnetic field could produce a low polarization fraction. However, this has not been confirmed by observations. \cite{ohashi2018} and \cite{cox2018alma} suggested that dust grain with small size \citep[$\lesssim$ 100$\mu$m;][]{ohashi2018} is more easily aligned with magnetic fields. It should be noted that it is difficult for large-sized dust grains ($>100\,\mu$m) to be aligned due to the magnetic field \citep{ohashi2018,Yang2021}. On small-scale (a few 100 au), in the presence of the toroidal magnetic field within the disk \citep{brandenburg1995}, the induced $E$-vectors are predicted to point in the radial direction for a face-on disk \citep{cho&lazarian2007} and more or less parallel to the disk minor axis for inclined and edge-on disks \citep{lazarian2007tracing} as shown in the top panel of Figure~\ref{origin}. 

{\bf \item Self-scattering of thermal dust emission:} Self-scattering exhibits a strong dependence on the wavelength and the polarization fraction is highest when $a_{\textnormal{\scriptsize{max}}}\sim\lambda/2\pi$ \citep{kataoka2015millimeter}. Theoretical studies have shown that self-scattering is more efficient with larger dust grains ($\gtrsim$100 $\mu$m) \citep[e.g.,][]{kataoka2015millimeter,kataoka2016grain}. Therefore, self-scattering is expected to be a dominant mechanism on disk-scale rather than on the cloud- or envelope-scale. Dust polarization due to self-scattering is also considered to be an effective tool for constraining grain sizes within the inner region of the disk \citep{kataoka2015millimeter}. In addition to the dependence on grain sizes and wavelengths, the polarization patterns also depend on the scattering angle. This means that the polarization configurations could possibly change with the disk inclination \citep[e.g.,][]{kataoka2015millimeter, yang2016inclination,pohl2016}. As shown in the middle panel of Figure~\ref{origin}, two radiative intensity rings will be produced for a face-on disk. Within the inner ring, the radiative intensity is larger in the azimuthal direction than in the radial direction. The scattering of the photon moving in an azimuthal direction towards the observer, will induce a radial orientation for the $E$-vectors. In contrast, on the outer ring, intensity flow radially is predominant over the azimuthal one. The scattering photons will induce an azimuthal orientation for the $E$-vectors \citep{kataoka2015millimeter,kataoka2016submillimeter}. \cite{kataoka2016submillimeter} explained that the light scattered at the optically thick region is isotropic, resulting in the cancellation of fluxes, which in turn produces little or no polarization, particularly at the center of the disk. In the optically thin region, the anisotropy of the radiation field causes the differences in the incoming fluxes, producing the polarization as a residual. As the flux becomes more anisotropic from the center to the edge, the polarization fraction is observed to increase outwards \citep{kataoka2015millimeter,kataoka2016submillimeter,ohashi2018}. For an inclined disk, \cite{yang2017scattering} demonstrated that self-scattering can produce a near-far side asymmetry in PI within the optically thick region. This results in a deviation between the PI peak and Stokes $I$ peak, meaning a positional shift of the PI peak relative to the Stokes $I$ peak. Such a phenomenon could be due to the larger grains on the surface layer of the disk not having fully settled to the mid-plane \citep{yang2017scattering}. \cite{yang2016inclination} and \cite{yang2017scattering} have analytically shown that the PI becomes elongating along the major axis and the $E$-vectors become aligning with the minor axis in the central disk region as the disk is tilted away from the observer. In some cases, the $E$-vectors remain in the azimuthal direction on the outer edge of the disk as the polarization from anisotropic radiation field is dominated in the azimuthal direction. In the case of an edge-on disk, similar to an inclined disk, an near-far side asymmetry arises, causing the $E$-vectors to align more parallel to the minor axis of the disk (see the middle panel of Figure~\ref{origin}).

{\bf\item Dust alignment due to radiative torque:} This mechanism has been proposed by \cite{Lazarian&Hoang2007RAT}, which is a direct dust grain alignment due to radiation field. \cite{tazaki2017radiative} applied this mechanism to investigate the protoplanetary disks around a T\,Tauri star. \cite{tazaki2017radiative} performed three-dimensional (3D) radiative transfer calculations, and demonstrated that the grains in a disk could align their minor axes with the direction of the radiative flux. As the radiative flux is expected to point radially in a disk, it will then induce the $E$-vectors in the azimuthal direction for a face-on disk, in an elliptical direction for an inclined disk \citep{kataoka2017evidence}, and parallel to the major axis for an edge-on disk as illustrated in the bottom panel of Figure~\ref{origin}. \cite{tazaki2017radiative} also suggested that the polarization fraction mainly depends on the maximum grain size $a_{\textnormal{max}}$ and the polarization fraction increases as $a_{\textnormal{max}}$ decreases. The radiative alignment is not easily to be detected by observations, except for the polarization of HL\,Tau in the ALMA 3.0\,mm observations. The dust polarization towards this source has been interpreted as originating from radiative torque because of the elliptical pattern of the $E$-vectors \citep{stephen2017,kataoka2017evidence}. However, such an interpretation still remains controversial. \cite{yang2019} suggested that the radiative torque will produce a circular pattern rather than an elliptical pattern for an inclined disk, which is inconsistent with the dust polarization pattern observed in HL\,Tau. Further observations might be necessary to validate this mechanism.
\end{enumerate}

In addition to the well-studied mechanisms mentioned above, theoretical studies have suggested new polarization mechanisms as shown below. However, these mechanisms have not yet been thoroughly tested by observations. In the following, we explain such processes:

\begin{enumerate}[resume]
{\bf\item Grain alignment by gas flow due to mechanical alignment:} This mechanism was originally proposed by \cite{gold1952MNRAS.112..215G}, which was widely discussed within the framework of interstellar medium. \cite{gold1952MNRAS.112..215G} proposed that polarization orientation is parallel to the gas flow, while \cite{lazarian2007mechanical} and \cite{hoang2018} proposed helical grain alignment where the polarization orientation is perpendicular to the gas flow. \cite{kataoka2019gasflow} opted for helical grain alignment as the possible mechanism that works in the protoplanetary disk, as the gas flow is supposedly subsonic. This mechanism will produce the $E$-vectors in an azimuthal direction for a face-on disk, in a circular direction for an inclined disk. In addition, \cite{kataoka2019gasflow} found that the polarization orientations could depend on the Stokes numbers of the dust grains, where the Stokes number $\rm{St}$ refers to the ratio of dust stopping time $t_{\rm s}$ to the Keplerian timescale $t_{\rm K}$ as $\rm{St}=t_{\rm s}/t_{\rm K}$
\footnote{
The stopping time is defined as $t_s=a_{\rm d} \rho_{\rm s}/(v_{\rm th}\rho)$ where $a_{\rm d}$, $\rho_{\rm s}$, $v_{\rm th}$ and $\rho$ are the dust grain size, material density,  thermal velocity of molecular gas and gas density, respectively. 
}. When the Stokes numbers are smaller than unity ($\rm{St}<1$), the dust grains are well coupled with the gas in the disk.
In this case, the $E$-vectors will be oriented in the azimuthal direction. 
As the Stokes numbers are approaching unity ($\rm{St}\sim1$), the coupling between the dust grains and gas becomes weaker and the gas drug force acting on dust grains cannot be ignored. In this case, the $E$-vectors will take on a spiral pattern. Once the Stokes numbers are larger than unity by a factor of 10 (i.e., $\rm{St}>10$), the $E$-vectors will be pointed in a radial direction.

{\bf\item Magnetic field alignment in the Mie regime:} \cite{guillet2020mie} found that $E$-vectors could align parallel to the magnetic field in the Mie regime when the grain size is $\sim$1\,mm. As a result, the $E$-vectors align themselves with toroidal magnetic field in the azimuthal direction for both face-on and inclined disks. This mechanism is an alternative way to explain the azimuthal $E$-vectors detected on the outer edge of the disk.
\end{enumerate}

There are cases where a polarization pattern could be explained by multiple polarization mechanisms. For example, the azimuthal pattern of $E$-vectors could be explained by either radiative torque, grain alignment by gas flow or magnetic field alignment in the Mie regime. Moreover, the morphology could appear as a mixture due to multiple mechanisms. In such cases, multi-wavelength observations are necessary to disentangle the polarization origins. However, since our data is only limited to a single wavelength, it is difficult to conclude the origins of the polarization for our sources. In the following subsections, we list the possible polarization mechanisms for each source. Some mechanisms that could explain most of the polarization features such as $E$-vector orientation, polarization fraction, and PI peak positional shift, whereas some mechanisms could only explain one of the features such as the $E$-vector orientation. We also discuss which mechanism is more likely to be the origin of the polarization for each source.

\subsubsection{MMS 2-North-A}\label{MMS2-North-A}
MMS\,2-North-A is a flat-spectrum source. We detected a compact disk-like structure around MMS\,2-North-A, which is nearly face-on with an inclination angle of $i \sim 25^{\circ}$. PI of MMS\,2-North-A shows a ring-like structure, and the $E$-vectors on the ring show an azimuthal pattern. In addition, a nearly pole-on outflow elongated along the north-south direction is possibly associated with MMS\,2-North-A. The mean polarization fraction is estimated to be $\sim$1.0\%.
 
\begin{figure*}[]
\includegraphics[scale=0.52]{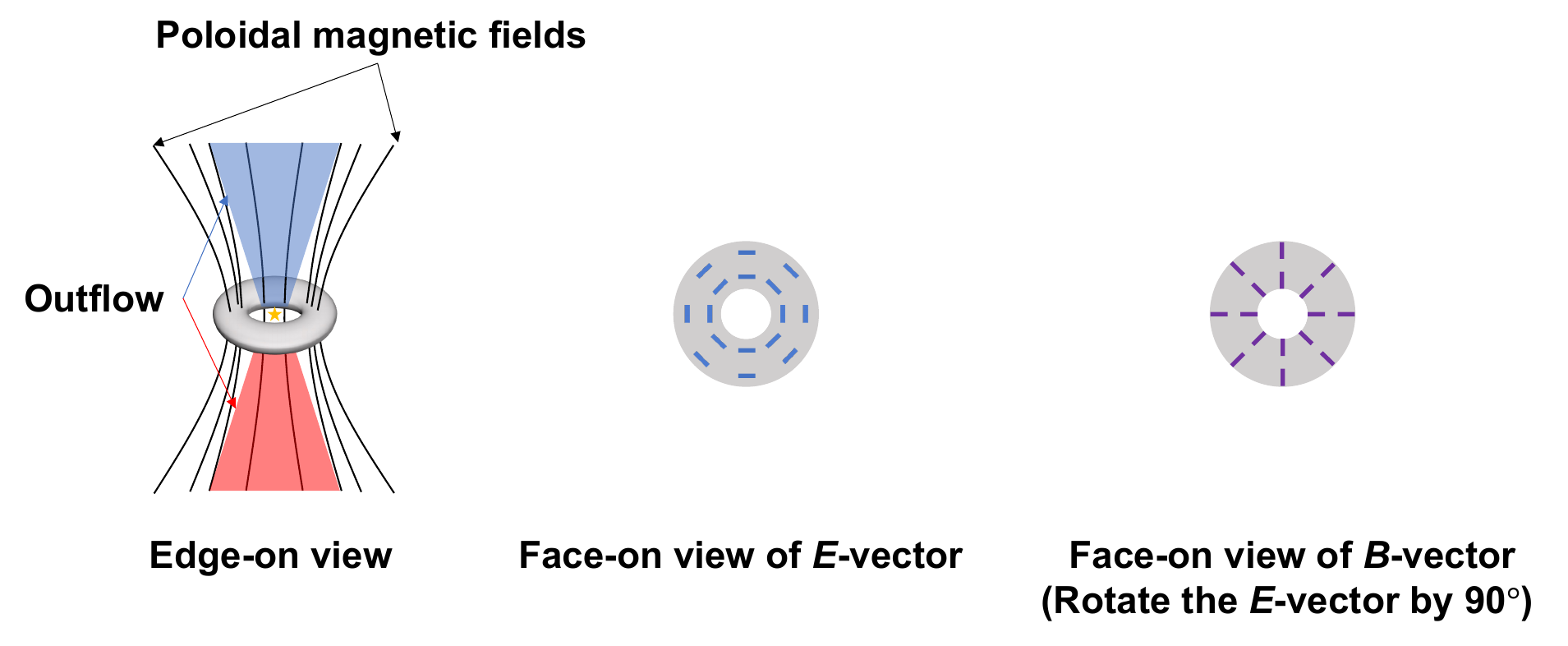}
\caption{Illustrations of MMS\,2-North-A with outflow in the presence of the poloidal magnetic field is shown in the left panel. The black thin solid lines indicate the poloidal magnetic field. The blue line segments on the middle panel indicate the $E$-vectors predicted by dust alignment due to poloidal magnetic field. The purple line segments in the right panel indicate the $B$-vectors, obtained by rotating the $E$-vectors by 90$^\circ$.}
\label{mms2-n-discussion}
\end{figure*}

The observed dust polarization could be explained by the poloidal magnetic field. The $B$-vectors point in the radial direction as shown in Figure~\ref{fig-fraction}\,(a). Considering that this disk is nearly face-on, the poloidal magnetic field may dominate over the toroidal one. The $E$-vectors are expected to be aligned perpendicular to the poloidal magnetic field, that is in the azimuthal direction. Thus, the $B$-vectors are expected to point in the radial direction as illustrated in Figure~\ref{mms2-n-discussion}. This is different from the case of dust alignment by toroidal magnetic field. In the synthetic polarization maps of protostellar source NGC\,1333\,IRAS\,4A, \cite{frau2011} has shown that the magnetic field tend to be aligned in the radial direction for face-on configurations. By comparing the Stokes $Q$ and $U$ images of MMS\,2-North-A with those of NGC\,1333\,IRAS\,4A (see the bottom panel of Fig.4 and Fig.5 in \cite{frau2011}), we found that the morphologies are similar between NGC\,1333\,IRAS\,4A and MMS\,2-North-A. Therefore, it is possible that the polarization pattern in MMS\,2-North-A is produced by poloidal magnetic field.

The observed dust polarization could also be attributed to self-scattering of thermal dust emission. The $E$-vectors trace the PI dust ring in an azimuthal direction. This configuration is consistent with that due to self-scattering for a face-on disk as shown in Figure~\ref{origin} middle panel \citep[e.g.,][]{kataoka2015millimeter,kataoka2016submillimeter,pohl2016}. The mean polarization fraction is $\sim$1.0\%, which can also be explained by self-scattering where a few percent is expected. Depolarization seems to occur at the center of the disk. Such a feature has also been seen in the disk of HD\,142527, which can be explained by the optical effects \citep[e.g.,][]{kataoka2015millimeter,kataoka2016submillimeter}.

Radiative torque could also explain the $E$-vector orientation, where the $E$-vectors are expected to point in an azimuthal direction for a face-on disk. The depolarization occurring at the center of the disk has also been seen in HL\,Tau at 3.1\,mm observations \citep{stephen2017}. Polarization patterns of HL\,Tau at 3.1\,mm are consistent with radiative torque model, and \cite{stephen2017} explained that the depolarization could be due to beam-averaging of the azimuthal $E$-vectors within the beam. Although the disk of MMS\,2-North-A is viewed slightly more face-on than HL\,Tau disk, since the difference in inclination is about $\sim$$10^{\circ}-$$20^{\circ}$, we expect the morphologies might be similar. 

Both grain alignment by gas flow and magnetic field alignment in the Mie regime could explain the azimuthal pattern of the $E$-vectors. However, for grain alignment by gas flow, this explanation holds true only when the Stokes numbers are smaller than unity. Since the studies associated with these two grain alignment mechanisms primarily focus on the morphology of the $E$-vectors \citep{kataoka2019gasflow,guillet2020mie}, we could not delve into the other polarization features like polarization fraction and PI morphology expected by these two grain alignment mechanisms.
  
Overall, self-scattering and radiative torque are the more likely origins of the polarization for MMS\,2-North-A, because the azimuthal $E$-vector orientation, depolarization, and the low polarization fraction are consistent with both mechanisms. However, the depolarization can also be due to smoothing over the complicated structure at the central beam, as seen in HL\,Tau. The unresolved polarization structures could lead to a decrease in the polarization fraction. Therefore, other mechanisms cannot be ruled out without verification from multi-wavelength observations at a higher angular resolution.

\subsubsection{MMS 2-South and MMS 7-North}
In this section, we will explain MMS\,2-South and MMS\,7-North together due to their similar polarization features such as $E$-vector orientation and PI peak positional shift. The brief summary for these two sources is as follows.

MMS\,2-South and MMS\,7-North are the flat-spectrum and Class I sources, respectively. We detected a compact and inclined disk-like structure around MMS\,2-South with an inclination angle of $i \sim 45^{\circ}$, and a compact and nearly edge-on disk-like structure around MMS\,7-North with an inclination angle of $i \sim 64^{\circ}$. For both sources, the PI is elongated along the major axis of the disk, and the peak of PI shifts towards the nearside of the disk with respect to the Stokes $I$ peak. The $E$-vectors in the central region are mostly parallel to the minor axis of the disk, while on the outer edge, the $E$-vectors are more azimuthal. The mean polarization fractions of MMS\,2-South and MMS\,7-North are estimated to be $\sim$1.2\% and $\sim$1.0\%, respectively.

The configurations of the $E$-vectors observed in MMS\,2-South and MMS\,7-North could be explained by dust alignment due to magnetic field. The dust polarization in the central disk regions is likely to be produced by the toroidal magnetic field as the $E$-vectors are mostly parallel to the minor axes of the disks. In contrast, the $E$-vectors are azimuthal near the outer disk edges (see Figure~\ref{fig-co-1}\,(b) and (h)). These patterns could be interpreted as the dust grains being aligned by a toroidal magnetic field in the central disk region and a poloidal magnetic field near the disk edge, which have been seen in HH\,111 \citep{lee2018alma}. However, the near-far side asymmetry in the central disk region could not be explained by this mechanism. 

For MMS\,2-South and MMS\,7-North, the observed dust polarization patterns are broadly consistent with the expected outcomes of self-scattering for inclined and nearly edge-on disks. In these two sources, the $E$-vectors in the central regions are predominantly aligned with the minor axes of both disks. Near the outer edges, the $E$-vectors point in the azimuthal direction. The orientation of $E$-vectors in the optically thick central region is mainly induced by the inclination effect, while on the outer edge is mostly dominated by the intrinsic polarization from the anisotropic radiation field \citep{yang2016inclination}. The directional change of the $E$-vectors can also be explained by a change in direction of the thermal radiative flux from inner disk to outer disk \citep{kataoka2016submillimeter}. Furthermore, the directional change of the $E$-vectors observed in MMS\,7-North mainly occurs on the nearside of its disk, unlike for MMS\,2-South, where it occurs on both the nearside and farside (see Figure~\ref{fig-co-1}\,(b) and (h)). We expect that this may come from the inclination effect because MMS\,7-North is viewed more edge-on than MMS\,2-South. A similar configuration has also been observed in HH 111 \citep{lee2018alma}, which is a nearly edge-on disk of a Class I protostar, exhibiting distinct $E$-vector orientations on the nearside and farside of its disk. The $E$-vectors on the nearside are parallel to the minor axis of the disk, whereas they are almost perpendicular to the minor axis on the farside. \cite{lee2018alma} suggested that this configuration could possibly be produced by a disk surrounded by rings and gaps that have marginally different inclinations compared to the disk. As a result, the $E$-vectors become less parallel to the minor axis of the disk. In addition, the PI peak shift towards the nearside has been observed for both sources, and this positional shift of the PI peak can be explained by the optical depth effect \citep{yang2017scattering}. These two sources have the low polarization fractions of $\sim$1.2\% and $\sim$1.0\%, which are consistent with the polarization fraction expected for the self-scattering, where a few percent ($\sim$1$-$2\%) is expected.

There is another possibility that the polarization could be caused by a combined mechanisms involving both self-scattering and radiative torque. We note that the some $E$-vectors in the central disk region are not completely parallel to the minor axis of the disk, but slightly rotate away from the minor axis. Near the edge, the $E$-vectors are distributed in the azimuthal direction. A similar feature has been seen in HL\,Tau at 1.3\,mm observations \citep{stephen2017}. The $E$-vectors that are not entirely aligned with the minor axis of the disk, along with the $E$-vectors in the azimuthal direction observed at the edge, could be interpreted as a mixed morphology caused by a combination of self-scattering and radiative torque.

Grain alignment by gas flow and magnetic field alignment in the Mie regime are unlikely to be the origins of the polarization for these two sources. The $E$-vectors are expected to be oriented elliptically for an inclined disk due to these two grain alignments, which are inconsistent with the $E$-vector orientations for both sources.

Overall, self-scattering is more likely to be the origin of the dust polarization observed from MMS\,2-South and MMS\,7-North than the dust alignment by magnetic field. Both mechanisms could explain the $E$-vector orientation. However, the PI peak positional shift and low polarization fraction are more consistent with self-scattering. In addition, dust polarization could also possibly be caused by the mixture of self-scattering and radiative torque, which needs to be verified with multi-wavelength observations.

\subsubsection{MMS 7-South}
MMS\,7-South is a Class I source. We detected a compact disk-like structure around MMS\,7-South, which is nearly edge-on with an inclination angle of $i \sim 60^{\circ}$. PI is only detected in the immediate central region around the Stokes $I$ peak. The detected PI is relatively weak and PI peak positional shift has not been seen from this source (see Figure~\ref{fig-co-1}\,(h)). Only a few $E$-vectors have been observed and these $E$-vectors appear to be more or less parallel to the minor axis of the disk. The mean polarization fraction is estimated to be $\sim$1.0\%.

There are only two features that could be used to discuss the origin of the polarization, and these are the $E$-vector orientation and the polarization fraction. The $E$-vector orientation could be explained by either self-scattering or magnetic field. However, the low polarization fraction is more consistent with self-scattering.

Overall, self-scattering is more likely to be the origin of dust polarization observed in MMS\,7-South since the $E$-vector orientation and low polarization fraction are consistent with those expected from self-scattering. However, the $E$-vector orientation is also consistent with dust alignment by magnetic field. Moreover, this source is possibly unresolved due to its compact size and weak polarization detection. Therefore, we could not rule out other mechanisms with the limited detection from this single wavelength observations.

\subsubsection{MMS 5}
MMS\,5 is a Class 0 source. We detected an inclined disk-like structure around MMS\,5 with an inclination angle of $i \sim 37^{\circ}$. PI is detected in the central region (within $\sim$100 au). The peak of the PI shifts towards the nearside of the disk with respect to the Stokes $I$ peak. The fitted disk has a P.A. $\sim$ 1.6$^\circ$, and the $E$-vectors peaked around 100$^{\circ}$. This suggests that the $E$-vectors could potentially be parallel to the minor axis of the disk. The mean polarization fraction is estimated to be $\sim$0.6\%. 

MMS\,5 is in the younger Class 0 stage and has a small disk size of $\sim$32 au, in which the disk size is about twice smaller than the synthesized beam. Therefore, the detection of a lower polarization fraction for MMS\,5 as compared with the other sources could be due to either the undeveloped disk or the beam dilution. If the $E$-vectors are indeed parallel to the minor axis of the disk as we expect, then this configuration can be explained by dust alignment due to toroidal magnetic field or self-scattering. The positional shift of the PI peak towards the nearside of the disk and the low polarization fraction, are consistent with self-scattering. 

The observed dust polarization in MMS\,5 contradicts the alignment mechanisms due to radiative torque, gas flow, and magnetic field in the Mie regime, as the $E$-vector orientation is inconsistent with expectations from these mechanisms. 

Overall, the dust polarization observed in MMS\,5 is more likely to be originated from self-scattering, as the $E$-vector orientation, polarization fraction and PI peak positional shift are consistent with those expected from self-scattering. However, we could not conclusively rule out the other mechanisms with the limited detection from this single wavelength observations. Multi-wavelength observations are needed to confirm the origin of the polarization for this source.

\vspace{1\baselineskip}
In summary, the dust polarization observed in our sources could be explained by multiple mechanisms. Self-scattering is likely the primary polarization origin. However, we could not definitely rule out other mechanisms due to their similarities in the polarization morphologies. It is essential to disentangle the polarization origins through multi-wavelength observations, which can provide valuable insights into better constraining the dust grain size. Our results have shown that dust polarization observed in Class 0/I sources at 1.1\,mm is possibly arising from self-scattering, indicating that there is dust growth at the early stages \citep[dust grains may have grown to as large as $\sim$170 $\mu$m, which is estimated by $a_{\textnormal{\scriptsize{max}}}\sim\lambda/2\pi$;][]{kataoka2015millimeter}. As for future work, we propose to observe these sources with multiple wavelengths. Previous studies have shown that polarization patterns observed at shorter wavelengths (e.g., Band 7) is highly consistent with self-scattering, some clear observational results such as the uniform $E$-vectors along the minor axis of the disk and PI peak positional shift \citep[e.g.,][]{stephen2017,lee2018alma}. We also expect to have more clear detection towards MMS\,5 at shorter wavelengths as the polarization detection towards this source at Band 6 is very limited. At longer wavelengths (e.g., Band 3), we predict the $E$-vectors will be distributed in the azimuthal direction if radiative torque is the dominant mechanism. If the maximum grain size has grown to as large as 500 $\mu$m, it might still be possible to observe $E$-vectors parallel to the minor axis of the disk due to self-scattering. However, the polarization fraction may be slightly higher compared to that detected at Band 6. If there is no significant change in the polarization properties, such as the $E$-vector orientation and polarization fraction, it suggests that the observed dust polarization has no strong dependence on either wavelength or maximum grain size, which is more consistent with dust alignment by magnetic field.

\subsection{Evolutionary tendency}\label{tendency}

\begin{figure*}[ht!]
\centering
\includegraphics[scale=0.47]{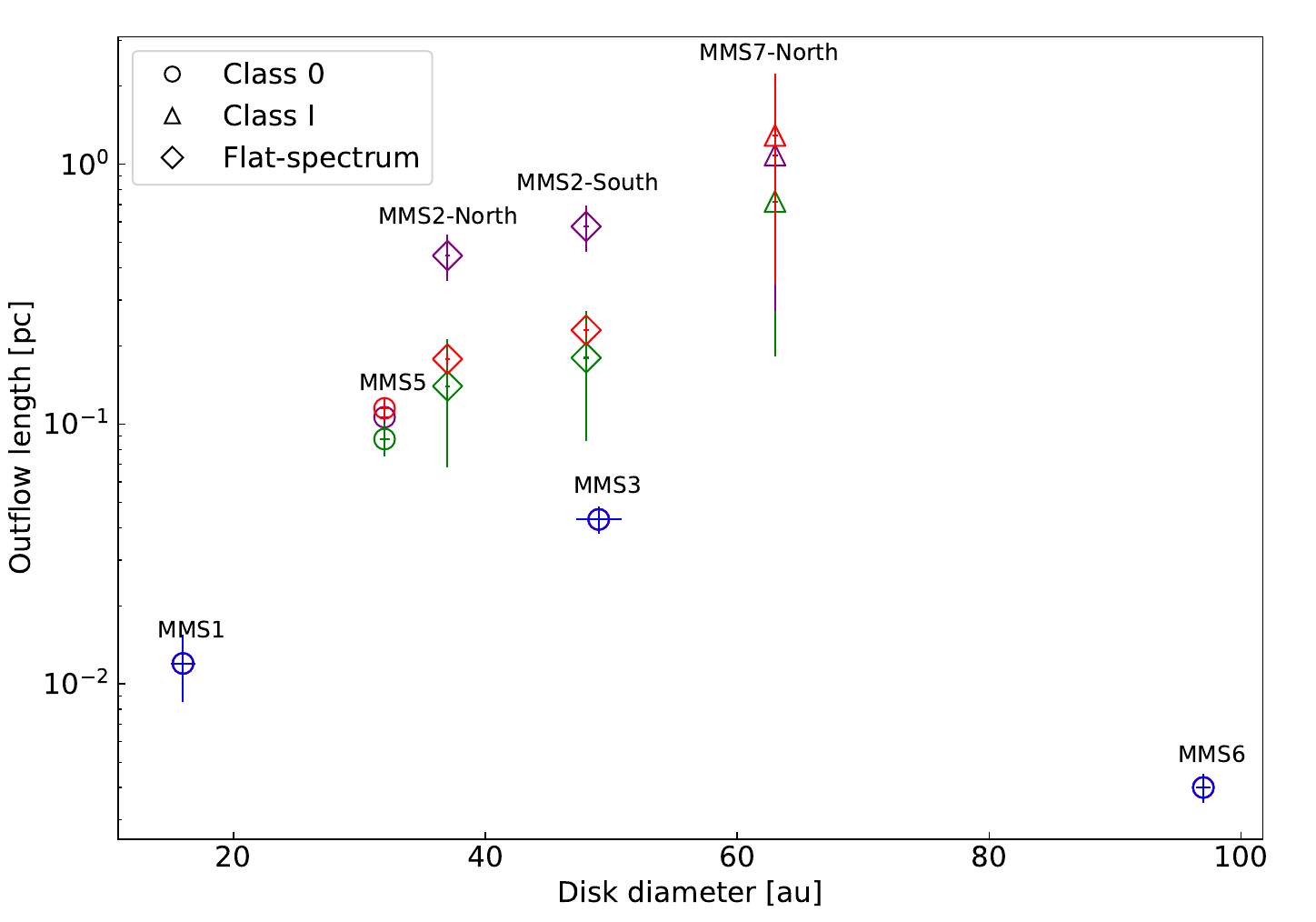}
\caption{The length of the CO outflow versus the disk diameter of the sources (excluding MMS\,4 and MMS\,7-South because no disk-like structure is detected towards MMS\,4 and no clear outflow is detected towards MMS\,7-South). The outflow length is estimated as the arithmetical average length of the blue- and red-shifted CO components. The outflow lengths for MMS\,1 (blue) is obtained from Takahashi et al.\,(2023a, submitted to ApJ), for MMS\,3 (blue) is obtained from \cite{morii2021revealing}, and for MMS\,6 (blue) is obtained from \cite{takahashi2019alma}, respectively. Among the sources, elongated outflows were detected towards MMS\,2, MMS\,5, and MMS\,7 from previous single-dish observations \citep{takahashi2008millimeter,tanabe2019nobeyama,feddersen2020carma}. Since the interferometric data might be affected by the missing flux, we plot the outflow lengths for MMS\,2-North, MMS\,2-South, MMS\,5, and MMS\,7-North with the lengths obtained from the single-dish observations by \cite{takahashi2008millimeter} denoted in red, \cite{tanabe2019nobeyama} denoted in green, and \cite{feddersen2020carma} denoted in purple, respectively. The disk diameters for all the sources are obtained from this paper. The circle, triangle, and diamond denote the Class 0, Class I, and flat-spectrum sources, respectively. The line in the horizontal direction represents the 2D Gaussian fitting errors for the fitted disk diameter. The line in vertical direction shows the minimum to maximum lengths of the blue- and red-shifted CO components obtained from each data set. The outflow length is corrected using the inclination angle $i$ for each source as listed in the last column of Table~\ref{tab:Summary of the fitting results}.}
\label{dia-co} 
\end{figure*}

In this section, we discuss the correlation between disk size, outflow size, polarization fraction, and their associations with the evolutionary stages of these protostellar sources. We define the disk size based on the most compact component fitted by the 2D Gaussian as listed in the second column of Table~\ref{tab:Summary of the Physical properties}. We classify the sources primarily based on the SED study by \cite{furlan2016herschel} as listed in the eighth column of Table~\ref{table:targets-summary}. MMS\,1 was not detected by \cite{furlan2016herschel}, and is classified as a Class 0 source since a very compact and collimated jet is detected by Takahashi et al.\,(2023a, submitted to ApJ). For the sources within the same evolutionary stage such as MMS\,1, MMS\,5, and MMS\,6, we compare the mean length of outflow to determine the evolution of each source, where the mean length is obtained by averaging the lengths of the blue- and red-shifted CO components.

Figure~\ref{dia-co} presents the mean length of outflow versus disk diameter along with the SED classification of each protostellar source. We found a clear positive correlation between the disk size and outflow size except for MMS\,6, in which MMS\,6 has the smallest outflow size but is measured with the largest disk size. The larger outflow size is likely to indicate a longer lifetime since the outflow velocities are typically comparable. The longer lifetime also allows the disks to grow in size in principle. Therefore, the outflow size and the disk size may be correlated, but this does not necessarily imply a direct causal relationship, such as a larger disk driving a larger outflow.

Figure~\ref{dia-co} also clearly shows that that the more evolved sources have the larger outflows. In addition, our results show a possible trend where the disk size appears to increase from the Class 0 to the Class I/flat-spectrum phase, which is consistent with the expected formation and evolution of disks in rotating and collapsing cloud cores \citep[e.g.,][]{Vorobyov2011,kengo2017}. Such a trend is also consistent with the observational results by \cite{Yen2017}, in which they showed that the radii of Keplerian disks grow from the Class 0 to the Class I phase.


On the contrary, \cite{tobin2020vla} showed that the disk size decreases from Class 0 to flat-spectrum phase. Such an opposite trend could be attributed to several factors. One of them could be due to the different Gaussian fitting methods. \cite{tobin2020vla} fitted each source with a single Gaussian component, while we fitted each source with multiple Gaussian components. Our disk sizes are estimated from the most compact components as shown in the second column of Table~\ref{tab:Summary of the Physical properties}. We found that if we fit the source with a single component, the radius for the compact component will become larger than when fitting with multiple components. Such an increase in size is especially prominent in the case of Class 0 sources. That is, the average increase in radius for our Class 0, Class I, and flat-spectrum sources is 110\%, 7\% and 13\%, respectively. \cite{tobin2020vla} also pointed out that the decrease of dust disk radii through evolutionary phase may be partially influenced by a systematic bias caused by measuring the radii using Gaussian fitting, in particular for Class 0 sources. In addition, \cite{tobin2020vla} expected that the decrease in the disk radii in the more evolved sources come from radial drift of the dust grains. In this process, the large grains in the outer disk are decoupled from the gas and gradually drift towards the central star, resulting in a loss of disk dust mass and a reduction in disk size \citep[e.g.,][]{weidenschilling1977,birnstiel2010,trapman2020}. Note that although \cite{Yen2017} has demonstrated a general trend in which Class I sources typically have larger disk sizes than Class 0 sources, there are some samples inconsistent with this trend, i.e., some Class 0 sources have larger disks than Class I sources. Since we have a limited number of samples, we need to expand the number of samples to make fair comparisons and confirm the trend in the future.

\begin{figure*}[]
\centering
\includegraphics[scale=0.47]{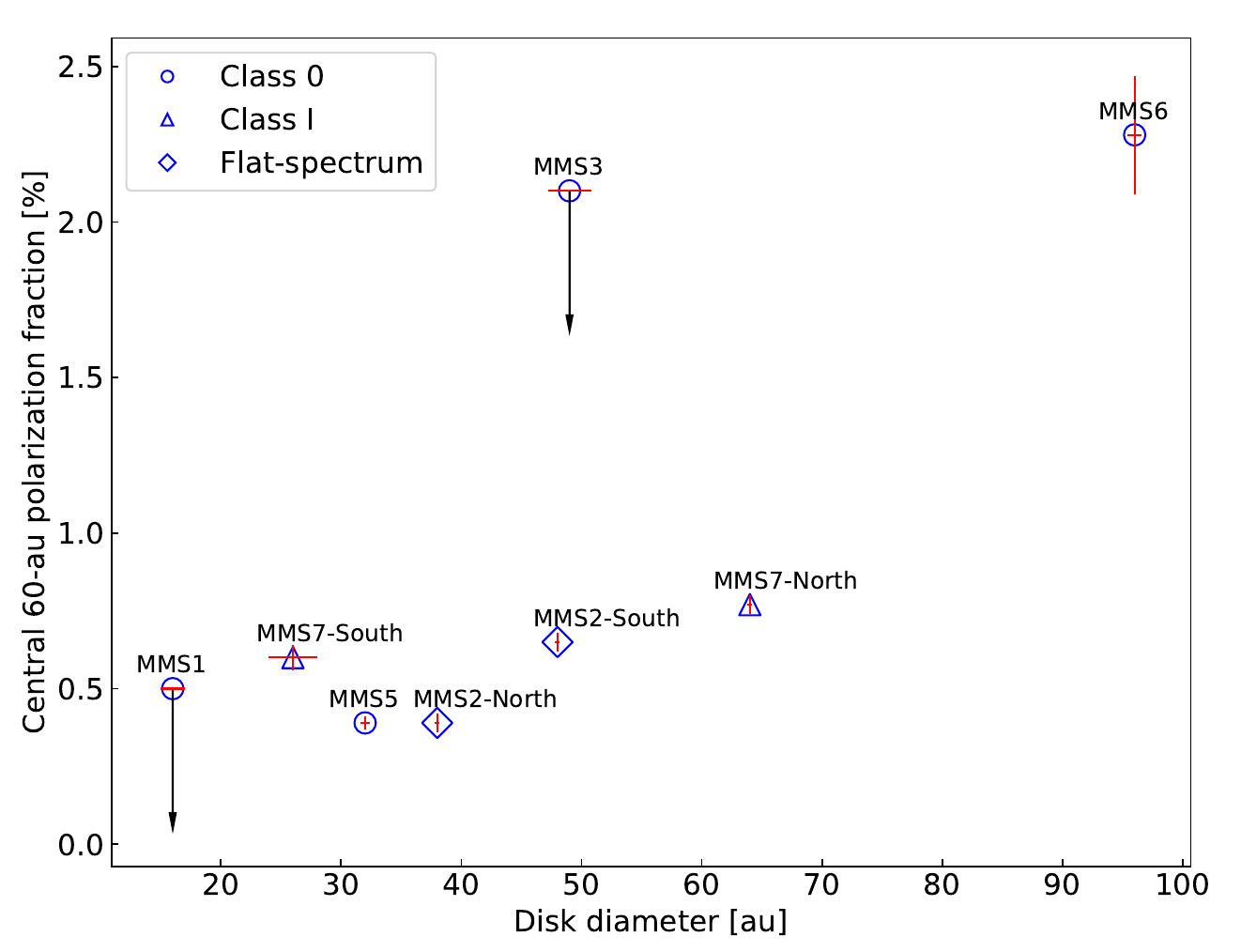}
dia-\caption{The central 60-au polarization fraction versus the disk diameter of the sources (excluding MMS\,4 because no disk-like structure is detected). The central 60-au polarization fraction is the mean polarization measured within central 60 au region of each disk. The circle, triangle, and diamond denote the Class 0, Class I, and flat-spectrum sources, respectively. Since there is no polarized emission detected towards MMS\,1 and MMS\,3, 3$\sigma$ upper limits of the polarization fractions are plotted with the black arrows for these two sources. The horizontal red line represents the 2D Gaussian fitting errors for the fitted disk diameter. The vertical red line represents the errors for the central 60-au polarization fraction. We estimate the central 60-au polarization fraction and its errors using the same method as described in Section~\ref{results} for the mean polarization fraction presented in Figure~\ref{hist-pfrac}.}
\label{dia-pfrac} 
\end{figure*}


Figure~\ref{dia-pfrac} shows the central 60-au polarization fraction measured over the same size scale towards each source as a function of disk diameter. For all sources that polarized emissions are detected, the mean polarization fraction is estimated within the central 60 au region to focus solely on the compact dust disk. For the non-polarized detected sources MMS\,1 and MMS\,3, we plot the 3$\sigma$ upper limit of the polarization fraction as the central 60-au polarization fraction. In general, we observe a central 60-au polarization fraction of $\lesssim$ 1\% for most of the sources, with the exception of MMS\,6. Such a low percentage is consistent with the expectation for the polarization resulting from self-scattering as discussed in Section~\ref{discuss-origin}. 

Furthermore, the central 60-au polarization fraction varies slightly among the sources. We expect that the variation of the polarization fraction could also be affected by the inclination of each source. It is important to highlight that, despite MMS\,2-North being in a more evolved evolutionary stage, it exhibits the lowest polarization fraction, which is comparable to that of MMS\,5. One explanation could be that the unresolved polarization results in a decrease in the polarization fraction as mentioned in Section~\ref{MMS2-North-A}. Another plausible explanation could be the inclination effect. Given that MMS\,2-North is viewed nearly face-on, it is likely that depolarization occurs at the central disk region in this face-on configuration, particularly if the polarization arises from self-scattering \citep{kataoka2016submillimeter}. Therefore, we detected less $E$-vectors and a much lower polarization fraction than the other sources, particularly in the central disk region. Similarly, by comparing the polarization fractions within the two binary systems, MMS\,2 and MMS\,7, we found that the polarization fractions of MMS\,2-North-A and MMS\,7-South are lower than that of MMS\,2-South and MMS\,7-North, respectively. The former two sources are viewed more face-on than the latter two sources, and the face-on sources are likely to have a lower polarization fraction than edge-on sources. This result is consistent with our expectation that the polarization fraction originating from self-scattering is likely to be influenced by the inclination of the disk.

Finally, among our protostellar samples, MMS\,6 stands out as an exceptional source due to its inconsistency with the observed trend, where the disk size and outflow length increases with the evolutionary stage as seen in the other sources. Although MMS\,6 has the most compact outflow and is at the earliest evolutionary stage (Class 0), it is brightest in Stokes $I$ \citep{takahashi2009,takahashi2012spatially,takahashi2012molecular,takahashi2013hierarchical,takahashi2019alma}. In fact, the measured Stokes $I$ peak flux is a factor of 2$-$34 brighter than those measured in the other protostellar sources among our samples. MMS\,6 is physically larger than other sources and contains more stuff around. The radiative flux is possibly affected by the amount of material and the dust temperature. The details will be discussed in Takahashi et al.\,(2023b, in preparation). Moreover, MMS\,6 has shown a significantly higher polarization fraction than other sources. We expect that the high polarization fraction is resulting from the polarization due to dust alignment by magnetic field \citep{takahashi2019alma,Gouellec2020,liubaobab2021,Gouellec2023}.

\section{Conclusion} \label{conslusion}
We present the results of ALMA 1.1\,mm dust linear polarization observations at ${\sim}0\dotarcsec14$ (55 au) and CO~($J=2-1$) observations at ${\sim}1\dotarcsec5$ (590 au) towards seven millimeter sources, MMS\,1$-$7 in OMC-3. The results are summarized as follows.

\begin{enumerate}
\item We have detected MMS\,1$-$7 in Stokes $I$. Except for the prestellar source MMS\,4, we have detected bright and compact disk-like structures towards six other sources. The disk size was fitted using the 2D Gaussian, ranging from 16 au to 97 au in linear size. The measured disk size is positively correlated with the evolutionary stage. Out of the seven sources, MMS\,2 (Flat-spectrum), MMS\,5 (Class 0), MMS\,6 (Class 0), and MMS\,7 (Class I) are detected in Stokes $Q$ and $U$. MMS\,1 (Class 0), MMS\,3 (Class 0), and MMS\,4 (Prestellar core) are not detected in Stokes $Q$ and $U$ with a 3$\sigma$ upper limit of 0.5\%, 2.1\%, and 4.7\%, respectively.

\item We detected clear CO outflows towards MMS\,1, MMS\,2, MMS\,3, MMS\,5, MMS\,6, and MMS\,7. The outflow axes in MMS\,2, MMS\,5, MMS\,6, and MMS\,7 are mostly aligned perpendicular to the major axes of the disks. The outflow size primarily has a positive correlation with the evolutionary stage and the disk size, with MMS\,6 as an exception. Regarding MMS\,4, we have detected a relatively compact millimeter source (geometric mean diameter $\sim$ 364 au) with a number density as high as $\sim$2 $\times 10^{9}$ cm$^{-3}$, while no clear bipolar outflow is detected. However, we have a marginal detection of the compact low-velocity ($v_{\textnormal {LSR}}$ = 6 to 7 km\,s$^{-1}$) blue-shifted CO emission elongated to the western direction with respect to this millimeter source, possibly implying the presence of the first hydrostatics core \citep{hirano2023}.  

\item The compact linearly polarized emissions associated with the dust disks are detected towards MMS\,2, MMS\,5, MMS\,6, and MMS\,7. $E$-vector orientations differ among individuals, such as the direction parallel to the minor axis of the disk and the azimuthal direction. There is no strong correlation between the cloud-scale and disk-scale $E$-vector orientations. The mean polarization fractions for MMS\,2, MMS\,5, and MMS\,7 are between 0.6$\%$ and 1.2$\%$, while MMS\,6 exhibited a relatively high mean polarization fraction of 3.5$\%$ around Stokes $I$ peak, and 15.7$\%$ on the arm-like structure. 

\item The dust polarization observed in these sources can be explained by multiple mechanisms, while self-scattering is the more likely polarization origin, as evident from PI peak positional shift towards the nearside of the disk and the low polarization fraction. However, other dust grain alignment mechanisms could not be conclusively ruled out. We will investigate the polarization origin by multi-wavelength observations in the future.



\item The central 60-au polarization fractions (mean polarization fraction measured within central 60 au of the disk) for MMS\,2, MMS\,5, and MMS\,7 are $\lesssim1\%$. The comparable values in the central 60-au polarization fraction and also the PI peak positional shifts might suggest a common polarization origin among these sources such as self-scattering. The central 60-au polarization fraction varies slightly among these sources, and this variation in the polarization fraction could be caused by the inclination effect of each dust disk. MMS\,6 has exhibited a significantly higher polarization fraction than the other sources, potentially suggesting an alternative mechanism, such as dust alignment by magnetic field.

\end{enumerate}

\vspace{1\baselineskip}
\noindent$Acknowledgements$. We thank the referee for the thorough review and for providing the helpful and detailed comments. This paper makes use of the following ALMA data: ADS/JAO.ALMA\#2015.1.00341.S. ALMA is a partnership of ESO (representing its member states), NSF (USA) and NINS (Japan), together with NRC (Canada), MOST and ASIAA (Taiwan), and KASI (Republic of Korea), in cooperation with the Republic of Chile. The Joint ALMA Observatory is operated by ESO, AUI/NRAO and NAOJ. Data analysis was carried out on the Multi-wavelength Data Analysis System operated by the Astronomy Data Center (ADC), National Astronomical Observatory of Japan. This work was supported by JST SPRING, Grant Number JPMJSP2136. This work was supported by Grants-in-Aid for Scientific Research (KAKENHI) of the Japan Society for the Promotion of Science (JSPS; grant Nos. JP21K03617 and JP21H00046).
J.M.G.\ acknowledges support by the grant PID2020-117710GB-I00 (MCI-AEI-FEDER, UE) and by the program Unidad de Excelencia Mar\'{\i}a de Maeztu CEX2020-001058-M. 


\begin{thebibliography}{}
\expandafter\ifx\csname natexlab\endcsname\relax\def\natexlab#1{#1}\fi
\providecommand{\url}[1]{\href{#1}{#1}}
\providecommand{\dodoi}[1]{doi:~\href{http://doi.org/#1}{\nolinkurl{#1}}}
\providecommand{\doeprint}[1]{\href{http://ascl.net/#1}{\nolinkurl{http://ascl.net/#1}}}
\providecommand{\doarXiv}[1]{\href{https://arxiv.org/abs/#1}{\nolinkurl{https://arxiv.org/abs/#1}}}

\bibitem[{{Andersson} {et~al.}(2015){Andersson}, {Lazarian}, \&
  {Vaillancourt}}]{andersson2015}
{Andersson}, B.~G., {Lazarian}, A., \& {Vaillancourt}, J.~E. 2015, \araa, 53,
  501, \dodoi{10.1146/annurev-astro-082214-122414}

\bibitem[{{Andre} \& {Montmerle}(1994)}]{andre1994}
{Andre}, P., \& {Montmerle}, T. 1994, \apj, 420, 837, \dodoi{10.1086/173608}

\bibitem[{{Andre} {et~al.}(1993){Andre}, {Ward-Thompson}, \&
  {Barsony}}]{andre1993submillimeter}
{Andre}, P., {Ward-Thompson}, D., \& {Barsony}, M. 1993, \apj, 406, 122,
  \dodoi{10.1086/172425}

\bibitem[{{Aso} {et~al.}(2000){Aso}, {Tatematsu}, {Sekimoto}, {Nakano},
  {Umemoto}, {Koyama}, \& {Yamamoto}}]{aso2000dense}
{Aso}, Y., {Tatematsu}, K., {Sekimoto}, Y., {et~al.} 2000, \apjs, 131, 465,
  \dodoi{10.1086/317378}

\bibitem[{{Birnstiel} {et~al.}(2010){Birnstiel}, {Dullemond}, \&
  {Brauer}}]{birnstiel2010}
{Birnstiel}, T., {Dullemond}, C.~P., \& {Brauer}, F. 2010, \aap, 513, A79,
  \dodoi{10.1051/0004-6361/200913731}

\bibitem[{{Brandenburg} {et~al.}(1995){Brandenburg}, {Nordlund}, {Stein}, \&
  {Torkelsson}}]{brandenburg1995}
{Brandenburg}, A., {Nordlund}, A., {Stein}, R.~F., \& {Torkelsson}, U. 1995,
  \apj, 446, 741, \dodoi{10.1086/175831}

\bibitem[{{CASA Team} {et~al.}(2022){CASA Team}, {Bean}, {Bhatnagar}, {Castro},
  {Donovan Meyer}, {Emonts}, {Garcia}, {Garwood}, {Golap}, {Gonzalez Villalba},
  {Harris}, {Hayashi}, {Hoskins}, {Hsieh}, {Jagannathan}, {Kawasaki},
  {Keimpema}, {Kettenis}, {Lopez}, {Marvil}, {Masters}, {McNichols},
  {Mehringer}, {Miel}, {Moellenbrock}, {Montesino}, {Nakazato}, {Ott}, {Petry},
  {Pokorny}, {Raba}, {Rau}, {Schiebel}, {Schweighart}, {Sekhar}, {Shimada},
  {Small}, {Steeb}, {Sugimoto}, {Suoranta}, {Tsutsumi}, {van Bemmel},
  {Verkouter}, {Wells}, {Xiong}, {Szomoru}, {Griffith}, {Glendenning}, \&
  {Kern}}]{bean2022casa}
{CASA Team}, {Bean}, B., {Bhatnagar}, S., {et~al.} 2022, \pasp, 134, 114501,
  \dodoi{10.1088/1538-3873/ac9642}

\bibitem[{{Caselli} {et~al.}(2019){Caselli}, {Pineda}, {Zhao}, {Walmsley},
  {Keto}, {Tafalla}, {Chac{\'o}n-Tanarro}, {Bourke}, {Friesen}, {Galli}, \&
  {Padovani}}]{caselli2019}
{Caselli}, P., {Pineda}, J.~E., {Zhao}, B., {et~al.} 2019, \apj, 874, 89,
  \dodoi{10.3847/1538-4357/ab0700}

\bibitem[{{Chini} {et~al.}(1997){Chini}, {Reipurth}, {Ward-Thompson}, {Bally},
  {Nyman}, {Sievers}, \& {Billawala}}]{chini1997dust}
{Chini}, R., {Reipurth}, B., {Ward-Thompson}, D., {et~al.} 1997, \apjl, 474,
  L135, \dodoi{10.1086/310436}

\bibitem[{{Cho} \& {Lazarian}(2007)}]{cho&lazarian2007}
{Cho}, J., \& {Lazarian}, A. 2007, \apj, 669, 1085, \dodoi{10.1086/521805}

\bibitem[{{Cox} {et~al.}(2018){Cox}, {Harris}, {Looney}, {Li}, {Yang}, {Tobin},
  \& {Stephens}}]{cox2018alma}
{Cox}, E.~G., {Harris}, R.~J., {Looney}, L.~W., {et~al.} 2018, \apj, 855, 92,
  \dodoi{10.3847/1538-4357/aaacd2}

\bibitem[{{Crutcher}(2012)}]{crutcher2012magnetic}
{Crutcher}, R.~M. 2012, \araa, 50, 29,
  \dodoi{10.1146/annurev-astro-081811-125514}

\bibitem[{{Draine} \& {Weingartner}(1996)}]{draine1996}
{Draine}, B.~T., \& {Weingartner}, J.~C. 1996, \apj, 470, 551,
  \dodoi{10.1086/177887}

\bibitem[{{Feddersen} {et~al.}(2020){Feddersen}, {Arce}, {Kong}, {Suri},
  {S{\'a}nchez-Monge}, {Ossenkopf-Okada}, {Dunham}, {Nakamura}, {Shimajiri}, \&
  {Bally}}]{feddersen2020carma}
{Feddersen}, J.~R., {Arce}, H.~G., {Kong}, S., {et~al.} 2020, \apj, 896, 11,
  \dodoi{10.3847/1538-4357/ab86a9}

\bibitem[{{Frau} {et~al.}(2011){Frau}, {Galli}, \& {Girart}}]{frau2011}
{Frau}, P., {Galli}, D., \& {Girart}, J.~M. 2011, \aap, 535, A44,
  \dodoi{10.1051/0004-6361/201117813}

\bibitem[{{Furlan} {et~al.}(2016){Furlan}, {Fischer}, {Ali}, {Stutz}, {Stanke},
  {Tobin}, {Megeath}, {Osorio}, {Hartmann}, {Calvet}, {Poteet}, {Booker},
  {Manoj}, {Watson}, \& {Allen}}]{furlan2016herschel}
{Furlan}, E., {Fischer}, W.~J., {Ali}, B., {et~al.} 2016, \apjs, 224, 5,
  \dodoi{10.3847/0067-0049/224/1/5}

\bibitem[{{Girart} {et~al.}(2013){Girart}, {Frau}, {Zhang}, {Koch}, {Qiu},
  {Tang}, {Lai}, \& {Ho}}]{girart2013sma}
{Girart}, J.~M., {Frau}, P., {Zhang}, Q., {et~al.} 2013, \apj, 772, 69,
  \dodoi{10.1088/0004-637X/772/1/69}

\bibitem[{{Girart} {et~al.}(2006){Girart}, {Rao}, \& {Marrone}}]{girart2006sma}
{Girart}, J.~M., {Rao}, R., \& {Marrone}, D.~P. 2006, Science, 313, 812,
  \dodoi{10.1126/science.112909310.48550/arXiv.astro-ph/0609177}

\bibitem[{{Gold}(1952)}]{gold1952MNRAS.112..215G}
{Gold}, T. 1952, \mnras, 112, 215, \dodoi{10.1093/mnras/112.2.215}

\bibitem[{{G{\'o}mez-Ruiz} {et~al.}(2019){G{\'o}mez-Ruiz}, {Gusdorf},
  {Leurini}, {Menten}, {Takahashi}, {Wyrowski}, \&
  {G{\"u}sten}}]{gomez2019warm}
{G{\'o}mez-Ruiz}, A.~I., {Gusdorf}, A., {Leurini}, S., {et~al.} 2019, \aap,
  629, A77, \dodoi{10.1051/0004-6361/201424156}

\bibitem[{{Greene} {et~al.}(1994){Greene}, {Wilking}, {Andre}, {Young}, \&
  {Lada}}]{greene1994}
{Greene}, T.~P., {Wilking}, B.~A., {Andre}, P., {Young}, E.~T., \& {Lada},
  C.~J. 1994, \apj, 434, 614, \dodoi{10.1086/174763}

\bibitem[{{Guillet} {et~al.}(2020){Guillet}, {Girart}, {Maury}, \&
  {Alves}}]{guillet2020mie}
{Guillet}, V., {Girart}, J.~M., {Maury}, A.~J., \& {Alves}, F.~O. 2020, \aap,
  634, L15, \dodoi{10.1051/0004-6361/201937314}

\bibitem[{{Hildebrand}(1988)}]{hildebrand1988magnetic}
{Hildebrand}, R.~H. 1988, \qjras, 29, 327

\bibitem[{{Hirano} {et~al.}(2023){Hirano}, {Sahu}, {Liu}, {Liu}, {Tatematsu},
  {Dutta}, {Li}, {Lee}, {Li}, {Hsu}, {Lin}, {Johnstone}, {Bronfman}, {Chen},
  {Eden}, {Kuan}, {Kwon}, {Lee}, {Liu}, {Rawlings}, {Ristorcelli}, \&
  {Traficante}}]{hirano2023}
{Hirano}, N., {Sahu}, D., {Liu}, S.-Y., {et~al.} 2023, arXiv e-prints,
  arXiv:2311.05308, \dodoi{10.48550/arXiv.2311.05308}

\bibitem[{{Hoang} {et~al.}(2018){Hoang}, {Cho}, \& {Lazarian}}]{hoang2018}
{Hoang}, T., {Cho}, J., \& {Lazarian}, A. 2018, \apj, 852, 129,
  \dodoi{10.3847/1538-4357/aa9edc}

\bibitem[{{Houde} {et~al.}(2004){Houde}, {Dowell}, {Hildebrand}, {Dotson},
  {Vaillancourt}, {Phillips}, {Peng}, \& {Bastien}}]{houde2004tracing}
{Houde}, M., {Dowell}, C.~D., {Hildebrand}, R.~H., {et~al.} 2004, \apj, 604,
  717, \dodoi{10.1086/382067}

\bibitem[{{Hull} \& {Zhang}(2019)}]{hull2019interferometric}
{Hull}, C. L.~H., \& {Zhang}, Q. 2019, Frontiers in Astronomy and Space
  Sciences, 6, 3, \dodoi{10.3389/fspas.2019.00003}

\bibitem[{{Hull} {et~al.}(2014){Hull}, {Plambeck}, {Kwon}, {Bower},
  {Carpenter}, {Crutcher}, {Fiege}, {Franzmann}, {Hakobian}, {Heiles}, {Houde},
  {Hughes}, {Lamb}, {Looney}, {Marrone}, {Matthews}, {Pillai}, {Pound},
  {Rahman}, {Sandell}, {Stephens}, {Tobin}, {Vaillancourt}, {Volgenau}, \&
  {Wright}}]{hull2014carma}
{Hull}, C. L.~H., {Plambeck}, R.~L., {Kwon}, W., {et~al.} 2014, \apjs, 213, 13,
  \dodoi{10.1088/0067-0049/213/1/13}

\bibitem[{{Hull} {et~al.}(2020){Hull}, {Cortes}, {Gouellec}, {Girart}, {Nagai},
  {Nakanishi}, {Kameno}, {Fomalont}, {Brogan}, {Moellenbrock}, {Paladino}, \&
  {Villard}}]{hull2020}
{Hull}, C. L.~H., {Cortes}, P.~C., {Gouellec}, V. J.~M.~L., {et~al.} 2020,
  \pasp, 132, 094501, \dodoi{10.1088/1538-3873/ab99cd}

\bibitem[{{Joos} {et~al.}(2012){Joos}, {Hennebelle}, \& {Ciardi}}]{joos2012}
{Joos}, M., {Hennebelle}, P., \& {Ciardi}, A. 2012, \aap, 543, A128,
  \dodoi{10.1051/0004-6361/201118730}

\bibitem[{{Kataoka} {et~al.}(2016{\natexlab{a}}){Kataoka}, {Muto}, {Momose},
  {Tsukagoshi}, \& {Dullemond}}]{kataoka2016grain}
{Kataoka}, A., {Muto}, T., {Momose}, M., {Tsukagoshi}, T., \& {Dullemond},
  C.~P. 2016{\natexlab{a}}, \apj, 820, 54,
  \dodoi{10.3847/0004-637X/820/1/5410.48550/arXiv.1507.08902}

\bibitem[{{Kataoka} {et~al.}(2019){Kataoka}, {Okuzumi}, \&
  {Tazaki}}]{kataoka2019gasflow}
{Kataoka}, A., {Okuzumi}, S., \& {Tazaki}, R. 2019, \apjl, 874, L6,
  \dodoi{10.3847/2041-8213/ab0c9a}

\bibitem[{{Kataoka} {et~al.}(2017){Kataoka}, {Tsukagoshi}, {Pohl}, {Muto},
  {Nagai}, {Stephens}, {Tomisaka}, \& {Momose}}]{kataoka2017evidence}
{Kataoka}, A., {Tsukagoshi}, T., {Pohl}, A., {et~al.} 2017, \apjl, 844, L5,
  \dodoi{10.3847/2041-8213/aa7e33}

\bibitem[{{Kataoka} {et~al.}(2015){Kataoka}, {Muto}, {Momose}, {Tsukagoshi},
  {Fukagawa}, {Shibai}, {Hanawa}, {Murakawa}, \&
  {Dullemond}}]{kataoka2015millimeter}
{Kataoka}, A., {Muto}, T., {Momose}, M., {et~al.} 2015, \apj, 809, 78,
  \dodoi{10.1088/0004-637X/809/1/78}

\bibitem[{{Kataoka} {et~al.}(2016{\natexlab{b}}){Kataoka}, {Tsukagoshi},
  {Momose}, {Nagai}, {Muto}, {Dullemond}, {Pohl}, {Fukagawa}, {Shibai},
  {Hanawa}, \& {Murakawa}}]{kataoka2016submillimeter}
{Kataoka}, A., {Tsukagoshi}, T., {Momose}, M., {et~al.} 2016{\natexlab{b}},
  \apjl, 831, L12, \dodoi{10.3847/2041-8205/831/2/L12}

\bibitem[{{Kauffmann} {et~al.}(2008){Kauffmann}, {Bertoldi}, {Bourke}, {Evans},
  \& {Lee}}]{kauffmann2008mambo}
{Kauffmann}, J., {Bertoldi}, F., {Bourke}, T.~L., {Evans}, N.~J., I., \& {Lee},
  C.~W. 2008, \aap, 487, 993, \dodoi{10.1051/0004-6361:200809481}

\bibitem[{{Koch} {et~al.}(2018){Koch}, {Tang}, {Ho}, {Yen}, {Su}, \&
  {Takakuwa}}]{koch2018alma}
{Koch}, P.~M., {Tang}, Y.-W., {Ho}, P. T.~P., {et~al.} 2018, \apj, 855, 39,
  \dodoi{10.3847/1538-4357/aaa4c1}

\bibitem[{{Koch} {et~al.}(2022){Koch}, {Tang}, {Ho}, {Hsieh}, {Wang}, {Yen},
  {Duarte-Cabral}, {Peretto}, \& {Su}}]{koch2022multiscale}
---. 2022, \apj, 940, 89, \dodoi{10.3847/1538-4357/ac96e3}

\bibitem[{{Lada}(1987)}]{lada1987}
{Lada}, C.~J. 1987, in Star Forming Regions, ed. M.~{Peimbert} \& J.~{Jugaku},
  Vol. 115, 1

\bibitem[{{Lai} {et~al.}(2001){Lai}, {Crutcher}, {Girart}, \&
  {Rao}}]{lai2001bima}
{Lai}, S.-P., {Crutcher}, R.~M., {Girart}, J.~M., \& {Rao}, R. 2001, \apj, 561,
  864, \dodoi{10.1086/32337210.48550/arXiv.astro-ph/0107322}

\bibitem[{{Larson}(1969)}]{larson1969}
{Larson}, R.~B. 1969, \mnras, 145, 271, \dodoi{10.1093/mnras/145.3.271}

\bibitem[{{Lazarian}(2007)}]{lazarian2007tracing}
{Lazarian}, A. 2007, \jqsrt, 106, 225, \dodoi{10.1016/j.jqsrt.2007.01.038}

\bibitem[{{Lazarian} \& {Hoang}(2007{\natexlab{a}})}]{Lazarian&Hoang2007RAT}
{Lazarian}, A., \& {Hoang}, T. 2007{\natexlab{a}}, \mnras, 378, 910,
  \dodoi{10.1111/j.1365-2966.2007.11817.x}

\bibitem[{{Lazarian} \& {Hoang}(2007{\natexlab{b}})}]{lazarian2007mechanical}
---. 2007{\natexlab{b}}, \apjl, 669, L77, \dodoi{10.1086/523849}

\bibitem[{{Le Gouellec} {et~al.}(2023){Le Gouellec}, {Maury}, {Hull},
  {Verliat}, {Hennebelle}, \& {Valdivia}}]{Gouellec2023}
{Le Gouellec}, V.~J.~M., {Maury}, A.~J., {Hull}, C.~L.~H., {et~al.} 2023, \aap,
  675, A133, \dodoi{10.1051/0004-6361/202245346}

\bibitem[{{Le Gouellec} {et~al.}(2020){Le Gouellec}, {Maury}, {Guillet},
  {Hull}, {Girart}, {Verliat}, {Mignon-Risse}, {Valdivia}, {Hennebelle},
  {Gonz{\'a}lez}, \& {Louvet}}]{Gouellec2020}
{Le Gouellec}, V.~J.~M., {Maury}, A.~J., {Guillet}, V., {et~al.} 2020, \aap,
  644, A11, \dodoi{10.1051/0004-6361/202038404}

\bibitem[{{Lee} {et~al.}(2018){Lee}, {Li}, {Ching}, {Lai}, \&
  {Yang}}]{lee2018alma}
{Lee}, C.-F., {Li}, Z.-Y., {Ching}, T.-C., {Lai}, S.-P., \& {Yang}, H. 2018,
  \apj, 854, 56, \dodoi{10.3847/1538-4357/aaa76910.48550/arXiv.1801.03802}

\bibitem[{{Lis} {et~al.}(1998){Lis}, {Serabyn}, {Keene}, {Dowell}, {Benford},
  {Phillips}, {Hunter}, \& {Wang}}]{lis1998350}
{Lis}, D.~C., {Serabyn}, E., {Keene}, J., {et~al.} 1998, \apj, 509, 299,
  \dodoi{10.1086/306500}

\bibitem[{{Liu}(2021)}]{liubaobab2021}
{Liu}, H.~B. 2021, \apj, 914, 25, \dodoi{10.3847/1538-4357/abf8b6}

\bibitem[{{Liu} {et~al.}(2020){Liu}, {Zhang}, {Qiu}, {Liu}, {Pillai}, {Girart},
  {Li}, \& {Wang}}]{liu2020alma}
{Liu}, J., {Zhang}, Q., {Qiu}, K., {et~al.} 2020, \apj, 895, 142,
  \dodoi{10.3847/1538-4357/ab9087}

\bibitem[{{Machida} {et~al.}(2008){Machida}, {Inutsuka}, \&
  {Matsumoto}}]{machida2008}
{Machida}, M.~N., {Inutsuka}, S.-i., \& {Matsumoto}, T. 2008, \apj, 676, 1088,
  \dodoi{10.1086/528364}

\bibitem[{{Masunaga} \& {Inutsuka}(2000)}]{matsunaga2000}
{Masunaga}, H., \& {Inutsuka}, S.-i. 2000, \apj, 531, 350,
  \dodoi{10.1086/308439}

\bibitem[{{Matsushita} {et~al.}(2019){Matsushita}, {Takahashi}, {Machida}, \&
  {Tomisaka}}]{matsushita2019very}
{Matsushita}, Y., {Takahashi}, S., {Machida}, M.~N., \& {Tomisaka}, K. 2019,
  \apj, 871, 221, \dodoi{10.3847/1538-4357/aaf1b6}

\bibitem[{{Matthews} {et~al.}(2005){Matthews}, {Lai}, {Crutcher}, \&
  {Wilson}}]{matthews2005multiscale}
{Matthews}, B.~C., {Lai}, S.-P., {Crutcher}, R.~M., \& {Wilson}, C.~D. 2005,
  \apj, 626, 959, \dodoi{10.1086/430127}

\bibitem[{{Matthews} \& {Wilson}(2000)}]{matthews2000magnetic}
{Matthews}, B.~C., \& {Wilson}, C.~D. 2000, \apj, 531, 868,
  \dodoi{10.1086/308523}

\bibitem[{{Matthews} {et~al.}(2001){Matthews}, {Wilson}, \&
  {Fiege}}]{matthews2001magnetic}
{Matthews}, B.~C., {Wilson}, C.~D., \& {Fiege}, J.~D. 2001, \apj, 562, 400,
  \dodoi{10.1086/323375}

\bibitem[{{Morii} {et~al.}(2021){Morii}, {Takahashi}, \&
  {Machida}}]{morii2021revealing}
{Morii}, K., {Takahashi}, S., \& {Machida}, M.~N. 2021, \apj, 910, 148,
  \dodoi{10.3847/1538-4357/abe61c}

\bibitem[{{Nielbock} {et~al.}(2003){Nielbock}, {Chini}, \&
  {M{\"u}ller}}]{nielbock2003stellar}
{Nielbock}, M., {Chini}, R., \& {M{\"u}ller}, S.~A.~H. 2003, \aap, 408, 245,
  \dodoi{10.1051/0004-6361:20030961}

\bibitem[{{Ohashi} {et~al.}(2018){Ohashi}, {Kataoka}, {Nagai}, {Momose},
  {Muto}, {Hanawa}, {Fukagawa}, {Tsukagoshi}, {Murakawa}, \&
  {Shibai}}]{ohashi2018}
{Ohashi}, S., {Kataoka}, A., {Nagai}, H., {et~al.} 2018, \apj, 864, 81,
  \dodoi{10.3847/1538-4357/aad632}

\bibitem[{{Ossenkopf} \& {Henning}(1994)}]{ossenkopf1994dust}
{Ossenkopf}, V., \& {Henning}, T. 1994, \aap, 291, 943

\bibitem[{{Pohl} {et~al.}(2016){Pohl}, {Kataoka}, {Pinilla}, {Dullemond},
  {Henning}, \& {Birnstiel}}]{pohl2016}
{Pohl}, A., {Kataoka}, A., {Pinilla}, P., {et~al.} 2016, \aap, 593, A12,
  \dodoi{10.1051/0004-6361/201628637}

\bibitem[{{Rao} {et~al.}(1998){Rao}, {Crutcher}, {Plambeck}, \&
  {Wright}}]{rao1998bima}
{Rao}, R., {Crutcher}, R.~M., {Plambeck}, R.~L., \& {Wright}, M.~C.~H. 1998,
  \apjl, 502, L75, \dodoi{10.1086/311485}

\bibitem[{{Rao} {et~al.}(2014){Rao}, {Girart}, {Lai}, \&
  {Marrone}}]{rao2014sma}
{Rao}, R., {Girart}, J.~M., {Lai}, S.-P., \& {Marrone}, D.~P. 2014, \apjl, 780,
  L6, \dodoi{10.1088/2041-8205/780/1/L6}

\bibitem[{{Rao} {et~al.}(2009){Rao}, {Girart}, {Marrone}, {Lai}, \&
  {Schnee}}]{rao2009sma}
{Rao}, R., {Girart}, J.~M., {Marrone}, D.~P., {Lai}, S.-P., \& {Schnee}, S.
  2009, \apj, 707, 921, \dodoi{10.1088/0004-637X/707/2/921}

\bibitem[{{Reipurth} {et~al.}(2004){Reipurth}, {Rodr{\'\i}guez}, {Anglada}, \&
  {Bally}}]{reipurth2004radio}
{Reipurth}, B., {Rodr{\'\i}guez}, L.~F., {Anglada}, G., \& {Bally}, J. 2004,
  \aj, 127, 1736, \dodoi{10.1086/381062}

\bibitem[{{Reipurth} {et~al.}(1999){Reipurth}, {Rodr{\'\i}guez}, \&
  {Chini}}]{reipurth1999vla}
{Reipurth}, B., {Rodr{\'\i}guez}, L.~F., \& {Chini}, R. 1999, \aj, 118, 983,
  \dodoi{10.1086/300958}

\bibitem[{{Ren} {et~al.}(2021){Ren}, {Zhu}, {Shi}, {Yue}, {Li}, {Zhang},
  {Mardones}, {Wu}, {Jiao}, {Liu}, {Luo}, {Xie}, {Zhang}, \&
  {Xu}}]{zhiyuan2021}
{Ren}, Z., {Zhu}, L., {Shi}, H., {et~al.} 2021, \mnras, 505, 5183,
  \dodoi{10.1093/mnras/stab1509}

\bibitem[{{Sadavoy} {et~al.}(2019){Sadavoy}, {Stephens}, {Myers}, {Looney},
  {Tobin}, {Kwon}, {Commer{\c{c}}on}, {Segura-Cox}, {Henning}, \&
  {Hennebelle}}]{sadavoy2019}
{Sadavoy}, S.~I., {Stephens}, I.~W., {Myers}, P.~C., {et~al.} 2019, \apjs, 245,
  2, \dodoi{10.3847/1538-4365/ab4257}

\bibitem[{{Shu} {et~al.}(1987){Shu}, {Adams}, \& {Lizano}}]{shu1987star}
{Shu}, F.~H., {Adams}, F.~C., \& {Lizano}, S. 1987, \araa, 25, 23,
  \dodoi{10.1146/annurev.aa.25.090187.000323}

\bibitem[{{Stanke} {et~al.}(2002){Stanke}, {McCaughrean}, \&
  {Zinnecker}}]{stanke2002unbiased}
{Stanke}, T., {McCaughrean}, M.~J., \& {Zinnecker}, H. 2002, \aap, 392, 239,
  \dodoi{10.1051/0004-6361:20020763}

\bibitem[{{Stephens} {et~al.}(2013){Stephens}, {Looney}, {Kwon}, {Hull},
  {Plambeck}, {Crutcher}, {Chapman}, {Novak}, {Davidson}, {Vaillancourt},
  {Shinnaga}, \& {Matthews}}]{stephens2013carma}
{Stephens}, I.~W., {Looney}, L.~W., {Kwon}, W., {et~al.} 2013, \apjl, 769, L15,
  \dodoi{10.1088/2041-8205/769/1/L1510.48550/arXiv.1304.6739}

\bibitem[{{Stephens} {et~al.}(2017){Stephens}, {Yang}, {Li}, {Looney},
  {Kataoka}, {Kwon}, {Fern{\'a}ndez-L{\'o}pez}, {Hull}, {Hughes}, {Segura-Cox},
  {Mundy}, {Crutcher}, \& {Rao}}]{stephen2017}
{Stephens}, I.~W., {Yang}, H., {Li}, Z.-Y., {et~al.} 2017, \apj, 851, 55,
  \dodoi{10.3847/1538-4357/aa998b}

\bibitem[{{Takahashi} \& {Ho}(2012)}]{takahashi2012molecular}
{Takahashi}, S., \& {Ho}, P. T.~P. 2012, \apjl, 745, L10,
  \dodoi{10.1088/2041-8205/745/1/L10}

\bibitem[{{Takahashi} {et~al.}(2009){Takahashi}, {Ho}, {Tang}, {Kawabe}, \&
  {Saito}}]{takahashi2009}
{Takahashi}, S., {Ho}, P. T.~P., {Tang}, Y.-W., {Kawabe}, R., \& {Saito}, M.
  2009, \apj, 704, 1459, \dodoi{10.1088/0004-637X/704/2/1459}

\bibitem[{{Takahashi} {et~al.}(2013){Takahashi}, {Ho}, {Teixeira}, {Zapata}, \&
  {Su}}]{takahashi2013hierarchical}
{Takahashi}, S., {Ho}, P. T.~P., {Teixeira}, P.~S., {Zapata}, L.~A., \& {Su},
  Y.-N. 2013, \apj, 763, 57, \dodoi{10.1088/0004-637X/763/1/57}

\bibitem[{{Takahashi} {et~al.}(2019){Takahashi}, {Machida}, {Tomisaka}, {Ho},
  {Fomalont}, {Nakanishi}, \& {Girart}}]{takahashi2019alma}
{Takahashi}, S., {Machida}, M.~N., {Tomisaka}, K., {et~al.} 2019, \apj, 872,
  70, \dodoi{10.3847/1538-4357/aaf6ed}

\bibitem[{{Takahashi} {et~al.}(2012){Takahashi}, {Saigo}, {Ho}, \&
  {Tomida}}]{takahashi2012spatially}
{Takahashi}, S., {Saigo}, K., {Ho}, P. T.~P., \& {Tomida}, K. 2012, \apj, 752,
  10, \dodoi{10.1088/0004-637X/752/1/10}

\bibitem[{{Takahashi} {et~al.}(2008){Takahashi}, {Saito}, {Ohashi}, {Kusakabe},
  {Takakuwa}, {Shimajiri}, {Tamura}, \& {Kawabe}}]{takahashi2008millimeter}
{Takahashi}, S., {Saito}, M., {Ohashi}, N., {et~al.} 2008, \apj, 688, 344,
  \dodoi{10.1086/592212}

\bibitem[{{Takahashi} {et~al.}(2006){Takahashi}, {Saito}, {Takakuwa}, \&
  {Kawabe}}]{takahashi2006millimeter}
{Takahashi}, S., {Saito}, M., {Takakuwa}, S., \& {Kawabe}, R. 2006, \apj, 651,
  933, \dodoi{10.1086/507482}

\bibitem[{{Tanabe} {et~al.}(2019){Tanabe}, {Nakamura}, {Tsukagoshi},
  {Shimajiri}, {Ishii}, {Kawabe}, {Feddersen}, {Kong}, {Arce}, {Bally},
  {Carpenter}, \& {Momose}}]{tanabe2019nobeyama}
{Tanabe}, Y., {Nakamura}, F., {Tsukagoshi}, T., {et~al.} 2019, \pasj, 71, S8,
  \dodoi{10.1093/pasj/psz100}

\bibitem[{{Tang} {et~al.}(2013){Tang}, {Ho}, {Koch}, {Guilloteau}, \&
  {Dutrey}}]{tang2013sma}
{Tang}, Y.-W., {Ho}, P. T.~P., {Koch}, P.~M., {Guilloteau}, S., \& {Dutrey}, A.
  2013, \apj, 763, 135, \dodoi{10.1088/0004-637X/763/2/135}

\bibitem[{{Tazaki} {et~al.}(2017){Tazaki}, {Lazarian}, \&
  {Nomura}}]{tazaki2017radiative}
{Tazaki}, R., {Lazarian}, A., \& {Nomura}, H. 2017, \apj, 839, 56,
  \dodoi{10.3847/1538-4357/839/1/56}

\bibitem[{{Tobin} {et~al.}(2020){Tobin}, {Sheehan}, {Megeath},
  {D{\'\i}az-Rodr{\'\i}guez}, {Offner}, {Murillo}, {van 't Hoff}, {van
  Dishoeck}, {Osorio}, {Anglada}, {Furlan}, {Stutz}, {Reynolds}, {Karnath},
  {Fischer}, {Persson}, {Looney}, {Li}, {Stephens}, {Chandler}, {Cox},
  {Dunham}, {Tychoniec}, {Kama}, {Kratter}, {Kounkel}, {Mazur}, {Maud},
  {Patel}, {Perez}, {Sadavoy}, {Segura-Cox}, {Sharma}, {Stephenson}, {Watson},
  \& {Wyrowski}}]{tobin2020vla}
{Tobin}, J.~J., {Sheehan}, P.~D., {Megeath}, S.~T., {et~al.} 2020, \apj, 890,
  130, \dodoi{10.3847/1538-4357/ab6f64}

\bibitem[{{Tomida} {et~al.}(2017){Tomida}, {Machida}, {Hosokawa}, {Sakurai}, \&
  {Lin}}]{kengo2017}
{Tomida}, K., {Machida}, M.~N., {Hosokawa}, T., {Sakurai}, Y., \& {Lin}, C.~H.
  2017, \apjl, 835, L11, \dodoi{10.3847/2041-8213/835/1/L11}

\bibitem[{{Trapman} {et~al.}(2020){Trapman}, {Ansdell}, {Hogerheijde},
  {Facchini}, {Manara}, {Miotello}, {Williams}, \& {Bruderer}}]{trapman2020}
{Trapman}, L., {Ansdell}, M., {Hogerheijde}, M.~R., {et~al.} 2020, \aap, 638,
  A38, \dodoi{10.1051/0004-6361/201834537}

\bibitem[{{Vorobyov}(2011)}]{Vorobyov2011}
{Vorobyov}, E.~I. 2011, \apj, 729, 146, \dodoi{10.1088/0004-637X/729/2/146}

\bibitem[{{Weidenschilling}(1977)}]{weidenschilling1977}
{Weidenschilling}, S.~J. 1977, \mnras, 180, 57, \dodoi{10.1093/mnras/180.2.57}

\bibitem[{{Williams} {et~al.}(2003){Williams}, {Plambeck}, \&
  {Heyer}}]{williams2003high}
{Williams}, J.~P., {Plambeck}, R.~L., \& {Heyer}, M.~H. 2003, \apj, 591, 1025,
  \dodoi{10.1086/375396}

\bibitem[{{Xu} \& {Kunz}(2021)}]{xu2021formation}
{Xu}, W., \& {Kunz}, M.~W. 2021, \mnras, 508, 2142,
  \dodoi{10.1093/mnras/stab2715}

\bibitem[{{Yang}(2021)}]{Yang2021}
{Yang}, H. 2021, \apj, 911, 125, \dodoi{10.3847/1538-4357/abebde}

\bibitem[{{Yang} \& {Li}(2020)}]{yang2020reversal}
{Yang}, H., \& {Li}, Z.-Y. 2020, \apj, 889, 15,
  \dodoi{10.3847/1538-4357/ab5f08}

\bibitem[{{Yang} {et~al.}(2016){Yang}, {Li}, {Looney}, \&
  {Stephens}}]{yang2016inclination}
{Yang}, H., {Li}, Z.-Y., {Looney}, L., \& {Stephens}, I. 2016, \mnras, 456,
  2794, \dodoi{10.1093/mnras/stv2633}

\bibitem[{{Yang} {et~al.}(2017){Yang}, {Li}, {Looney}, {Girart}, \&
  {Stephens}}]{yang2017scattering}
{Yang}, H., {Li}, Z.-Y., {Looney}, L.~W., {Girart}, J.~M., \& {Stephens}, I.~W.
  2017, \mnras, 472, 373, \dodoi{10.1093/mnras/stx1951}

\bibitem[{{Yang} {et~al.}(2019){Yang}, {Li}, {Stephens}, {Kataoka}, \&
  {Looney}}]{yang2019}
{Yang}, H., {Li}, Z.-Y., {Stephens}, I.~W., {Kataoka}, A., \& {Looney}, L.
  2019, \mnras, 483, 2371, \dodoi{10.1093/mnras/sty3263}

\bibitem[{{Yen} {et~al.}(2017){Yen}, {Koch}, {Takakuwa}, {Krasnopolsky},
  {Ohashi}, \& {Aso}}]{Yen2017}
{Yen}, H.-W., {Koch}, P.~M., {Takakuwa}, S., {et~al.} 2017, \apj, 834, 178,
  \dodoi{10.3847/1538-4357/834/2/178}

\bibitem[{{Yu} {et~al.}(1997){Yu}, {Bally}, \& {Devine}}]{yu1997shock}
{Yu}, K.~C., {Bally}, J., \& {Devine}, D. 1997, \apjl, 485, L45,
  \dodoi{10.1086/310799}

\bibitem[{{Zhang} {et~al.}(2014){Zhang}, {Qiu}, {Girart}, {Liu}, {Tang},
  {Koch}, {Li}, {Keto}, {Ho}, {Rao}, {Lai}, {Ching}, {Frau}, {Chen}, {Li},
  {Padovani}, {Bontemps}, {Csengeri}, \& {Ju{\'a}rez}}]{zhang2014sma}
{Zhang}, Q., {Qiu}, K., {Girart}, J.~M., {et~al.} 2014, \apj, 792, 116,
  \dodoi{10.1088/0004-637X/792/2/116}

\bibitem[{{Zucconi} {et~al.}(2001){Zucconi}, {Walmsley}, \&
  {Galli}}]{Zucconi2001}
{Zucconi}, A., {Walmsley}, C.~M., \& {Galli}, D. 2001, \aap, 376, 650,
  \dodoi{10.1051/0004-6361:20010778}

\end{thebibliography}

\begin{appendix}

\section{Reliability of size estimation based on Gaussian fitting}\label{fitting}
To ensure that the fitted size and the associated fitting error are reasonable (and more or less comparable to those given by the CASA task ``\texttt{imfit}''), we performed independent experiments as described below. 
We discuss the reliability of fitting the compact sources in \S\ref{sec:A1}.
Then, we discuss the reliability of fitting the sources with large inclination angles in \S\ref{sec:A2}.

\subsection{Fit for small sized source and fitting error}
\label{sec:A1}
We performed the following experiments to (1) test how much size we can fit and deconvolve the compact component with respect to the synthesised beam size and (2) investigate how much uncertainty is introduced by the fit. As an example, we refer to the most compact resolved source MMS\,1. The fitted deconvolved size of MMS\,1 is about 1/3 of the synthesised beam size, and the fitting errors are about 1/40 of the synthesised beam size. 

We constructed a Gaussian model of the compact component similar to MMS\,1, to test the validity of fitting to the small source and its error. 
The model has a deconvolved size of $0\dotarcsec048 \times 0\dotarcsec034$ and the same total flux density as MMS\,1. 
In other words, we artificially created a small Gaussian object (MMS1 model) with a size of $0\dotarcsec048 \times 0\dotarcsec034$. 
We convolved it to our synthesized beam size. 
To mimic the observational data, we also added noise, which is obtained from the residual map of the clean process. 
Finally, we performed a Gaussian fitting 100 times (or 100 cases) for this object, varying the location and position angle (P.A.) on the residual map, in which the location and P.A. were arbitrarily set. 
All 100 trials were successfully done and the results are plotted in Figure~\ref{mms1fit}. In each trial, we performed the deconvolution process to investigate the deviation of the source size.

Figure~\ref{mms1fit} shows that the mean fitting size ($0\dotarcsec0481 \times 0\dotarcsec0343$) is very close to the size of the model we constructed ($0\dotarcsec048 \times 0\dotarcsec034$). 
For all cases, we calculated the standard deviation of the size.
The standard deviation is as small as $0\dotarcsec002$ for the major and minor axes. These values are comparable to the fitting errors that the CASA task ``\texttt{imfit}'' gave us (0\dotarcsec003) for MMS\,1. 
This experiment shows that even when the source size is as small as $\sim$1/3 of the synthesised beam size, we can successfully fit and deconvolve the source with an error of $\sim$1/70 of the synthesised beam size. 
Thus, we consider that for the compact sources, our fitting sizes smaller than the synthesised beam size are reliable, and the error given by the CASA task ``\texttt{imfit}'' is also reliable since the fitting size and error are comparable to those estimated from our experiments. 

\setcounter{figure}{0}
\renewcommand{\thefigure}{A\arabic{figure}}
\renewcommand{\theHfigure}{A\arabic{figure}}
\begin{figure*}[ht!]
\gridline{\hspace{-2\baselineskip}
          \fig{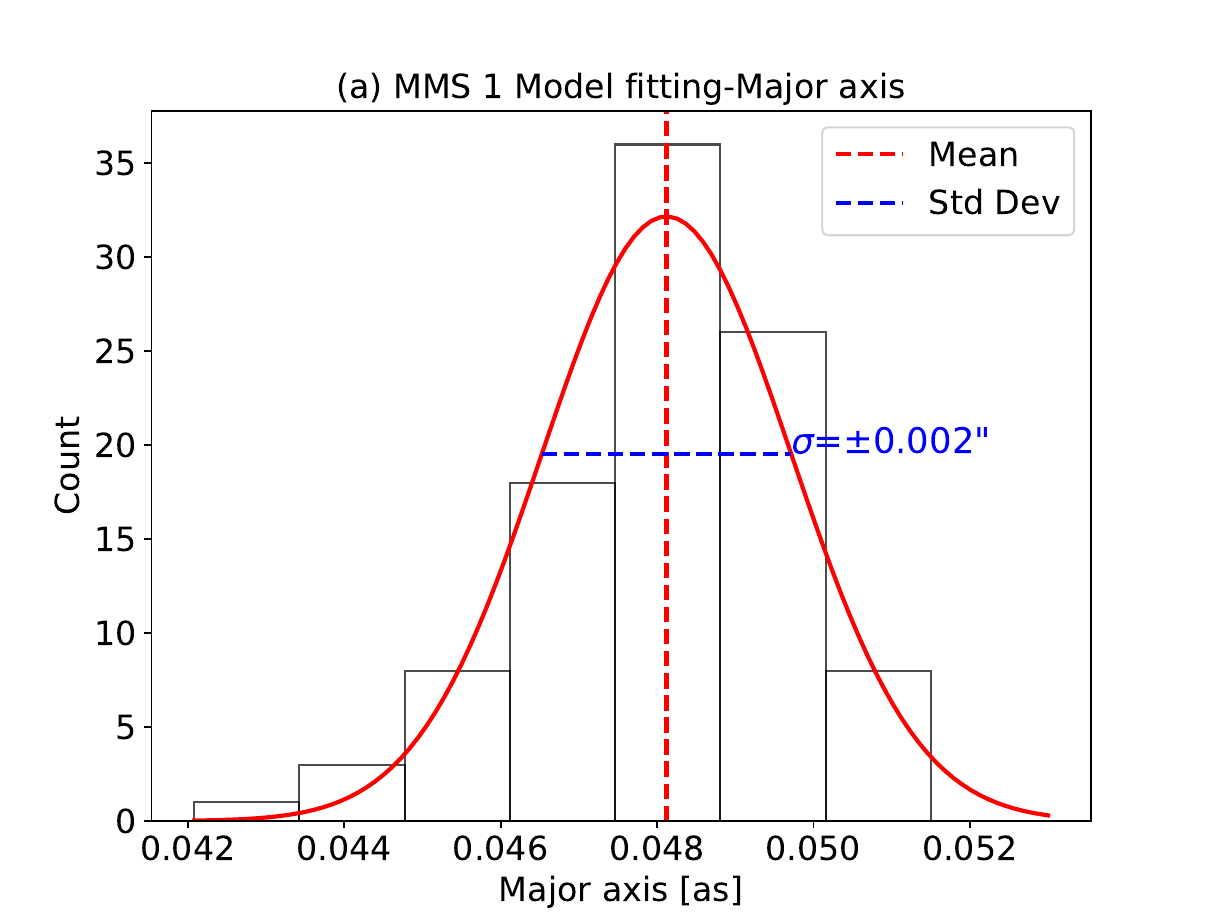}{0.43\textwidth}{}
          {}\hspace{-15\baselineskip}
          \fig{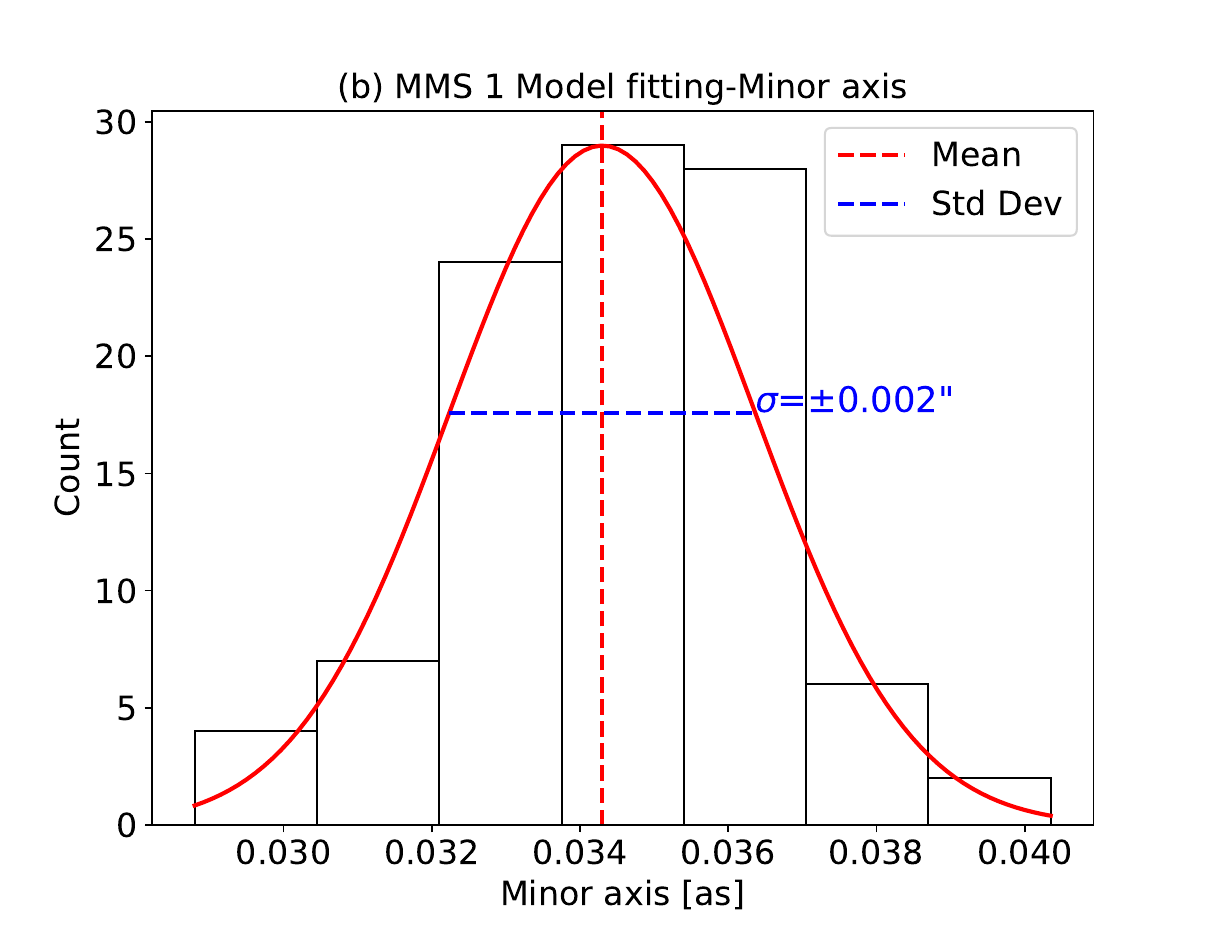}{0.43\textwidth}{}
          {}\hspace{-5\baselineskip}
          }
\caption{Histogram of the Gaussian fitted sizes for (a) major axis and (b) minor axis of the constructed MMS\,1 model. 
In each panel, the red curve, red dashed line, and black dashed line are the Gaussian  fitting result, mean fitted value, and standard deviation, respectively. 
}
\label{mms1fit}
\end{figure*}

\subsection{Large inclination angle with small minor axis} 
\label{sec:A2}

For sources with a large inclination angle and a minor axis much smaller than the beam size, we believe that the sizes of these objects are also reliable as long as the minor axis of the beam size is not extremely small, for example, $<$1/10 or $<$1/20 of the synthesized beam size. 
We performed experiments similar to those described in \S\ref{sec:A1}. 
We created two different models, model A and model B. 
For model A, we constructed a model with an inclination angle of 80$^{\circ}$ where the minor axis is set to 1/3 of the synthesised beam size (0$\dotarcsec$043) with a total flux density of 10 mJy\,beam$^{-1}$. For model B, we constructed a model with the same inclination angle (80$^{\circ}$) as in model A, but we set the minor axis to 1/20 of the synthesised beam size (0$\dotarcsec$007) with a total flux density same as MMS\,2-North-B. 
Thus, for model B, the minor axis is significantly smaller than the synthesised beam size. We performed the Gaussian fitting and deconvolution process 100 times in the same way as \S\ref{sec:A1}. 
We then examined the fitting success rate, source size, and standard deviation of the models. 

\begin{figure*}[ht!]
\gridline{\hspace{-2\baselineskip}
          \fig{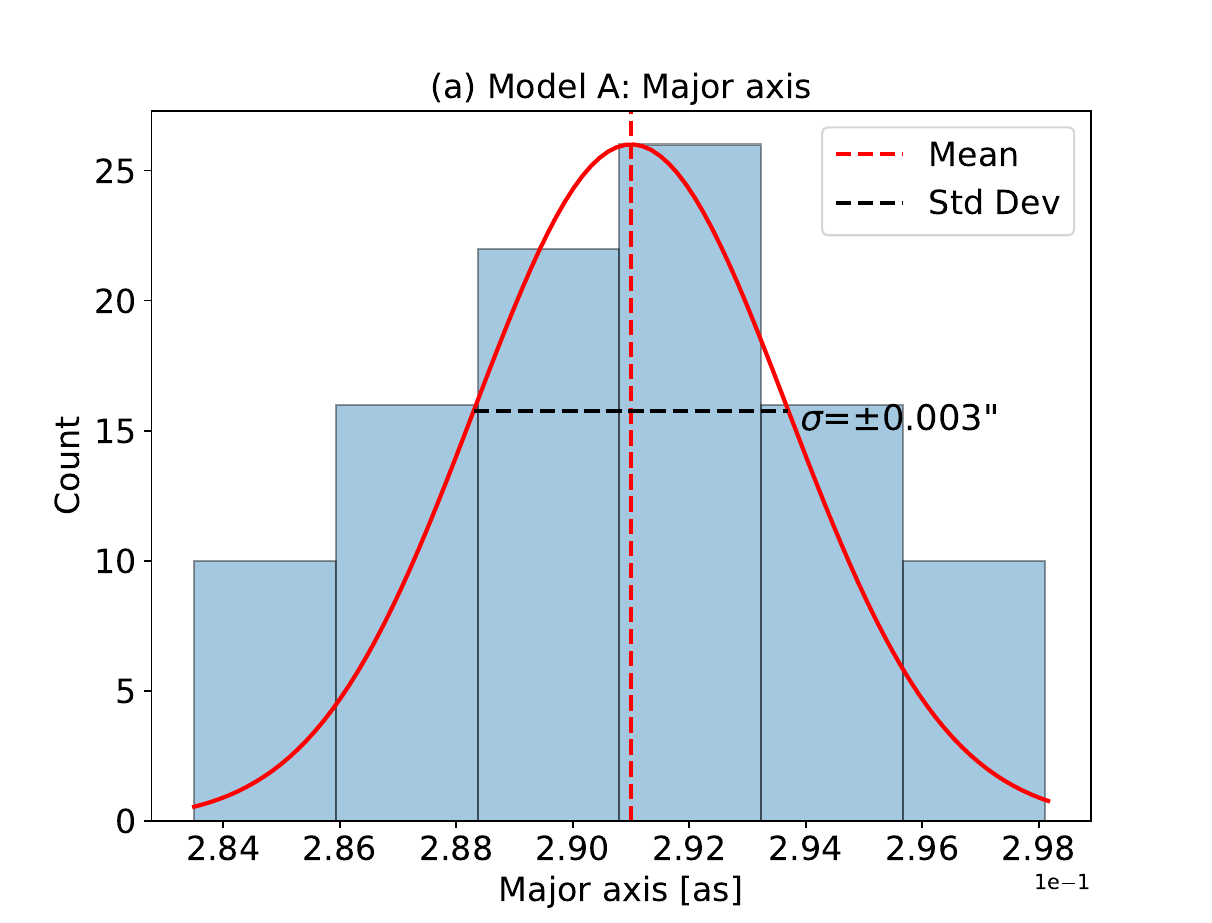}{0.43\textwidth}{}
          {}\hspace{-15\baselineskip}
          \fig{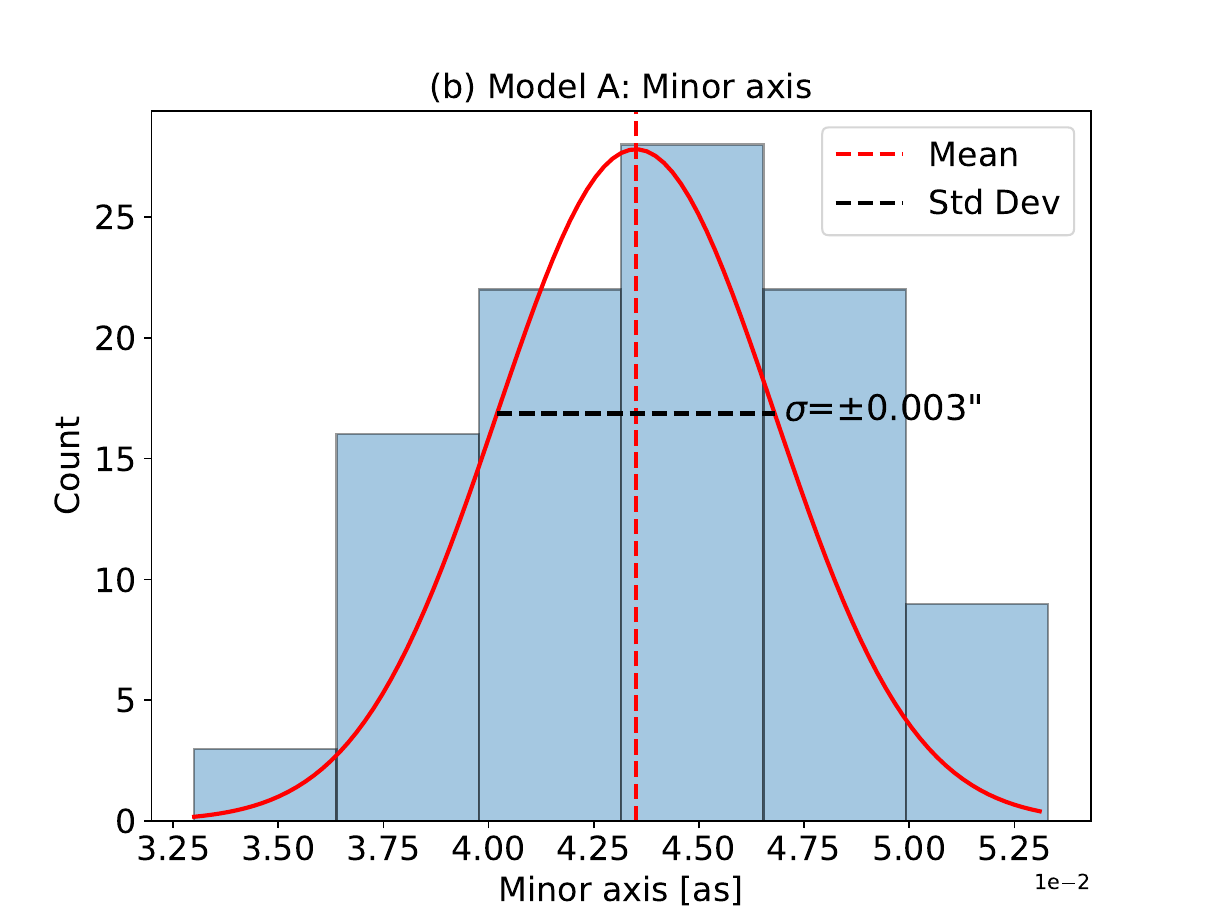}{0.43\textwidth}{}
          {}\hspace{-5\baselineskip}}
\gridline{\hspace{-2\baselineskip}          
          \fig{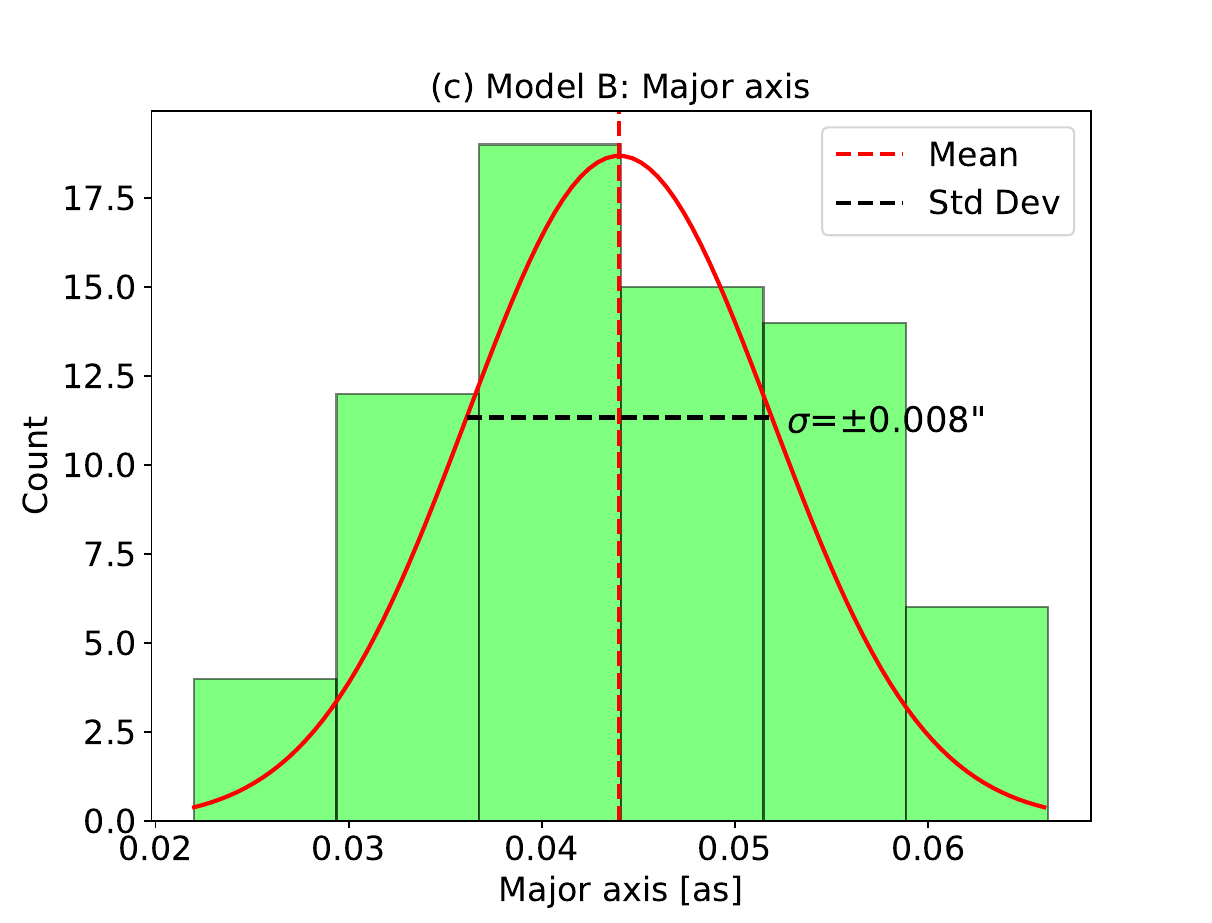}{0.43\textwidth}{}
          {}\hspace{-15\baselineskip}
          \fig{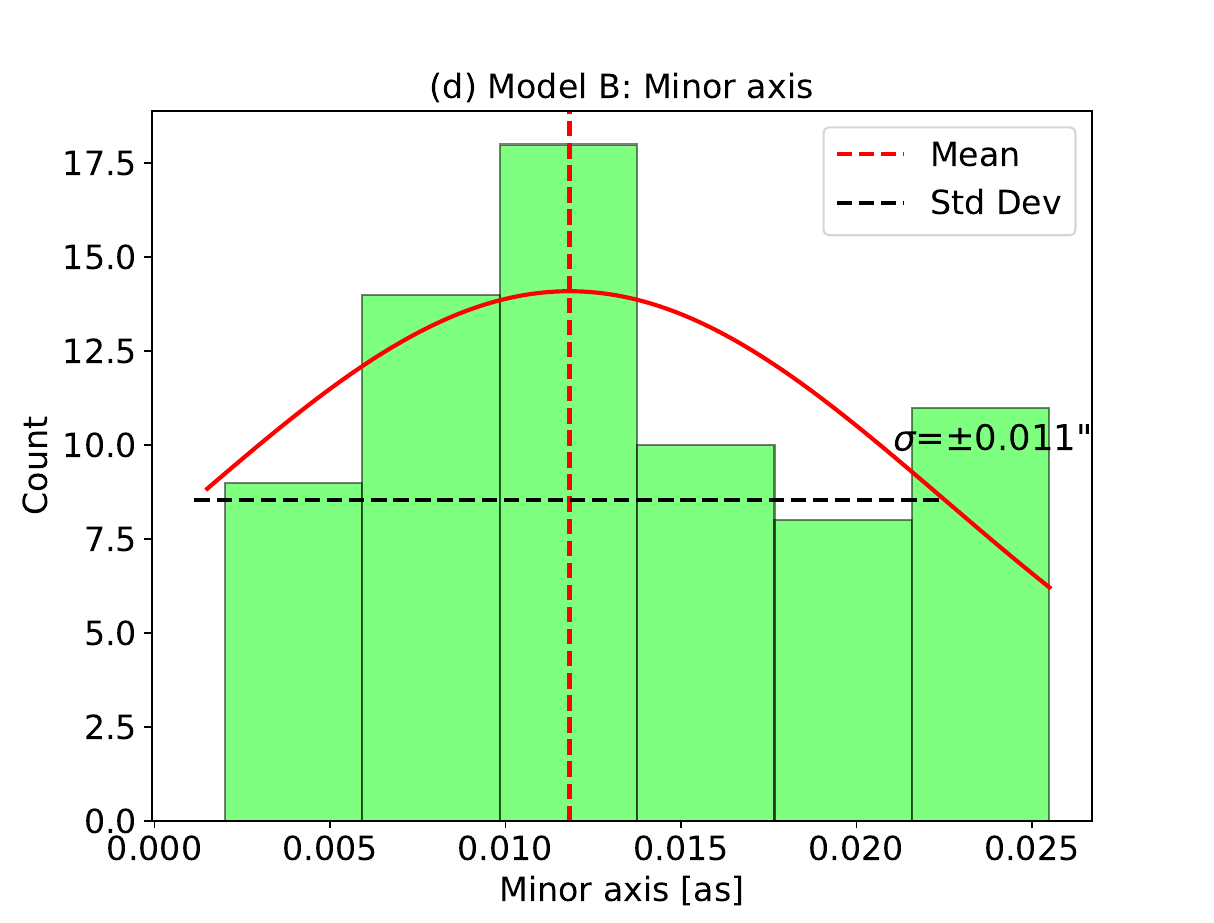}{0.43\textwidth}{}
          {}\hspace{-5\baselineskip}}
\caption{(top panels) Histogram of the Gaussian fitted sizes for the (a) major axis and (b) minor axis of model A. 
(bottom panels) Histogram of the Gaussian fitted sizes for (c) major axis and (d) minor axis of model B. 
In each panel, the red curve, red dashed line, and black dashed line are the Gaussian  fitting result, mean fitted value, and standard deviation, respectively. 
}
\label{modelA}
\end{figure*}

For model A ($0\dotarcsec29 \times 0\dotarcsec043$), all the 100 fittings and deconvolution processes are successful (i.e., successful rate $= 100\%$). 
The top panels of Figure~\ref{modelA} show that the mean fitting size ($0\dotarcsec291 \times 0\dotarcsec044$) is consistent with or very close to the size of model A ($0\dotarcsec29 \times 0\dotarcsec043$). In these panels, we also found that the standard deviation of the source size is as small as 0$\dotarcsec$003 for the major axis and 0$\dotarcsec$003 for the minor axis, which is comparable to the fitting errors given by the CASA task “\texttt{imfit}” ($\sim$1/40 of the synthesised beam size).

For model B ($0\dotarcsec048\times0\dotarcsec07$), we performed the Gaussian fitting 100 times as for model A. 
We found that 35\% (35/100) of the fitting cases failed to derive the source size, meaning they are unresolved sources. 
Thus, we could not determine the upper limit of the size by Gaussian fitting for these cases. 
The bottom panels of Figure~\ref{modelA} show that, among the 65 successful fitting cases, the derived mean values of both the major and minor axes ($0\dotarcsec044\times0\dotarcsec12$) are inconsistent with the original model size ($0\dotarcsec048\times0\dotarcsec07$). In particular, for the minor axis, the derived mean value is larger than the original size by a factor of $\sim$2. 
In addition, we derived the standard deviation of the source size of $0\dotarcsec008$ for the major axis and $0\dotarcsec011$ for the minor axis, which are a factor of $\sim$3$-$4 larger than those of model A. 
Thus, this experiment indicates that we cannot reasonably fit the source when the minor axis of the source is as small as 1/20 of the synthesised beam size.

The two experiments clearly show that even at the larger inclination angle, we can reasonably fit the source as long as the minor axis is larger than or comparable to 1/3 of the synthesised beam size. On the other hand, we cannot reliably fit the source when the minor axis is smaller than 1/20 of the synthesised beam size. 

It should be noted that among the sources listed in Table \ref{tab:Summary of the fitting results}, most of the sources have a minor axis either comparable to or larger than 1/3 of the beam size, except for the most compact source MMS\,2-North-B. 
MMS\,2-North-B has a minor axis of about 1/20 of the synthesized beam size and is the only unresolved source among all our sources. 
In addition, MMS\,2-North-B is the only source showing fitting errors larger than the fitted deconvolved minor axis. In our fitting, MMS\,2-North-B has an extremely small minor axis (1/20 of the synthesized beam size), resulting in a very large inclination angle. 
Thus, MMS\,2-North-B is the only source we consider that the fitted sizes cannot be reliable. We added a footnote indicating that this source is not reliably resolved in Table \ref{tab:Summary of the fitting results}. Based on our experiments, all the other sources except for MMS\,2-North-B have reliable fitting values. 

\clearpage
\section{Image of CO~($J$ = 2$-$1) emission}\label{co-image}
\setcounter{figure}{0}
\renewcommand{\thefigure}{B\arabic{figure}}
\renewcommand{\theHfigure}{B\arabic{figure}}

\begin{figure*}[ht!]
\centering
\includegraphics[scale=0.37]{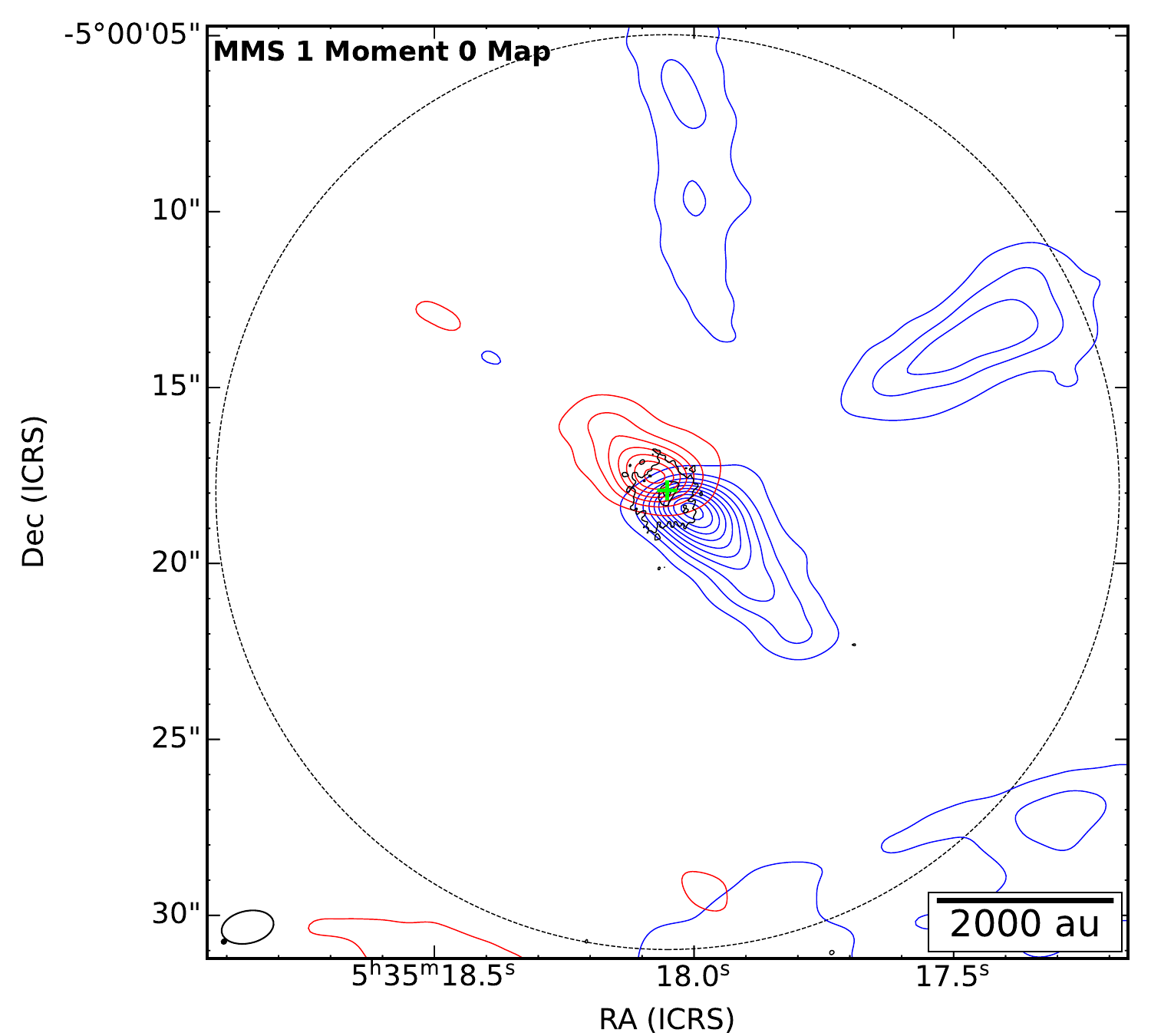}
\caption{Integrated intensity image of CO~($J$ = 2$-$1) emission for MMS\,1. The blue and red contours show the integrated intensity of the CO emission obtained from Takahashi et al.\,(2023a, submitted to ApJ). The CO emission is integrated from $v_{\textnormal{LRS}}=-45$ to $-5$ km\,s$^{-1}$ (blue) and from $v_{\textnormal{LRS}}=20$ to 75 km\,s$^{-1}$ (red), respectively. The blue and red contour levels are [5, 15, 25, 35, 45, 55, 65, 75, 85, 95]$\times\sigma$ (1$\sigma$ = 0.12 Jy\,beam$^{-1}$). The black contours show the Stokes $I$ obtained from the ALMA 1.1 mm dust continuum data and the contour levels are [5, 25]$\times\sigma$ (1$\sigma$ = 23.4 $\mu$Jy\,beam$^{-1}$). The green cross denotes the peak position of Stokes $I$ for MMS\,1. The circle in black dashed line shows the primary beam size (FWHM) of $\sim$26$\arcsec$ for the CO observations. The synthesized beam size for continuum is denoted by a filled dot, and for
CO, it is denoted by an empty ellipse in the bottom left corner. }
\label{mom0-mms1}
\end{figure*}

\begin{figure*}[ht!]
\centering
\includegraphics[scale=1]{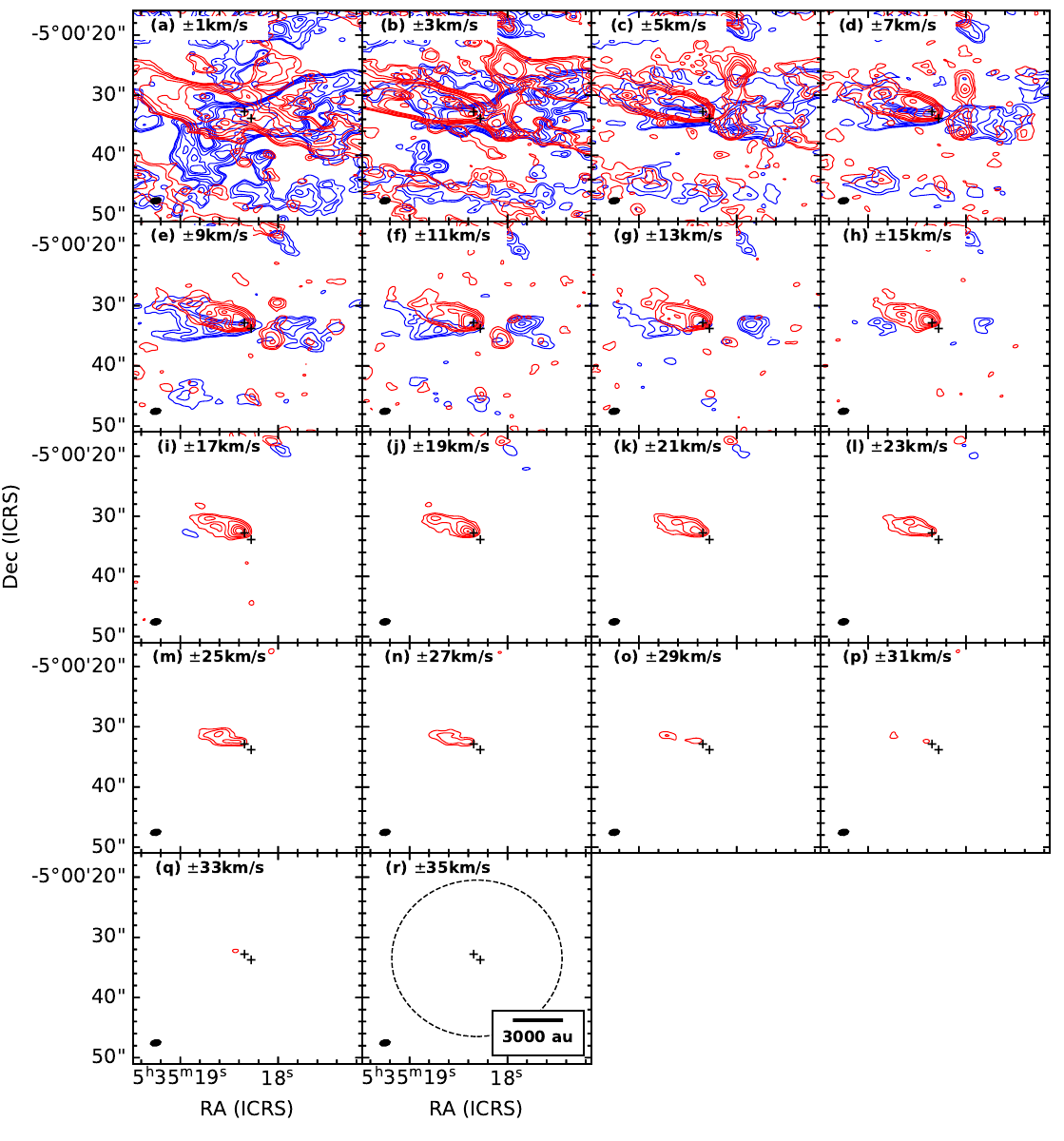}
\caption{Channel maps of CO~($J$ = 2$-$1) emission for MMS\,2. The absolute velocity of $v_{\textnormal {LSR}}-v_{\textnormal {sys}}$ is shown in the upper left corner of each panel, where the $v_{\textnormal{sys}}$ = 11\,km\,s$^{-1}$ \citep[$v_{\textnormal{sys}}$ = 9$-$12.5 km\,s$^{-1}$ is obtained from][and we adopted an average value of $v_{\textnormal{sys}}$ = 11\,km\,s$^{-1}$]{williams2003high}. The two black crosses indicate the peak positions of Stokes $I$ for MMS\,2-North-A and MMS\,2-South, respectively. The blue and red contour levels are [5, 10, 20, 40, 80, 160, 320, 360, 400]$\times\sigma$ (1$\sigma$= 5.8 mJy\,beam$^{-1}$). The synthesized beam size is denoted by a filled ellipse in the bottom left corner of each panel. The circle in panel (r) shows the primary beam size (FWHM) of $\sim$26$\arcsec$.}
\label{channel-mms2}
\end{figure*}

\begin{figure*}[ht!]
\centering
\includegraphics[scale=0.37]{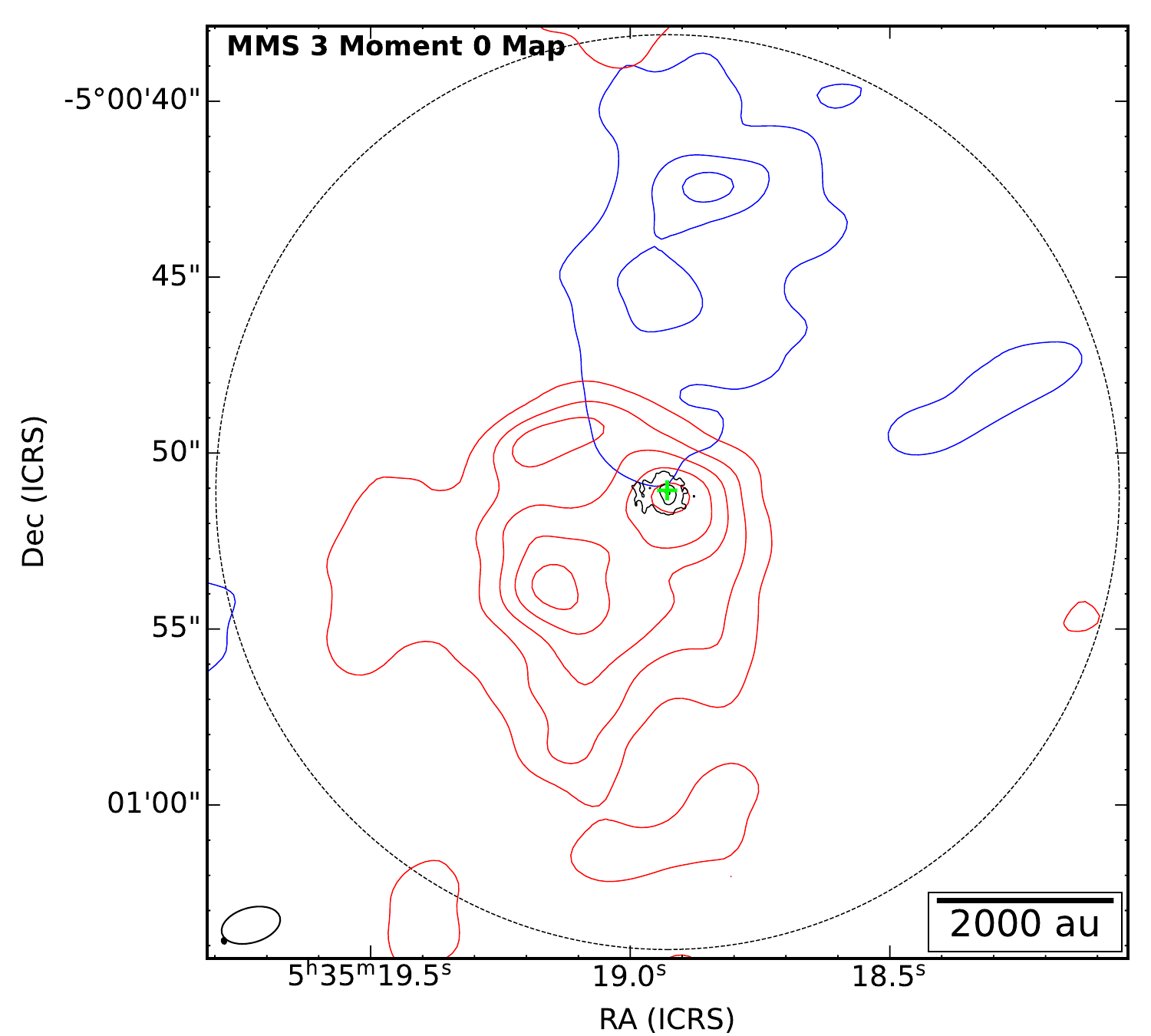}
\caption{Integrated intensity image of CO~($J$ = 2$-$1) emission for MMS\,3. The blue and red contours denote the integrated intensity of the CO emission obtained from \cite{morii2021revealing}. The CO emission is integrated from $v_{\textnormal{LRS}}=-10.8$ to 7.2 km\,s$^{-1}$ (blue) and from $v_{\textnormal{LRS}}=$ 13.2 to 41.2 km\,s$^{-1}$ (red), respectively. The blue and red contour levels are [5, 15, 25, 35, 45]$\times\sigma$ (1$\sigma$ = 0.145 Jy\,beam$^{-1}$). The black contours denote the Stokes $I$ obtained from the ALMA 1.1 mm dust continuum data and the contour levels are [5, 25]$\times\sigma$ (1$\sigma$ = 25.0 $\mu$Jy\,beam$^{-1}$). The green cross denotes the peak position of Stokes $I$ for MMS\,3. The circle in black dashed line shows the primary beam size (FWHM) of $\sim$26$\arcsec$ for the CO observations. The synthesized beam size for continuum is denoted by a filled dot, and for CO, it is denoted by an empty ellipse in the bottom left corner. }
\label{mom0-mms3}
\end{figure*}

\begin{figure*}[ht!]
\centering
\includegraphics[scale=0.37]{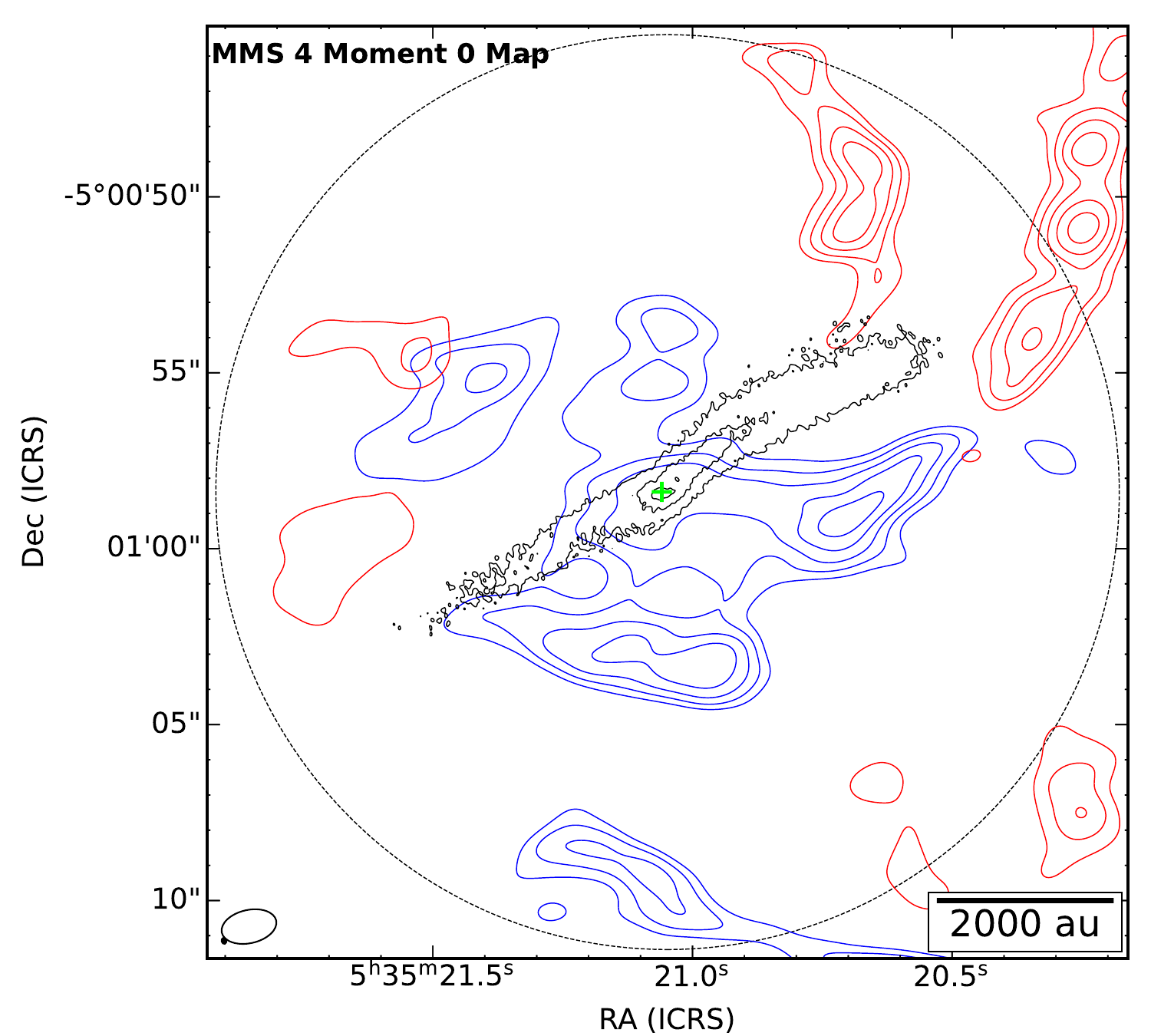}
\caption{Integrated intensity image of CO~($J$ = 2$-$1) emission for MMS\,4. The blue and red contours denote the integrated intensity of the CO emission. The CO emission is integrated from $v_{\textnormal{LRS}}$ = 6 to 7 km\,s$^{-1}$ (blue) and from $v_{\textnormal{LRS}}$ = 14 to 15 km\,s$^{-1}$ (red), respectively. The blue and red contour levels are [5, 7, 9, 11]$\times\sigma$ (1$\sigma$ = 0.12 Jy\,beam$^{-1}$). The black contours denote the Stokes $I$ and the contour levels are [5, 15, 25]$\times\sigma$ (1$\sigma$ = 23.4 $\mu$Jy\,beam$^{-1}$). The green cross denotes the peak position of Stokes $I$ obtained from the ALMA 1.1 mm dust continuum data for MMS\,4. The circle in black dashed line shows the primary beam size (FWHM) of $\sim$26$\arcsec$ for the CO observations. The synthesized beam size for continuum is denoted by a filled dot, and for CO, it is denoted by an empty ellipse in the bottom left corner.}
\label{mom0-mms4}
\end{figure*}

\begin{figure*}[ht!]
\includegraphics[scale=1]{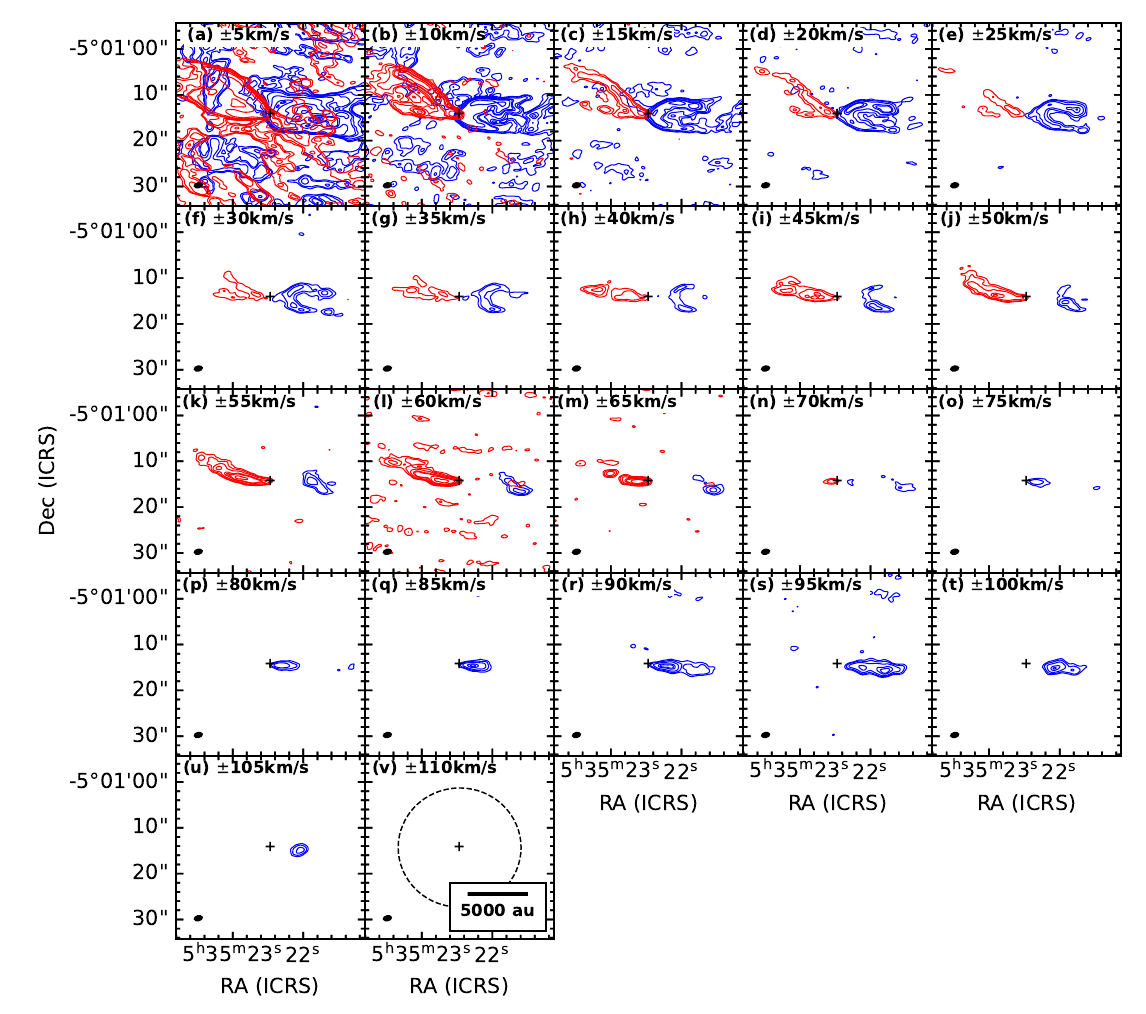}
\caption{Channel maps of CO~($J$ = 2$-$1) emission for MMS\,5. The absolute velocity of $v_{\textnormal {LSR}}-v_{\textnormal {sys}}$ is shown in the upper left corner of each panel, where $v_{\textnormal{sys}}$ = 11 km\,s$^{-1}$ \citep{matsushita2019very}. The black cross indicates the peak position of Stokes $I$ for MMS\,5. The blue and red contour levels are [5, 10, 20, 40, 80, 160, 320, 360, 400]$\times\sigma$ (1$\sigma$= 3.9 mJy\,beam$^{-1}$). The synthesized beam size is denoted by a filled ellipse in the bottom left corner of each panel. The circle in panel (v) shows the primary beam size (FWHM) of $\sim$26$\arcsec$. }
\label{channel-mms5}
\end{figure*}

\begin{figure*}[ht!]
\hspace{-1.5\baselineskip}
\includegraphics[scale=0.65]{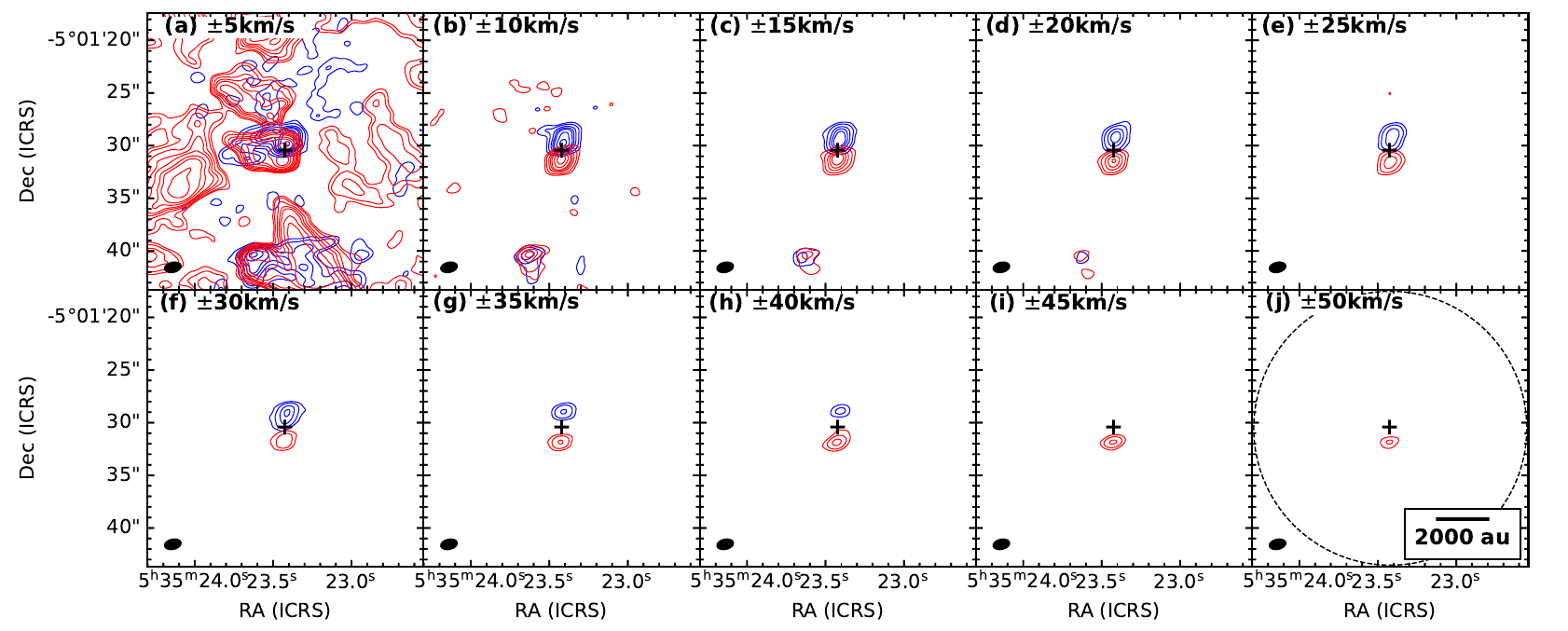}
\caption{Channel maps of CO~($J$ = 2$-$1) emission for MMS\,6. The absolute velocity of $v_{\textnormal {LSR}}-v_{\textnormal {sys}}$ is shown in the upper left corner of each panel, where $v_{\textnormal {sys}}=10.3$ km\,s$^{-1}$ \citep{takahashi2012molecular}. The black cross indicates the peak position of Stokes $I$ for MMS\,6. The blue and red contour levels are [5, 10, 20, 25, 30, 40, 80, 120, 160]$\times\sigma$ (1$\sigma$= 3.9 mJy\,beam$^{-1}$). The synthesized beam size is denoted by a filled ellipse in the bottom left corner of each panel. The circle in panel (j) shows the primary beam size (FWHM) of $\sim$26$\arcsec$.}
\label{channel-mms6}
\end{figure*}

\begin{figure*}[ht!]
\includegraphics[scale=0.8]{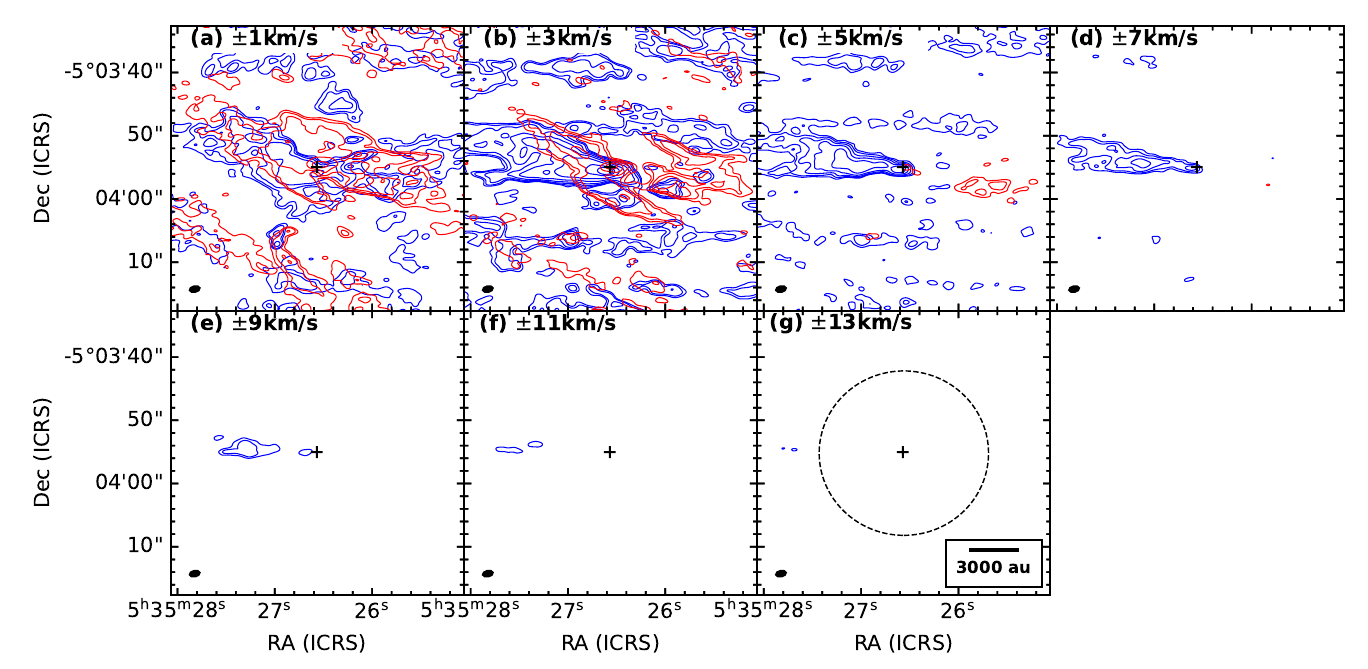}
\caption{Channel maps of CO~($J$ = 2$-$1) emission for MMS\,7. The absolute velocity of $v_{\textnormal {LSR}}-v_{\textnormal {sys}}$ is shown in the upper left corner of each panel, where $v_{\textnormal {sys}}=10.6$ km\,s$^{-1}$ \citep[e.g.,][]{reipurth1999vla,takahashi2006millimeter}. The black cross indicates peak position of Stokes $I$ for MMS\,7-North. The blue and red contour levels are [5, 10, 20, 40, 80, 160, 200, 320, 400]$\times\sigma$ (1$\sigma$= 5.8 mJy\,beam$^{-1}$). The synthesized beam size is denoted by a filled ellipse in the bottom left corner of each panel. The circle in panel (g) shows the primary beam size (FWHM) of $\sim$26$\arcsec$.}
\label{channel-mms7}
\end{figure*}

We present CO integrated intensity images for the non-polarized detected sources MMS\,1, MMS\,3, and MMS\,4 in Figure~\ref{mom0-mms1}, Figure~\ref{mom0-mms3}, and Figure~\ref{mom0-mms4}, respectively. Figure~\ref{mom0-mms3} for MMS\,3 and Figure~\ref{mom0-mms1} for MMS\,1 are obtained from \cite{morii2021revealing} and Takahashi et al. (2023a, submitted to ApJ), respectively. We also present the CO channel maps for MMS\,2, MMS\,5, MMS\,6, and MMS\,7 in Figure~\ref{channel-mms2}, Figure~\ref{channel-mms5}, Figure~\ref{channel-mms6}, and Figure~\ref{channel-mms7}, respectively. The CO integrated intensity images for these four sources are presented in Section~\ref{results}. The placement of these figures is determined by their first mention in Section~\ref{results}.

\clearpage
\section{Extended polarized emission associated with MMS 1}\label{mms1poli}
In order to verify if we could detect the polarized emission associated with the extended component (envelope). We optimize the S/N of extended emission towards MMS\,1 by using a parameter $uv$-taper in the CASA task ``\texttt{tclean}'', where a taper of 300 k$\lambda$ has been applied. Figure~\ref{mms1-poli} shows that PI is detected from the envelope in the southwest with respect to the Stokes $I$ peak. The partial detection of the polarization in the envelope may indicate that current sensitivity allows us to detect a part of extended polarized emission likely associated with the envelope. 

\setcounter{figure}{0}
\renewcommand{\thefigure}{C\arabic{figure}}
\renewcommand{\theHfigure}{C\arabic{figure}}
\begin{figure*}[ht!]
\centering
\includegraphics[scale=0.36]{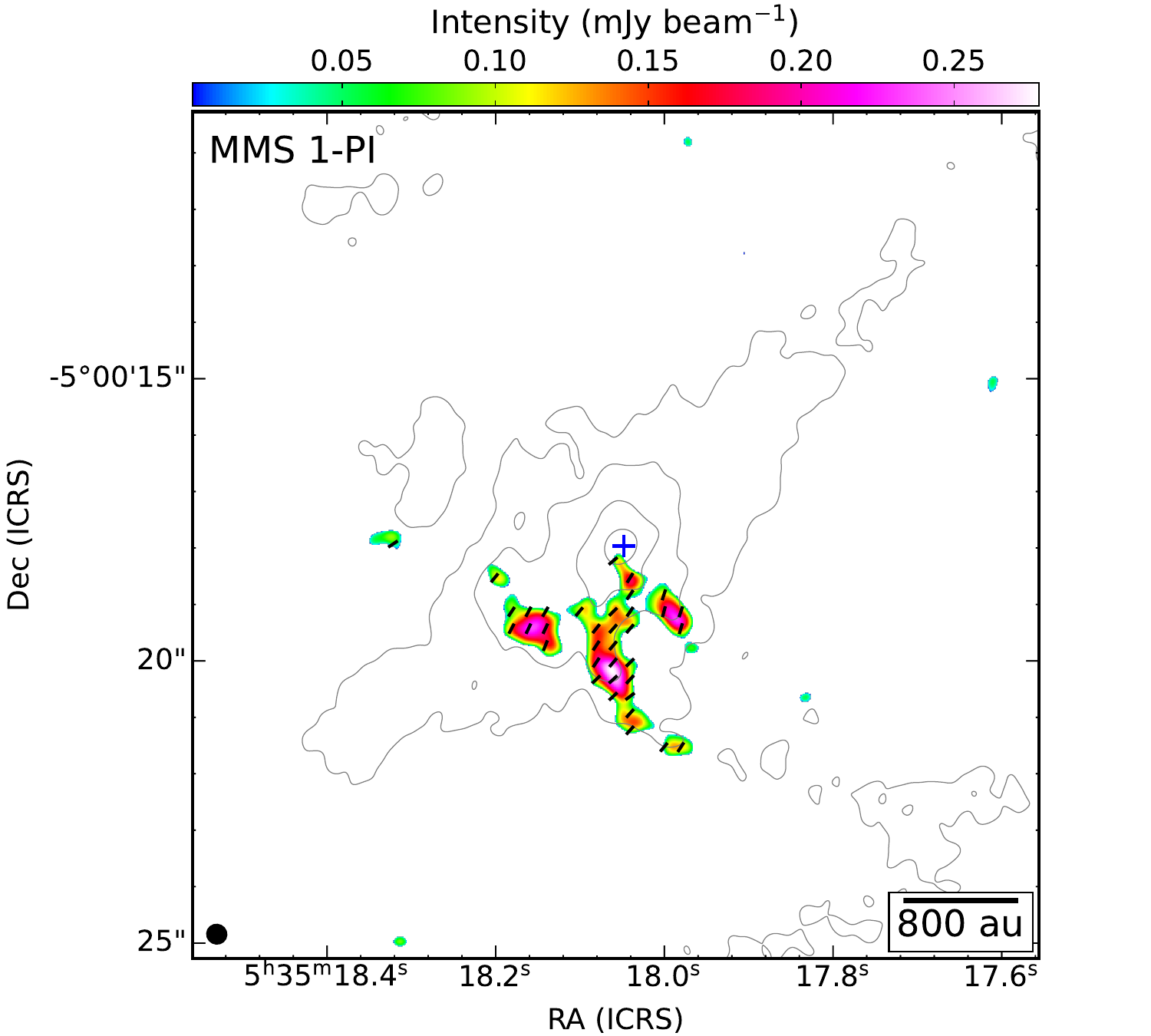}
\caption{PI (colour) overlaid with $E$-vectors (black line segments) and contours of Stokes $I$ (grey) for MMS\,1 from the ALMA 1.1\,mm dust continuum data. The blue cross indicates the peak position of Stokes $I$. The contour levels are [3, 9, 27, 81]$\times\sigma$ (1$\sigma$ = 0.11 mJy\,beam$^{-1}$). The synthesized beam size of $0\dotarcsec35\times0\dotarcsec34$ is denoted by a filled dot in the bottom left corner.}
\label{mms1-poli}
\end{figure*}

\clearpage
\section{Bridging structure associated with mms 2}\label{mms2natural}
We made the image with the natural weighting in CASA task ``\texttt{tclean}'' to improve the S/N of the extended structures towards MMS\,2. The spiral arm-like structures are detected at 3$\sigma$ level of the Stokes $I$ emission from both MMS\,2-North-A and MMS\,2-South as shown in Figure~\ref{fig-stokesIQU-3}\,(d). These two spiral arm-like structures are connected as a bridge as shown in the Figure~\ref{mms2-natural}. This bridge may associated with the gas flow and material feeding between these two sources.

\setcounter{figure}{0}
\renewcommand{\thefigure}{D\arabic{figure}}
\renewcommand{\theHfigure}{D\arabic{figure}}
\begin{figure*}[ht!]
\centering
\includegraphics[scale=0.37]{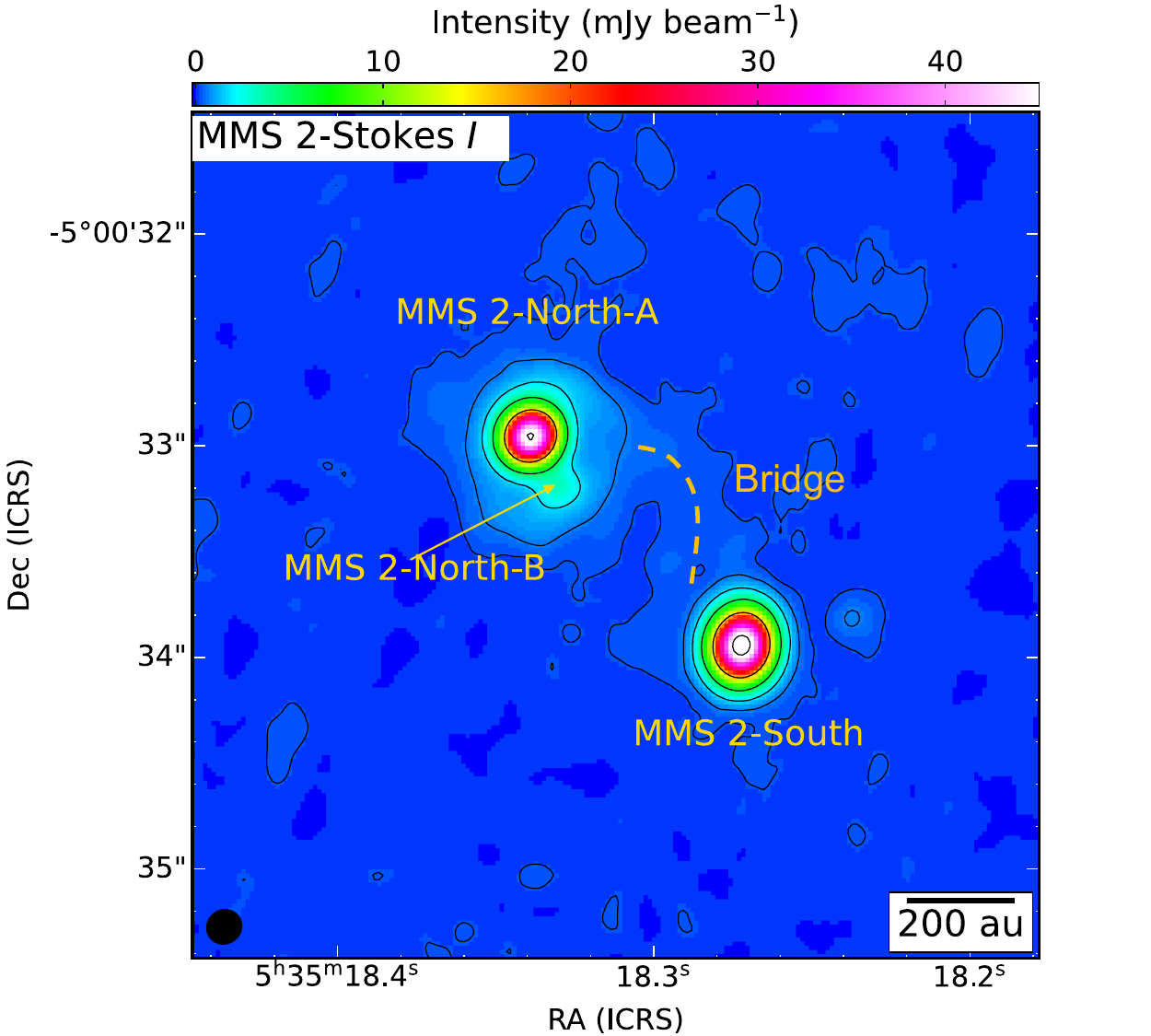}
\caption{The colour and black contours show the Stokes $I$ of MMS\,2 from the ALMA 1.1\,mm dust continuum data. The contour levels are [3, 9, 27, 81, 243, 729]$\times\sigma$ (1$\sigma$ = 6.6 $\mu$Jy\,beam$^{-1}$). The synthesized beam size of $0\dotarcsec17\times0\dotarcsec16$ is denoted by a filled dot in the bottom left corner.}
\label{mms2-natural}
\end{figure*}

\end{appendix}



\end{document}